\documentclass[aip,jcp,reprint,noshowkeys,superscriptaddress]{revtex4-1}
\usepackage{graphicx,dcolumn,bm,xcolor,microtype,multirow,amscd,amsmath,amssymb,amsfonts,physics,wrapfig,txfonts,siunitx}
\usepackage[version=4]{mhchem}
\usepackage{comment}

\usepackage[normalem]{ulem}

\newcommand{\SupInf}{\textcolor{blue}{Supporting Information}}

\newcommand{\mc}{\multicolumn}

\newcommand{\QP}{\textsc{quantum package}}

\newcommand{\Na}{N^{\uparrow}}
\newcommand{\Nb}{N^{\downarrow}}
\newcommand{\Nd}{N_\text{d}}
\newcommand{\Nv}{N_\text{v}}
\newcommand{\Nas}{N_\text{s}^{\uparrow}}
\newcommand{\Nbs}{N_\text{s}^{\downarrow}}

\newcommand{\Ndet}{N_\text{det}}

\usepackage[
	colorlinks=true,
    citecolor=blue,
    breaklinks=true
	]{hyperref}
\begin{document}

\newcommand{\LCPQ}{Laboratoire de Chimie et Physique Quantiques (UMR 5626), Universit\'e de Toulouse, CNRS, UPS, France}

\title{Seniority and Hierarchy Configuration Interaction for Radicals and \\ Excited States}

\author{F\'abris Kossoski}
\email{fkossoski@irsamc.ups-tlse.fr}
\affiliation{\LCPQ}
\author{Pierre-Fran\c{c}ois Loos}
\email{loos@irsamc.ups-tlse.fr}
\affiliation{\LCPQ}

\begin{abstract}
Hierarchy configuration interaction (hCI) has been recently introduced as an alternative configuration interaction (CI) route combining excitation degree and seniority number, which showed to efficiently recover both dynamic and static correlations for closed-shell molecular systems [\href{https://doi.org/10.1021/acs.jpclett.2c00730}{\textit{J.~Phys.~Chem.~Lett.}~\textbf{2022}, \textit{13}, 4342}].
Here, we generalize hCI for an arbitrary reference determinant, allowing calculations for radicals and for excited states in a state-specific way.
We gauge this route against excitation-based CI (eCI) and seniority-based CI (sCI) by evaluating how different ground-state properties of radicals converge to the full CI limit.
We find that hCI outperforms or matches eCI, whereas sCI is far less accurate, in line with previous observations for closed-shell molecules.
Employing the second-order Epstein-Nesbet (EN2) perturbation theory as a correction significantly accelerates the convergence of hCI and eCI.
We further explore various hCI and sCI models to calculate excitation energies of closed- and open-shell systems.
Our results underline that both the choice of the reference determinant and the set of orbitals drive the fine balance between correlation of ground and excited states.
State-specific hCI2 and higher order models perform similarly to their eCI counterparts, whereas lower orders of hCI deliver poor results,
unless supplemented by the EN2 correction, which substantially improves their accuracy.
In turn, sCI1 produces decent excitation energies for radicals, encouraging the development of related seniority-based coupled-cluster methods.
\bigskip
\begin{center}
        \boxed{\includegraphics[keepaspectratio,width=3.25in]{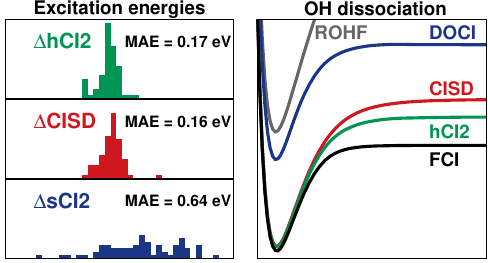}}
\end{center}
\bigskip
\end{abstract}

\maketitle

\section{Introduction}
\label{sec:intro}

Configuration interaction (CI) offers a systematic way to solve the many-body electronic structure problem. \cite{SzaboBook,Helgakerbook}
By including progressively more determinants in the Hilbert space, the wave function becomes increasingly closer to the exact one, and so does the electronic energy.
In full CI (FCI), all determinants are accounted for and the problem is solved exactly (for a given basis set).
In practice, however, one resorts to approximate CI models, where only the determinants that satisfy a given criterion are included in the truncated Hilbert space.

The most well-known CI route is based on the excitation degree $e$.
Starting from a reference determinant, typically the Hartree-Fock (HF) determinant, one generates all connected determinants by exciting at most $e$ electrons.
The excitation degree thus defines the order of the approximate excitation-based CI (eCI) model:
CI with single excitations (CIS), CI with single and double excitations (CISD), CI with single, double, and triple excitations (CISDT), etc.
The eCI route rather quickly captures dynamic (weak) correlation, though it struggles with the description of static (strong) correlation.

A different CI route is based on the seniority number $s$ (the number of unpaired electrons in a given determinant).
In seniority-based CI (sCI), \cite{Bytautas_2011,Allen_1962,Smith_1965,Veillard_1967} there is no reference determinant and one accounts for all determinants having seniority equal or less than $s$.
In contrast to eCI, sCI recovers static correlation more efficiently and works well to describe molecular dissociation, \cite{Bytautas_2015,Alcoba_2014,Alcoba_2014b}
at the expense of a poorer account of dynamic correlation and a higher computational cost.
For systems with an even number of electrons, the first approximate model is defined by $s=0$ (sCI0), usually referred to as doubly-occupied CI (DOCI),
which is followed by the higher-order models: sCI2, sCI4, \ldots.
For odd numbers of electrons, the seniority route follows along odd numbers of $s$: sCI1, sCI3, and so on.

We have recently introduced a third CI route, hierarchy CI (hCI), \cite{Kossoski_2022}
where the Hilbert space is partitioned according to a hierarchy parameter $h$ that combines the excitation degree $e$ and the seniority number $s$ defined as $h = (e+s/2)/2$.
This definition ensures that all classes of determinants whose number share the same scaling with system size are included at the same hierarchy $h$.
This key feature distinguishes hCI from previous schemes combining excitation and seniority. \cite{Alcoba_2014,Raemdonck_2015,Alcoba_2018}
By allowing for higher-order excitations of paired electrons (as explained in detail below), hCI is reminiscent of perfect pairing models for closed-shell systems. \cite{Hurley_1953,Cullen_1996,VanVoorhis_2000,Parkhill_2009,Parkhill_2010,Lehtola_2016,Lehtola_2018}
For different properties and closed-shell molecular systems,
hCI was found to display an overall faster convergence toward the FCI results than the traditional eCI route. \cite{Kossoski_2022}
In this sense, it was able to recover both static and dynamic correlations more effectively than either eCI or sCI.
However, hCI has only been defined for a closed-shell reference, thus being limited to the ground state of systems with an even number of electrons.

There are two possible approaches to target excited states with CI methods.
One can employ the ground-state HF orbitals and obtain excitation energies from the higher-lying eigenvalues of the CI matrix, which we refer to as the ground-state-based approach.
Instead, one may optimize the orbitals for the excited state of interest (described with an appropriate reference), followed by a separate CI calculation,
in a so-called state-specific approach ($\Delta$CI).
There has been a recent surge in the development of state-specific methods, covering
single-reference and multiconfigurational self-consistent field,
\cite{Ziegler_1977,Burton_2021,Shea_2018,Tran_2019,Tran_2020,Hardikar_2020,Shea_2020,Burton_2021,Burton_2022,Hanscam_2022,Kossoski_2023,Marie_2023,Tran_2023}
density-functional theory,
\cite{Filatov_1999,Kowalczyk_2011,Kowalczyk_2013,Gilbert_2008,Barca_2018,Hait_2020,Hait_2021,Zhao_2020,Levi_2020,Carter-Fenk_2020,Toffoli_2022,Schmerwitz_2022,Schmerwitz_2023}
perturbation theory, \cite{Clune_2020,Zhao_2020b,Clune_2023}
quantum Monte Carlo, \cite{Scemama_2018a,Scemama_2018b,Dash_2018,Dash_2019,Dash_2021,Cuzzocrea_2022,Shepard_2022,Otis_2020,Otis_2023}
and coupled-cluster (CC)
\cite{Piecuch_2000,Mayhall_2010,Lee_2019,Kossoski_2021,Marie_2021,Rishi_2023,Tuckman_2023}
methods.
In particular, by employing a minimal configuration state function (CSF) reference,
we have recently shown that excitation-based $\Delta$CI models deliver far more accurate excitation energies than their ground-state-based analogs. \cite{Kossoski_2023}

Here, our first goal is to generalize hCI for an arbitrary type of reference, thus extending its applicability from ground-state closed-shells \cite{Kossoski_2022}
to radicals and state-specific excited-state calculations.
This is done in Sec.~\ref{sec:hCI}.
The computational details about the implementation and the specific calculations performed here are presented in Sec.~\ref{sec:compdet}.
In Sec.~\ref{sec:res_A}, we assess the performance of hCI, eCI, and sCI models by calculating various properties of four ground-state radicals, which comprises our second goal.
Our third goal is to evaluate how perturbation theory, more precisely the second-order Epstein-Nesbet (EN2) perturbative correction (both standard and renormalized), \cite{Garniron_2019}
impacts various properties of ground-state closed-shell systems and radicals, for hCI, eCI, and sCI models.
This part is presented in Sec.~\ref{sec:res_B}.
Inspired by the results of hCI for ground-state closed-shell systems, \cite{Kossoski_2022}
and by the promising set of excitation-based $\Delta$CI, \cite{Kossoski_2023} here we explore hCI models for excited states, following both the ground-state-based and state-specific approaches.
In this sense, our fourth goal is to assess the accuracy of hCI models for excited states, which is the subject of Sec.~\ref{sec:res_C}.
Furthermore, to the best of our knowledge, sCI models have not yet been directly used to target excited states,
despite the growing number of methods that exploit the concept of seniority,
for both ground
\cite{Limacher_2013,Limacher_2014,Tecmer_2014,Boguslawski_2014a,Boguslawski_2015,Boguslawski_2014b,Boguslawski_2014c,Johnson_2020,Henderson_2014,Stein_2014,Henderson_2015,Chen_2015,Bytautas_2018,Marie_2021,Boguslawski_2021,Tecmer_2022,Mamache_2023,Fecteau_2022}
and excited states.
\cite{Boguslawski_2016b,Boguslawski_2016c,Boguslawski_2019,Nowak_2019,Kossoski_2021,Marie_2021,Tecmer_2022,Rishi_2023,Nowak_2023,Fecteau_2022}
Our fifth goal, detailed in Sec.~\ref{sec:res_D}, is therefore to define and gauge ground-state-based and state-specific sCI models for excited states.
Finally, in Sec.~\ref{sec:res_E}, we assess how the excitation energies are impacted by the EN2 perturbative correction to hCI and sCI models, our sixth and last goal.
Section \ref{sec:conclusion} closes the present contribution with the main conclusions and perspectives.

\section{Hierarchy configuration interaction}
\label{sec:hCI}

\begin{figure*}
\includegraphics[width=0.05\linewidth]{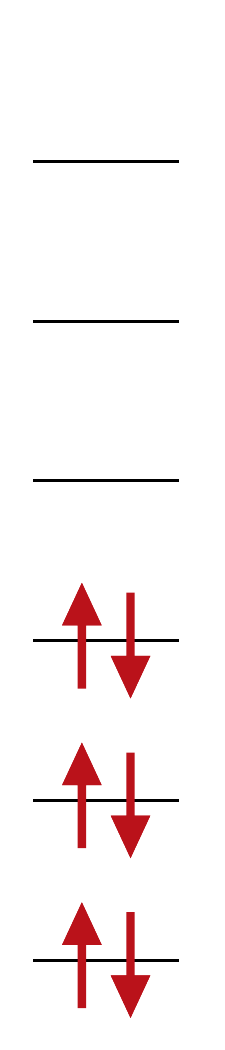}
\includegraphics[width=0.25\linewidth]{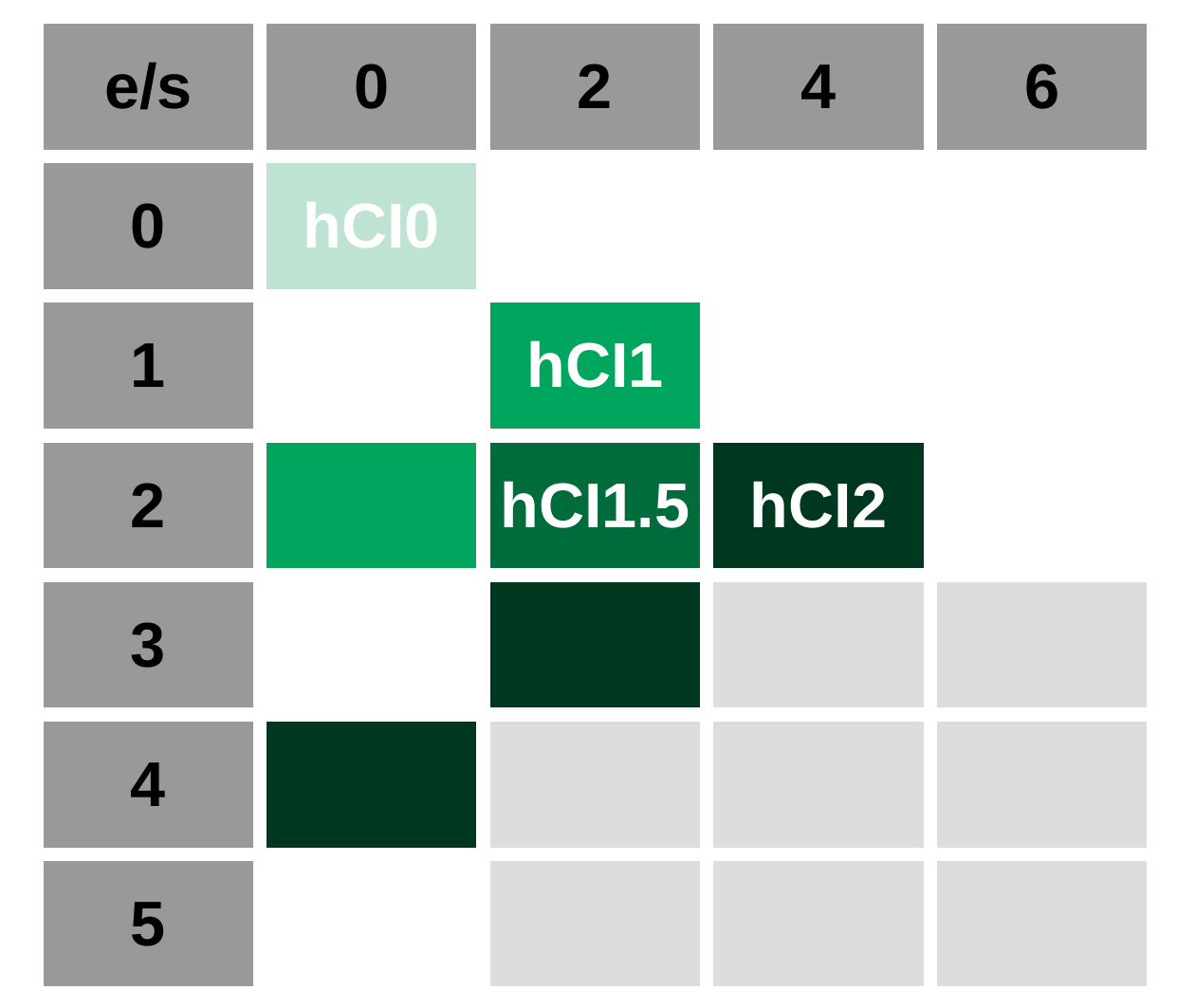}
\hspace{0.03\linewidth}
\includegraphics[width=0.05\linewidth]{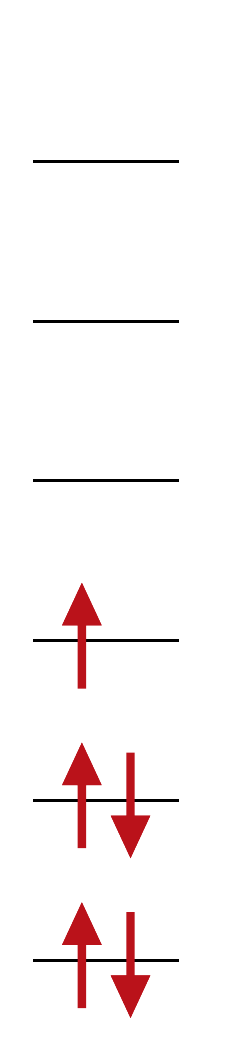}
\includegraphics[width=0.25\linewidth]{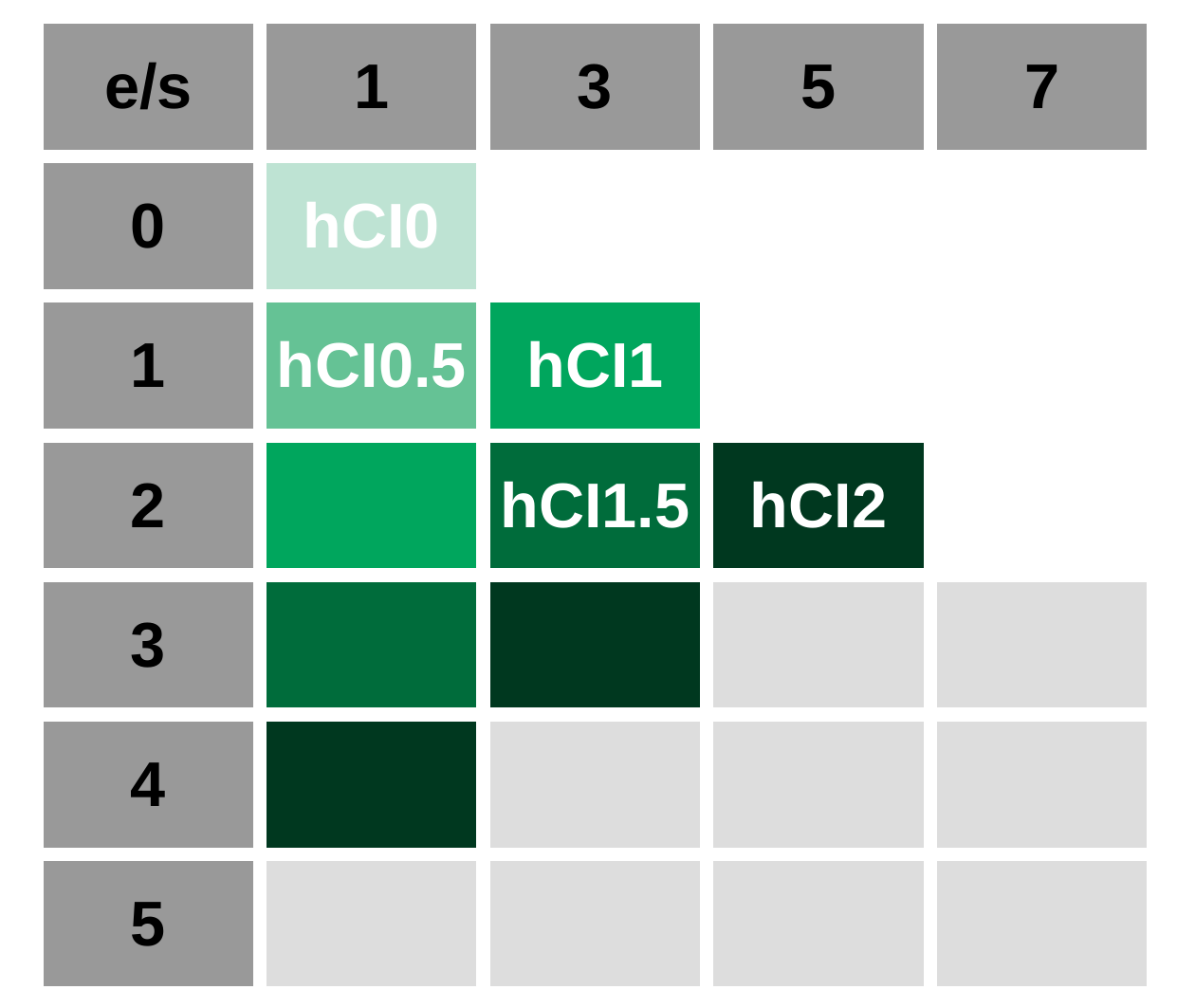}
\hspace{0.03\linewidth}
\includegraphics[width=0.05\linewidth]{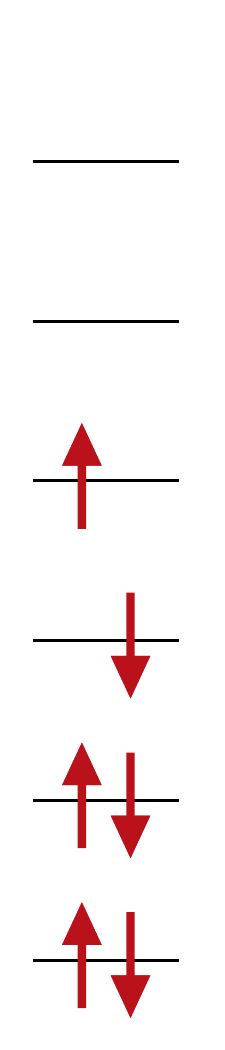}
\includegraphics[width=0.25\linewidth]{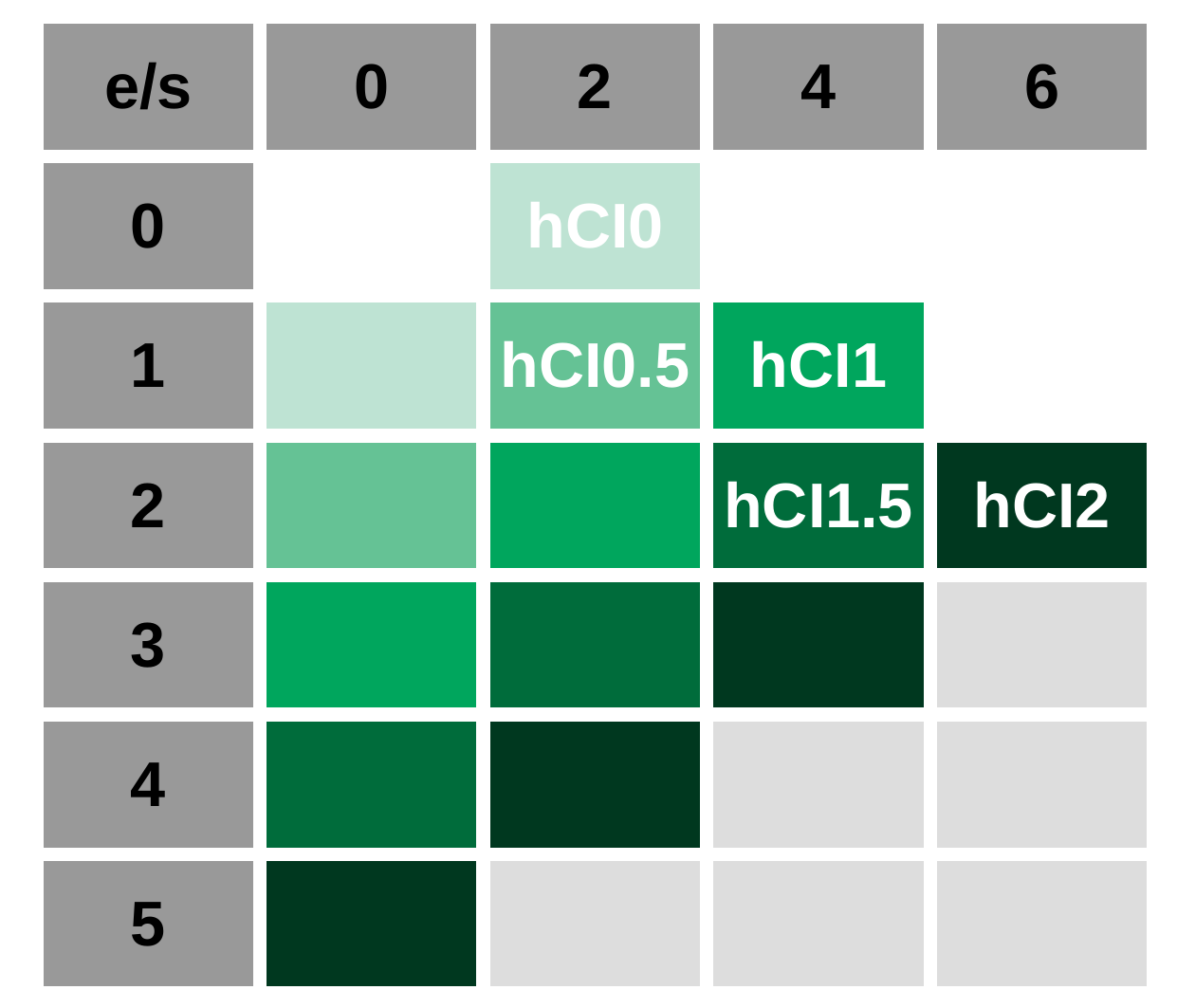}
\caption{Partitioning of the Hilbert space according to the seniority number $s$ and the excitation degree $e$ with respect to a given reference determinant (shown by its side),
for a closed-shell reference (left), an open-shell reference with one unpaired electron (center) and an open-shell reference with two unpaired electrons (right).
The color tones represent the determinants that are included at a given hCI model. hCI0 reduces to the reference determinant (usually the HF solution) in the former two cases, where $s_0 = 0$ and $s_0 = 1$.}
\label{fig:hCI}
\end{figure*}

We introduced hCI \cite{Kossoski_2022} as a particular truncation of the Hilbert space,
viewed as a two-dimensional map of determinants built from their seniority $s$ and their excitation degree $e$ with respect to a reference determinant, as shown in Fig.~\ref{fig:hCI}.
By defining orbital subspaces according to the occupancy in the reference determinant (doubly-occupied, singly-occupied, or unoccupied),
we refer to the class of an excited determinant as the combination of the number of electrons and seniority in each orbital subspace.
The number of determinants $\Ndet$ within a given class scales polynomially with the number of basis functions $N$, with the exponent depending on the specific class.
For instance, for the class of doubly-excited determinants with no unpaired electrons ($e=2$ and $s=0$), $\Ndet = \order*{N^2}$.
hCI was defined such that, at a given hierarchy $h$, all classes of determinants presenting a scaling of $\Ndet = \order*{N^{2h}}$ or less are accounted for.
This means moving diagonally in the seniority-excitation map (denoted by the color tones in Fig.~\ref{fig:hCI}).
In comparison, eCI spans the map horizontally (top to bottom), whereas sCI does it vertically (left to right).

hCI was initially introduced for a closed-shell reference (left panel of Fig.~\ref{fig:hCI}). \cite{Kossoski_2022}
Here, we generalize it for an arbitrary Slater determinant reference, including systems with an odd number of electrons.
With respect to a given reference determinant, we define the hierarchy $h$ of a candidate determinant to be included in the truncated hCI model as
\begin{equation}
  \label{eq:h}
  h = \frac{e+ (s-s_0)/2}{2},
\end{equation}
where $s$ and $s_0$ denote the seniority of the candidate and reference determinants, respectively, and $e$ represents the excitation degree that connects them.
The definition in Eq.~\eqref{eq:h} guarantees the sought-after relation between the classes of determinants and their scaling.
Namely, all classes whose number of determinants $\Ndet$ share the same scaling with $N$ enter at the hierarchy $h$.
Had we employed the absolute value of $s-s_0$ or discarded $s_0$ in the definition of $h$, this property would not hold.
The term $(s-s_0)/2$ is always an integer, with $s$ and $s_0$ being even (odd) for systems with even (odd) numbers of electrons. The excitation degree $e$ is also an integer.
Therefore, $h$ assumes integer or half-integer values, for any type of reference determinant.
For a given hCI model defined by $h$, we include all the candidate determinants having a hierarchy less than or equal to $h$ with respect to any determinant in the reference.
This definition allows us to build multireference hCI models as well.
Notice that Eq.~\eqref{eq:h} simplifies to the previous definition \cite{Kossoski_2022} for the case of a closed-shell reference determinant where $s_0 = 0$.
hCI can be viewed as a CI model that includes increasingly more dissimilar determinants from a given reference determinant.
The level of dissimilarity is represented by the hierarchy $h$, which accounts for differences in orbital occupation (through the excitation degree $e$)
and differences in the number of unpaired electrons [through the term $(s-s_0)/2$].

\begin{figure*}
\includegraphics[width=1.0\linewidth]{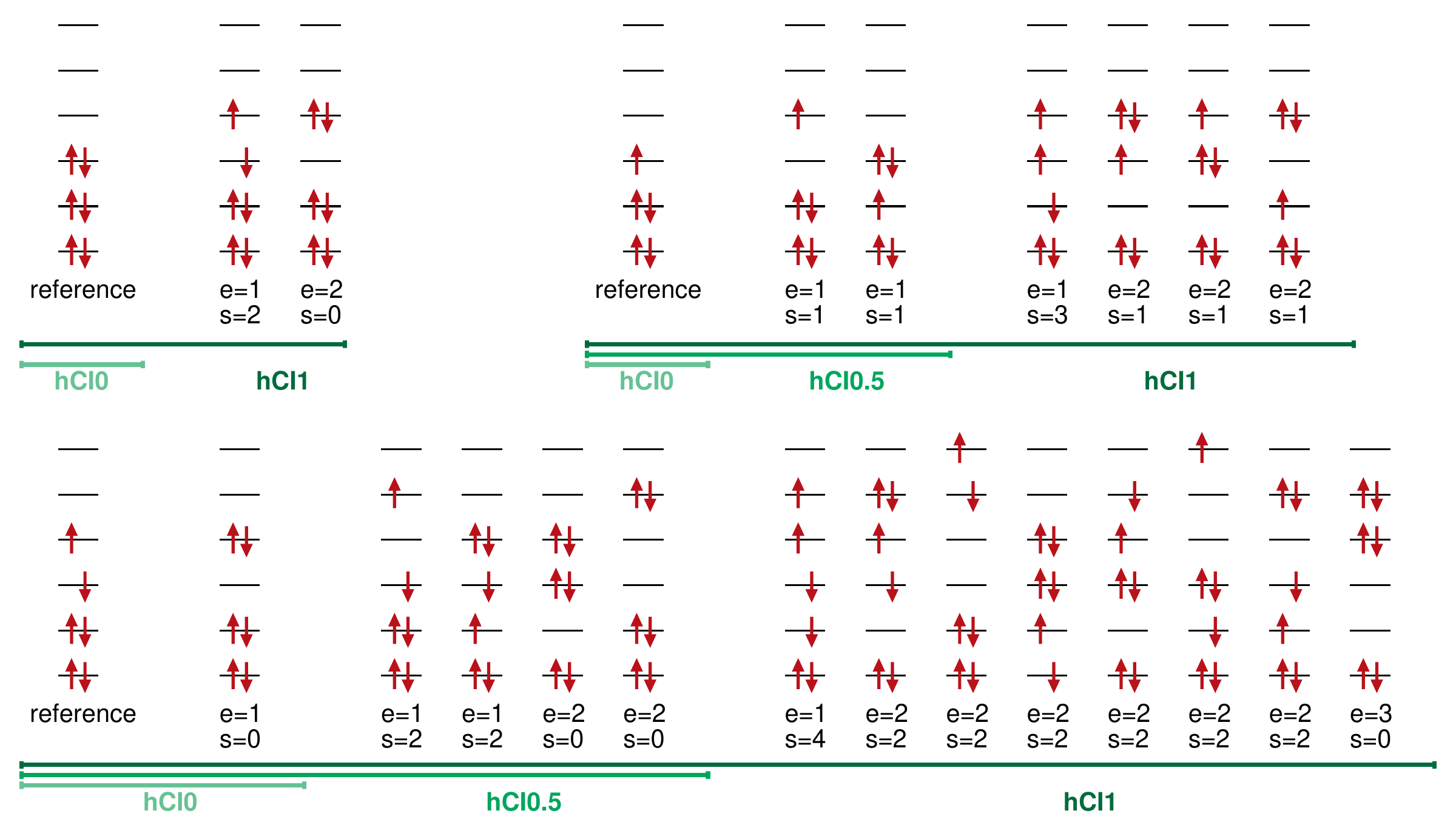}
\caption{Classes of determinants generated in hCI0, hCI0.5, and hCI1, for three different reference determinants:
closed-shell (top left), open-shell with one unpaired electron (top right), and open-shell with two unpaired electrons (bottom).
By dividing the orbitals of the reference determinant into doubly-occupied, singly-occupied, and unoccupied subspaces,
the class of an excited determinant is defined by the number of electrons and seniority in each orbital subspace.}
\label{fig:determinants}
\end{figure*}

We show in Fig.~\ref{fig:hCI} how the seniority-excitation map of determinants is partitioned for a closed-shell reference ($s_0 = 0$), and for open-shell references with one ($s_0 = 1$) or two ($s_0 = 2$) unpaired electrons.
For these three types of references, the classes of determinants included up to hCI1 are presented in Fig.~\ref{fig:determinants}.
The closed-shell reference determinant is the natural choice for describing the ground state of a closed-shell system, as well as their doubly-excited states. \cite{Barca_2018,Lee_2019,Hait_2020,Kossoski_2021,Marie_2021,Kossoski_2023}
Meanwhile, a determinant having one unpaired electron is suitable for an open-shell system with doublet ground and excited states.
Singlet and triplet singly-excited states of closed-shell systems as well as diradicals \cite{Krylov_2006,Horbatenko_2021,Monino_2022} would require
a reference determinant with two unpaired electrons.

In eCI, the number of determinants $\Ndet$ scales polynomially with the number of basis functions $N$ as $N^{2e}$. \cite{SzaboBook}
Likewise, in hCI, $\Ndet$ scales as $N^{2h}$.
The actual computational cost scales as $\order*{N^{2e+2}}$ for eCI, and similarly for hCI, scaling as $\order*{N^{2h+2}}$.
Both the scaling of $\Ndet$ and the computational scaling along the two routes are shown in Table \ref{tab:scaling}.
As for sCI models, the scaling of $\Ndet$ with respect to $N$ is exponential at all orders. \cite{Bytautas_2011,Allen_1962,Smith_1965,Veillard_1967}
This is because excitations of all degrees are included, even at lower orders, such as sCI0 and sCI1.

\begin{table}[ht!]
\caption{Scaling of $\Ndet$ and Computational Scaling (Cost) in Terms of the Number of Basis Functions $N$ for the Hierarchy- and Excitation-Based CI Routes, With and Without the EN2 Perturbative Correction.}
\label{tab:scaling}
\begin{ruledtabular}
\begin{tabular}{ll|ll|cc}
\mc{2}{c|}{Hierarchy-based}   &  \mc{2}{c|}{Excitation-based} &  $\Ndet$ & Cost \\
\hline
hCI1   &            & CIS    &           & $\order*{N^2}$ & $\order*{N^4}$ \\
hCI1.5 &            &        &           & $\order*{N^3}$ & $\order*{N^5}$ \\
\hline
hCI2   & hCI1+EN2   & CISD   & CIS+EN2   & $\order*{N^4}$ & $\order*{N^6}$ \\
hCI2.5 & hCI1.5+EN2 &        &           & $\order*{N^5}$ & $\order*{N^7}$ \\
\hline
hCI3   & hCI2+EN2   & CISDT  & CISD+EN2  & $\order*{N^6}$ & $\order*{N^8}$ \\
hCI3.5 & hCI2.5+EN2 &        &           & $\order*{N^7}$ & $\order*{N^9}$ \\
\hline
hCI4   & hCI3+EN2   & CISDTQ & CISDT+EN2 & $\order*{N^8}$ & $\order*{N^{10}}$ \\
\end{tabular}
\end{ruledtabular}
\end{table}

For a closed-shell reference determinant ($s_0=0$), hCI0 reduces to the reference determinant, usually chosen as the HF one.
The first non-trivial order is hCI1 (see left panel of Fig.~\ref{fig:hCI}), which accounts for all single excitations (as CIS)
plus all paired double excitations (two electrons promoted from the same occupied orbital into the same virtual orbital),
as shown in the top left panel of Fig.~\ref{fig:determinants}.
For both classes of determinants, their number scales as $N^2$, and are thus taken into account at the same hierarchy of hCI ($h=1$ in this case).
An odd number of excitations from a closed-shell reference always leave unpaired electrons, hence the empty blocks at $s=0$ for odd $e$.
hCI1.5 augments the set of hCI1 determinants with the set of double excitations where two electrons are unpaired.
At the next integer order, hCI2 incorporates all classes of determinants where $\Ndet$ scales as $N^4$.
In total, it accounts for all single and double excitations (as CISD), plus the subset of triple excitations that leave only two unpaired electrons,
plus the subset of quadruple excitations where no electrons are unpaired.

For an odd number of electrons, $s$ assumes odd values starting from $s=1$.
The simplest reference is an open-shell determinant with one unpaired electron ($s_0=1$), shown in the center panel of Fig.~\ref{fig:hCI} and in the top right panel of Fig.~\ref{fig:determinants}.
hCI0 also reduces to the reference determinant (the restricted open-shell HF solution being a natural choice).
In contrast to the closed-shell case, here there are no empty blocks in the seniority-excitation map.
Actually, the hCI series displays the hCI0.5 level, accounting only for the single excitations from and into the singly-occupied orbital.
When HF orbitals are employed, these excitations do not connect with the reference, and in this case, hCI0.5 provides the same energy as restricted open-shell HF.
More generally, since there are more types of orbitals for the open-shell reference (doubly-occupied, singly-occupied, and unoccupied) than for the closed-shell reference (doubly-occupied or unoccupied),
there are correspondingly more classes of determinants at a given level of hCI.
This can be appreciated from the comparison of hCI1 for closed- and open-shell references, shown in Fig.~\ref{fig:determinants}.

Finally, one can employ a reference determinant having more unpaired electrons to define hCI models.
The case of two unpaired electrons ($s_0=2$) is shown in the right panel of Fig.~\ref{fig:hCI}.
The key difference in the hCI sequence, with respect to the closed-shell case, lies in the displacement by one block to the right in the seniority-excitation map,
reflecting the shift from the seniority-zero to the seniority-two reference.
In addition, in the two previous cases, the hCI0 level only included the reference determinant
whereas, for $s_0=2$, hCI0 does not only account for the reference determinant
but also for the two closed-shell determinants produced by the single excitation that pairs the two unpaired electrons.
The hCI1 classes of determinants for the $s_0=2$ reference can be seen in the bottom of Fig.~\ref{fig:determinants},
which clearly outnumbers the fewer classes associated with the $s_0=0$ and $s_0=1$ cases.

One could adopt references with even larger $s_0$, further displacing the hCI sequence to the right in the seniority-excitation map.
In this case, hCI0 would include all the lower spin determinants obtained by partially or totally pairing the unpaired electrons.
Notice that the number of such determinants does not depend on the system size (scaling as $N^0$), and are thus included at the $h=0$ level, in line with the spirit of hCI.
In a similar fashion, hCI can be built on top of a larger $s_0$ reference for odd numbers of electrons.
The deduction for the number of determinants in a given hCI model, for the $s_0=0$ closed-shell, $s_0=1$ open-shell, as well as for an arbitrary reference determinant, can be found in Appendix \ref{app:appendix}.

For a reference containing unpaired electrons, an approximate CI model generally produces spin-contaminated states.
One can impose the solutions to have a well-defined spin by including additional determinants generated via higher-order hierarchies, which account for the missing spin-flip configurations. \cite{Chilkuri_2021}
The spin-contamination problem and its solution are therefore equivalent to that encountered in eCI. \cite{Maurice_1996}
Here, we have employed this procedure and considered pure spin states.
For the calculation of excitation energies, however, we have also assessed the effect of not imposing this condition.

For the $s_0 = 2$ reference, one could alternatively employ a high-spin triplet determinant (two unpaired spin-up electrons),
rather than the low-spin determinant shown in Fig.~\ref{fig:hCI}.
Regardless of the choice, hCI (as eCI) produces the same energies for the triplet states, provided that a spin eigenstate is imposed.

\section{Computational details}
\label{sec:compdet}

The hCI models introduced here were implemented in {\QP} \cite{Garniron_2019} through a straightforward modification of the
\textit{configuration interaction using a perturbative selection made iteratively} (CIPSI) algorithm. \cite{Huron_1973,Giner_2013,Giner_2015,Garniron_2018}
By allowing only for the determinants that are connected with the reference determinant(s) up to a given maximum hierarchy $h$,
the CIPSI algorithm is restricted to the truncated Hilbert space specified by the reference determinant(s), and the value of $h$.
{\QP} \cite{Garniron_2019} was also employed to perform all the present eCI, sCI, and FCI calculations.
In a given calculation, the energies are considered to be converged when the (largest) EN2 correction computed in the truncated Hilbert space
lies below \SI{0.01}{\milli\hartree}. \cite{Garniron_2017}
This selected CI procedure requires considerably fewer determinants than the total number of determinants in the truncated Hilbert space
while delivering fairly converged energies.
The ground- and excited-state CI energies are obtained with the Davidson iterative algorithm. \cite{Davidson_1975}

For a given approximate CI model, we further evaluated the standard and the renormalized EN2 perturbative correction. \cite{Garniron_2019}
This calculation involves a single loop over the determinants left outside the truncated (internal) space but connected to it via at most double excitations.
Looping over these external doubly excited determinants has a computational scaling equal to $\Ndet$,
thus $\order*{N^{2e+4}}$ for eCI and $\order*{N^{2h+4}}$ for hCI, where $e$ or $h$ define the internal CI space.
For example, hCI2 and CISD present a $\order*{N^6}$ computational scaling, whereas hCI2+EN2 and CISD+EN2 scale as $\order*{N^8}$,
though with a small prefactor stemming from the EN2 calculation that employs a very efficient semistochastic algorithm. \cite{Garniron_2017}
The computational scaling associated with the CI+EN2 calculations is also presented in Table \ref{tab:scaling}.

To gauge the performance of hCI, eCI, and sCI for radicals, we have calculated the ground-state potential energy curves (PECs) for the dissociation of four radicals:
\ce{OH}, \ce{CN}, vinyl (\ce{C2H3}), and \ce{H7}.
The CI calculations employed the ground-state restricted open-shell HF orbitals, described with the cc-pVDZ basis set, and within the frozen-core approximation.
For such small systems and basis sets, FCI is attainable and provides the reference results for gauging the approximate CI models.
The equilibrium geometry of vinyl was taken from Ref.~\onlinecite{Loos_2020} and is also reproduced in the {\SupInf}.
Their PECs were computed along the \ce{C=C} double bond breaking coordinate, with the remaining internal coordinates kept frozen.
For \ce{H7}, we considered equally spaced and linearly arranged hydrogen atoms and the PECs were computed along the symmetric dissociation coordinate.

The results were analyzed along the same lines as our previous report on hCI for closed-shell systems. \cite{Kossoski_2022}
Namely, for the different CI models considered here,
we evaluated the convergence of the non-parallelity error (NPE), the distance error, the harmonic vibrational frequencies, and the equilibrium bond lengths, as functions of $\Ndet$.
The NPE of a given level of theory is defined as the maximum minus the minimum energy differences between its corresponding PEC and the FCI PEC, for a given range of coordinates.
Here, we redefine the previous definition of the distance error \cite{Kossoski_2022} to accommodate for the fact the approximate PEC might appear below the FCI one when perturbative corrections are employed.
In such cases and with the previous definition, undesired negative values could be attained.
The distance error is redefined based on the signed differences between two PECs, as the absolute value of their maximum difference plus the absolute value of their minimum difference,
evaluated at a given coordinate interval.
This new definition measures how close are two PECs, remaining always non-negative.
From here on, we employ equilibrium properties when referring to both the equilibrium geometry and the harmonic vibrational frequency.
Details about how the equilibrium properties were obtained from the calculated PECs,
along with the ranges defining the NPE and distance errors can be found in the \SupInf.

The various CI models introduced here were further assessed based on calculated vertical excitation energies for 69 electronic states,
from 17 closed-shell systems and 13 radicals (shown in Table~\ref{tab:systems}), with geometries extracted from the QUEST database. \cite{Veril_2021}
Our set includes mostly small systems, ranging from \ce{BH} to glyoxal, displaying a mix of valence and Rydberg singly-excited states,
and 2 doubly-excited states (glyoxal and nitroxyl). It does not include large molecules nor charge transfer states.
The full set of excited states and calculated excitation energies, for the various CI models, are provided in the {\SupInf}.

\begin{table}[ht!]
\caption{Systems Considered in the Excited State Calculations.}
\label{tab:systems}
\begin{ruledtabular}
\begin{tabular}{llllll}
\mc{3}{c}{Closed-shells}           & \mc{3}{c}{Open-shells} \\
\cline{1-3} \cline{4-6}
glyoxal      & \ce{HCF}  & \ce{H2O} & \ce{BeH} & \ce{NH2} & \ce{HCO} \\
acetaldehyde & \ce{HCCl} & \ce{H2S} & \ce{BeF} & \ce{PH2} & \ce{HOC} \\
silylidene   & \ce{HPO}  & \ce{N2}  & \ce{BH2} & vinyl    & \ce{OH}  \\
nitroxyl     & \ce{CF2}  & \ce{NH3} & \ce{CN}  & allyl    & \ce{CO+} \\
ethylene     & \ce{BH}   & \ce{HCl} & \ce{CH3} &          & \\
methanimine  & \ce{BF}   &          &          &          & \\
\end{tabular}
\end{ruledtabular}
\end{table}

We employed the aug-cc-pVDZ basis set for systems having up to three non-hydrogen atoms and the 6-31+G(d) basis set for the larger ones.
Core orbitals were frozen systematically.
We impose the CI solutions to be eigenstates of the spin angular momentum operator, which implies accounting for a set of appropriate spin-flipped determinants stemming from higher-order excitations or hierarchies. \cite{Chilkuri_2021}
In this sense, our CIS calculations for radicals actually correspond to the so-called extended CIS \cite{Maurice_1996} for instance, and equivalently for the other hCI, eCI, and sCI models.
Notice that spin-contaminated solutions would have different energies than the spin eigenstates considered here.

We performed calculations following both the standard ground-state-based CI route and the state-specific CI route. \cite{Kossoski_2023}
For the latter, we employed the state-specific orbitals obtained in Ref.~\onlinecite{Kossoski_2023}.
Notice that, in contrast to eCI, the energies obtained with sCI and hCI models are not invariant under rotations within the occupied and virtual subspaces.
The restricted HF solution (restricted open-shell HF for the open-shell systems) was taken as the reference determinant for the hCI and eCI ground-state calculations
(left and middle panels of Fig.~\ref{fig:hCI}).
For the state-specific hCI and eCI calculations, we employed a minimal CSF reference: \cite{Kossoski_2023}
a single open-shell determinant for the doublet excited states and a single open-shell CSF for the excited states from closed-shell systems
(middle and right panels of Fig.~\ref{fig:hCI}, respectively).
The computed excitation energies were benchmarked against the reference theoretical values provided in the QUEST database. \cite{Veril_2021}
For each excited state considered here, the reference value and the method used to obtain it (of high-order CC or extrapolated FCI quality) can be found in the {\SupInf}.

As for the sCI models,
we considered sCI1 for both ground and (state-specific) excited-state calculations for the doublet open-shell systems.
For the closed-shell systems and their excited states, we explored two models.
In the sCI2/sCI0 model, the ground state was described with sCI0 and the excited state with sCI2, which is the minimal sCI calculation for computing singly-excited states of closed-shell systems.
Further including the seniority-two sector for the ground-state calculations defines the sCI2/sCI2 model.

From here on, CI models carrying the $\Delta$ symbol denote a state-specific approach, whereas those without it correspond to a ground-state-based approach.

\section{Results and discussion}
\label{sec:res}

\subsection{hCI for radicals}
\label{sec:res_A}

\begin{figure*}
\includegraphics[width=1.0\linewidth]{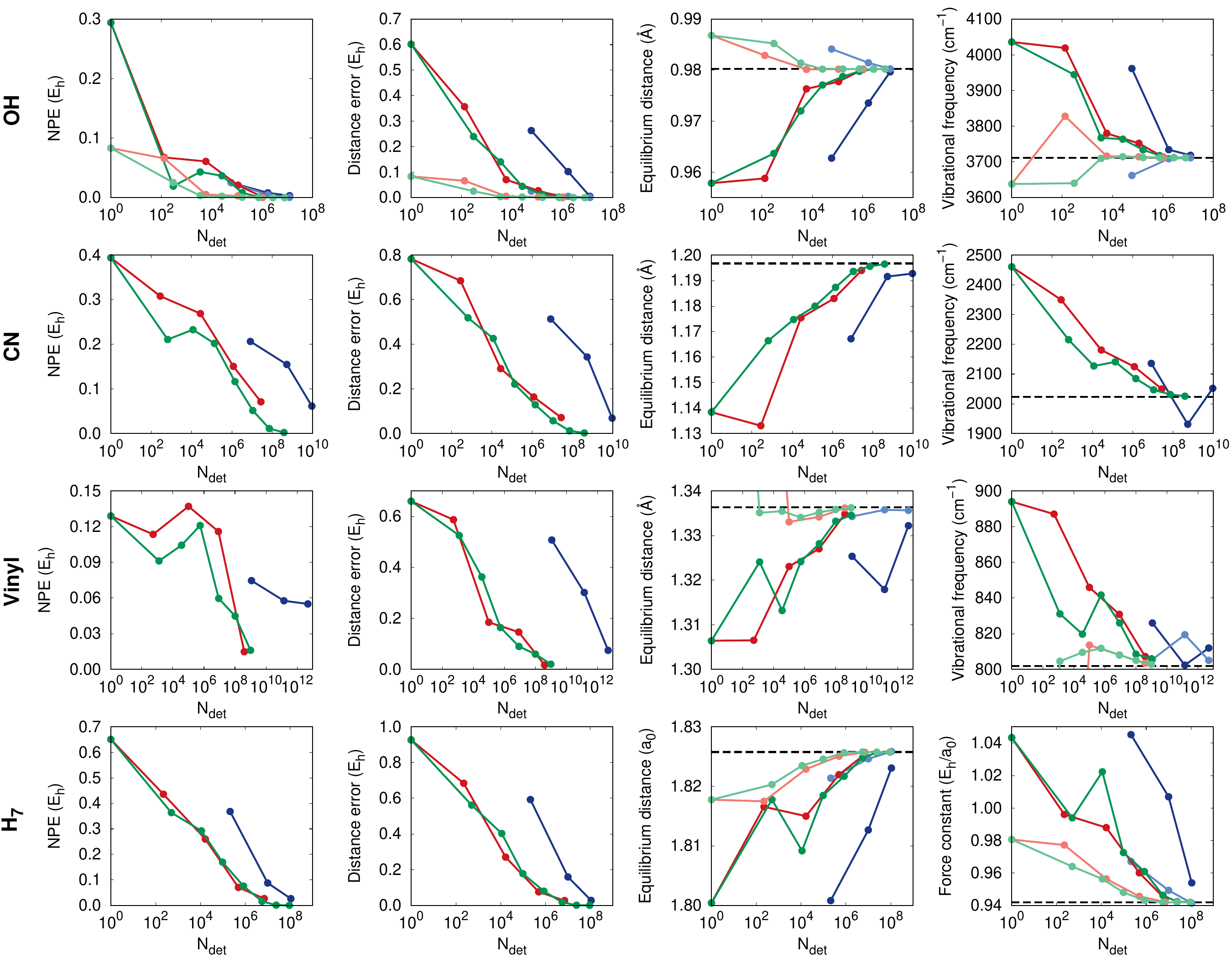}
\caption{
Non-parallelity error (NPE), distance error, equilibrium distance, and vibrational frequency (or force constant), for \ce{OH}, \ce{CN}, vinyl, and \ce{H7},
as functions of the number of determinants ($\Ndet$), according to the hCI (green), eCI (red), and sCI (blue) routes,
with (light tone crosses) and without (dark tone circles) the standard EN2 perturbative correction.
Each point denotes one CI model, according to the sequences: HF, hCI1, hCI1.5, hCI2, etc. (green); HF, CIS, CISD, etc. (red); and sCI1, sCI3, and sCI5 (blue).
The dashed lines represent the FCI results.}
\label{fig:plot_all}
\end{figure*}

\begin{figure*}
\includegraphics[width=1.0\linewidth]{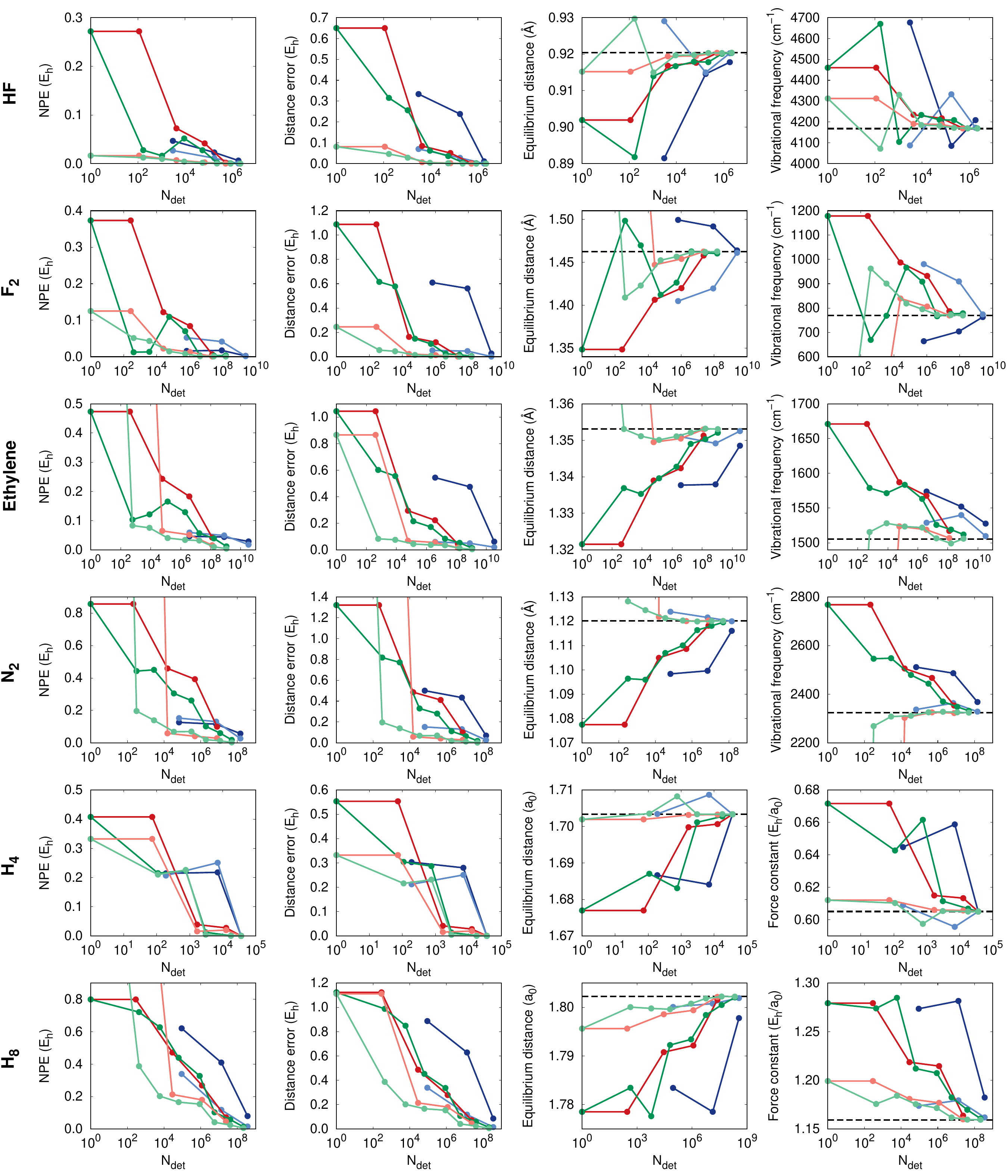}
\caption{
Non-parallelity error (NPE), distance error, equilibrium distance, and vibrational frequency (or force constant), for \ce{HF}, \ce{F2}, ethylene, \ce{N2}, \ce{H4}, and \ce{H8}, as functions of the number of determinants ($\Ndet$), according to the hCI (green), eCI (red) and sCI (blue) routes,
with (light tone crosses) and without (dark tone circles) the standard EN2 perturbative correction.
Each point denotes one CI model, according to the sequences: HF, hCI1, hCI1.5, hCI2, etc. (green); HF, CIS, CISD, etc. (red); and sCI0, sCI2, and sCI4 (blue).
The dashed lines represent the FCI results.}
\label{fig:plot_all_closed}
\end{figure*}

The full set of PECs for the open-shell systems (\ce{OH}, \ce{CN}, \ce{H7}, and vinyl) are presented in the {\SupInf}.
From these PECs, we obtain the NPEs, distance errors, equilibrium bond lengths, and harmonic vibrational frequencies, which are plotted as functions of $\Ndet$ in Fig.~\ref{fig:plot_all}.
Each point in the figure denotes one CI model.
For instance, the hCI route is represented by the dark green line, with the first point being hCI0 (which corresponds to HF in the present cases), the second being hCI1, the third being hCI1.5, etc.
The corresponding results for the closed-shell systems (\ce{HF}, \ce{F2}, \ce{N2}, ethylene, \ce{H4}, and \ce{H8}) are shown in Fig.~\ref{fig:plot_all_closed}.
When we first introduced hCI, \cite{Kossoski_2022} we surveyed the same molecules, though only for the bare CI models (without a perturbative correction).
These previous results are reproduced in Fig.~\ref{fig:plot_all_closed} together with the present results, which include such correction.
First, we discuss the results for the bare CI models (hCI, eCI, and sCI) for the open-shell systems (represented by the dark tones in the figures),
leaving the discussion about the perturbative correction (light tones) for both open-shell and closed-shell systems to subsection Sec.~\ref{sec:res_B}.

We find that hCI typically improves the convergence with respect to $\Ndet$ when compared to eCI, for the four properties surveyed here.
The present finding for open-shell systems, therefore, parallels the previously reported superiority of hCI for closed-shell systems. \cite{Kossoski_2022}

The smaller NPEs obtained with hCI are related to the better description of the PECs at dissociation.
This is due to a larger fraction of static correlation being recovered,
stemming from the classes of determinants appearing lower in the seniority-excitation map (see Fig.~\ref{fig:hCI}),
namely, determinants high in excitation degree $e$ and low in seniority $s$, absent in eCI of the same order.
The smaller NPEs attained with hCI are clear for \ce{OH} (involving a single bond breaking),
vinyl (double bond breaking), and \ce{CN} (triple bond breaking),
whereas for \ce{H7} (multiple bond breaking), hCI and eCI are comparable.
This dependence on the order of bond breaking had also been observed for closed-shell systems \cite{Kossoski_2022} (as can be seen in Fig.~\ref{fig:plot_all_closed}),
and would be expected, as describing dissociation becomes increasingly more challenging as the bond order increases.
Even though hCI manages to recover more static correlation than eCI,
the advantage of the former is more striking for single bond breaking, whereas multiple bond breaking (like \ce{H7} and \ce{H8}) inevitably requires higher-order excited determinants.

The distance errors obtained with hCI and eCI are overall comparable for open-shells, whereas the former presents slightly better convergence for the closed-shell systems.
In comparison to eCI, hCI leads to either comparable or somewhat faster (for \ce{CN}) convergence of the equilibrium geometries for the open-shell systems.
Finally, hCI systematically outperforms eCI in the calculation of vibrational frequencies, except for \ce{H7}, where no big difference is found.
For the \ce{CN} radical, CIS, sCI1, and sCI3 produce crossings between ground and excited states around equilibrium, hence non-smooth adiabatic PECs
and somewhat less reliable equilibrium properties in these specific cases.
Given the overall slight superiority of hCI for the equilibrium properties, it also manages to account for more dynamic correlation than eCI,
for both open-shell and closed-shell systems.

In great contrast to hCI and eCI, sCI models deliver a poor convergence of all observables.
The single exception is the NPE for \ce{OH} obtained from sCI0, in between the NPEs obtained from hCI2 and CISDT, methods having a comparable computational cost in this particular case.
Going to larger basis sets or bigger systems, the computational burden of sCI models would increase considerably more than hCI or eCI models,
due to its formal exponential scaling.
As far as configuration interaction methods are concerned, we conclude that the sCI route is unattractive to open-shell systems,
in line with similar findings for closed-shell systems \cite{Kossoski_2022} (also see Fig.~\ref{fig:plot_all_closed}).

We recall that ground-state restricted open-shell HF orbitals were employed in all calculations for the ground-state radicals.
In contrast to eCI, methods that rely on the seniority to generate excited determinants (like the hCI and sCI models addressed here)
are not invariant under orbital rotations within the occupied and virtual subspaces. \cite{Bytautas_2011,Limacher_2014,Stein_2014}
One could exploit the rotations within each subspace (by means of orbital localization for example) to hopefully render more suitable orbitals for hCI and sCI calculations.
In our first contribution on hCI to closed-shell systems, \cite{Kossoski_2022} we went one step further
and variationally optimized the orbitals at the correlated CI level, thus allowing rotations between occupied and virtual subspaces.
Except for the lower-order CI models, the cost and complications of orbital optimization outweigh the mixed improvement in the computed properties. \cite{Kossoski_2022}
For the closed-shell systems (surveyed both here and in the previous study \cite{Kossoski_2022}),
the results become significantly more accurate with the EN2 perturbative correction rather than by variationally optimizing the orbitals.

\subsection{hCI plus EN2 correction}
\label{sec:res_B}

The results for the CI models corrected by the standard EN2 energy are represented by the light tones in Figs.~\ref{fig:plot_all} and \ref{fig:plot_all_closed}.
We excluded instances where the PECs present important discontinuities.
These discontinuities reflect crossings between states of different symmetries in the unperturbed CI calculation,
appearing as kinks in the computed ground-state PEC.
Due to the abrupt change of character, the EN2 correction is not uniform and gives rise to the observed discontinuities.
This happens for the open-shell systems at dissociation (except for \ce{OH}), for which we do not present NPEs and distance errors.
For \ce{OH} and the six closed-shell systems, the EN2 correction produced smooth PECs at dissociation.
Around the equilibrium geometry, the EN2 correction also leads to well-behaved PECs, for all CI models, and for all systems, except for \ce{CN}
(thus no equilibrium properties are presented for this radical).
This is simply due to its several close-lying excited states.
If we were to follow the PEC of a given symmetry (and not the lowest-lying state as we did here), there would be no discontinuities stemming from the EN2 contribution.

We found that the EN2 perturbative correction significantly reduces the errors of the bare CI calculations.
This is observed for all the CI models, systems, and observables considered here.
For \ce{OH}, in particular, the improvement is massive, as even the lower levels of CI+EN2 deliver very close results to FCI.
Importantly, the hCI+EN2 route outperforms its eCI+EN2 counterpart, thus preserving the advantage observed without the perturbative correction.
Similarly, the correction typically maintains the comparable performance of the two routes when this is the case according to the unperturbed CI calculations.

The improvement brought by the perturbative correction is such that a given CI+EN2 model often provides more accurate results than the higher-order bare CI model sharing the same computational scaling.
For instance, hCI1+EN2 and hCI2 display the same $\order*{N^6}$ computational scaling, and the former outperforms the latter in many cases.
In Figs.~\ref{fig:plot_all} and \ref{fig:plot_all_closed}, hCI1+EN2 correlates with the second light green mark, and hCI2 with the fourth dark green mark.
The former model provides smaller NPEs and distance errors for \ce{OH}, \ce{HF}, \ce{F2} (all single-bond breakings),
the opposite being true for multiple bond breaking.
The advantage of the EN2 correction is more striking for the equilibrium properties,
where hCI1+EN2 is more accurate than hCI2 for \ce{OH}, vinyl, ethylene, and \ce{N2},
whereas both models perform similarly for the remaining systems.
It is worth recalling that the prefactor associated with the EN2 calculation is smaller than the one associated with the higher-order CI calculation,
which, in the above example, makes hCI1+EN2 cheaper than hCI2.
In many cases, the accuracy of hCI1+EN2 is actually better or comparable to that of hCI2.5 or CISDT, considerably more expensive models.

A similar impact on the effect of the EN2 correction also holds for eCI.
For that, we compare CISD+EN2 and CISDT, both sharing a $\order*{N^8}$ scaling.
CISD+EN2 correlates with the third dark red mark and CISDT with the fourth light red mark in Figs.~\ref{fig:plot_all} and ~\ref{fig:plot_all_closed}.
The errors for the four observables are systematically smaller than the former model,
except for the NPE and distance error of \ce{H8} and the equilibrium properties of \ce{H7}, where they are comparable.

The EN2 correction usually ameliorates the performance of the sCI route, though to a lesser extent than observed for hCI and eCI.
This reflects the poorer reference provided by sCI, from which perturbation theory struggles to recover from.
In a handful of cases, the EN2 correction does not improve the bare sCI results,
for both NPEs (\ce{OH}, \ce{N2}, \ce{F2}, and ethylene)
and equilibrium properties (vinyl, \ce{HF}, \ce{F2}, and ethylene).
Overall, the sCI+EN2 route is too expensive for the attained accuracy.

In very few cases, the EN2 correction does not improve the bare CI results, suggesting important cancellation of errors in the latter.
This is seen when employing the hCI1 and hCI1+EN2 models to compute the NPE of \ce{OH} and \ce{F2}, and to some extent the equilibrium properties of \ce{F2}.
These are exceptions though, as the EN2 correction practically always improves the accuracy of the computed observables.
It further leads to an overall more monotonic convergence of the observables.
In this sense, it regularizes oscillations seen in the unperturbed case, probably related to the cancellation of errors at the lower orders.
This is clearly seen in the equilibrium properties of vinyl, \ce{H7}, and \ce{F2},
and in the NPE of \ce{OH}, \ce{HF}, \ce{F2}, and ethylene.

Instead of the usual EN2 correction, discussed so far and shown in Figs.~\ref{fig:plot_all} and \ref{fig:plot_all_closed},
one can compute the renormalized EN2 correction. \cite{Garniron_2019}
Analogous results showing the convergence of observables for both usual and renormalized EN2 corrections are shown in the {\SupInf}.
Significant differences can be encountered at the lower orders of CI, becoming negligible at higher orders.
More often than not, the usual correction performs better.
The difference is noticeable for ethylene, \ce{N2}, \ce{H7}, and \ce{H8}, and to a lesser extent for vinyl,
whereas for \ce{H4} the results are mixed.
The renormalized correction is slightly more accurate for \ce{OH}, \ce{HF}, and \ce{F2}.
These results suggest that renormalizing the EN2 energy may only be helpful for single bond breaking, though by a small amount,
whereas it worsens the results for multiple bond breaking.
Overall, the usual EN2 correction should probably be favored when employed in combination with the approximate CI models.

We further explored hCI to describe the automerization barrier in the ground state of cyclobutadiene,
which connects the two equivalent rectangular $D_{2h}$ equilibrium geometries through the square $D_{4h}$ transition state geometry. \cite{Bally_1980,Tallarico_1996}
This is a well-known and challenging problem, requiring high-level calculations to achieve quantitative values for the height of the barrier (see Ref.~\onlinecite{Monino_2022} and references therein).
Here we employ the geometries presented in Ref.~\onlinecite{Monino_2022}, the 6-31+G(d) basis set, and the frozen-core approximation.
We optimize the orbitals for two closed-shell determinants in the $D_{2h}$ geometry, and for two spin-flipped open-shell determinants in the $D_{4h}$ geometry,
from which the excited determinants were generated in the subsequent CI calculations.
Even though the calculations target the ground state, they are labelled $\Delta$CI because different references are employed for each geometry.
For this basis set, the reference value of \SI{7.51}{kcal/mol} was obtained at CC with single, double, triple, and quadruple excitations. \cite{Monino_2022}
With respect to this reference value, the errors for the automerization barrier are
\SI{+3.55}{kcal/mol} with our two-determinant calculations, \SI{-2.28}{kcal/mol} with $\Delta$hCI2, \SI{-1.07}{kcal/mol} with $\Delta$hCI2+EN2, 
\SI{+0.61}{kcal/mol} with $\Delta$CISD, and \SI{-0.49}{kcal/mol} with $\Delta$CISD+EN2.
These values are comparable to those obtained with other methods, extensively discussed in Ref.~\onlinecite{Monino_2022}.
Here we just mention the errors of CC with singles and doubles, \SI{+0.80}{kcal/mol}, and of spin-flip equation-of-motion CC (EOM-CC) with singles and doubles, \SI{-1.65}{kcal/mol},
which share the same $\order*{N^6}$ computational scaling as $\Delta$CISD and $\Delta$hCI2.
This specific example serves to illustrate that the accuracy of hCI models can be similar to that of other methods.
As such, hCI can be taken into consideration in future systematic studies that tackle challenging chemical problems where strong correlation comes into play.

\subsection{hCI for excited states}
\label{sec:res_C}

For each CI model considered here, we evaluate the mean signed error (MSE), mean absolute error (MAE), root-mean-square error (RMSE), and standard deviation of the errors (SDE).
with respect to the reference theoretical values for the excitation energies.
For completeness, the defition of these statistical measures can be found in the {\SupInf}.
For the lower-order models, the calculations were performed for 50 (closed-shell) and 19 (open-shell) excited states.
In turn, subsets of 16 (closed-shell) and 6 (open-shell) excited states were considered for the higher-order CI models, given the more intensive computational cost.
Even though these subsets are too small for meaningful absolute statistics, they should be enough to reveal main trends.
A detailed comparison of eCI based on ground-state and state-specific approaches can be found elsewhere. \cite{Kossoski_2023}
The focus of the present discussion lies on the comparison between hCI and eCI, and sCI,
and on a similar comparison between ground-state and state-specific approaches for hCI and sCI.
The bare (unperturbed) hCI and sCI models are discussed in this section and in Sec.~\ref{sec:res_D}, respectively,
whereas their EN2 perturbed analogues are left for Sec.~\ref{sec:res_E}.

We start the discussion with the excitations from closed-shell systems, with corresponding statistical errors shown in Table \ref{tab:stat1}.
The first level of hCI, hCI1, produces poor excitation energies, with a MAE of \SI{1.16}{\eV}.
This model shares the same computational scaling with CIS, but includes the paired double excitations,
which clearly worsen the decent CIS results (MAE of \SI{0.61}{\eV}).
Moving one rank up, to hCI1.5, the MAE increases to \SI{1.95}{\eV}, and then reaches a maximum of \SI{3.53}{\eV} at the hCI2 level.
We notice that hCI2 delivers somewhat smaller errors than CISD (MAE of \SI{4.09}{\eV}), even though both are way too large.
From that point on, the next hCI models generate progressively better results.
Despite the improvement with respect to hCI2, hCI2.5 is still as inaccurate (MAE of \SI{1.95}{\eV}) as hCI1.5.
At the hCI3 level, significantly smaller errors are finally achieved (MAE of \SI{0.19}{\eV}).
This is close to that obtained with the eCI model of the same order, CISDT (\SI{0.17}{\eV}).
Even smaller errors are produced at the hCI3.5 level (MAE of \SI{0.11}{\eV}), but at a considerable computational cost.

\begin{table*}[ht!]
\caption{Mean Signed Error (MSE), Mean Absolute Error (MAE), Root-Mean Square Error (RMSE), and Standard Deviation of the Errors (SDE), in Units of eV, with Respect to Reference Theoretical Values, for the Set of
Singly Excited States of Closed-Shell Systems Listed in the {\SupInf}.
}
\label{tab:stat1}
\begin{ruledtabular}
\begin{tabular}{ldddddlddddd}
method            & \mc{1}{c}{count} & \mc{1}{c}{MSE} & \mc{1}{c}{MAE} & \mc{1}{c}{RMSE} & \mc{1}{c}{SDE} & method & \mc{1}{c}{count} & \mc{1}{c}{MSE} & \mc{1}{c}{MAE} & \mc{1}{c}{RMSE} & \mc{1}{c}{SDE} \\
\hline
CIS               & 48  & +0.03 & 0.61 & 0.59 & 0.77  & & & & & & \\
CISD              & 16  & +4.09 & 4.09 & 4.18 & 0.84  & & & & & & \\
CISDT             & 16  & +0.12 & 0.17 & 0.18 & 0.14  & & & & & & \\
CISDTQ            & 16  & +0.15 & 0.15 & 0.17 & 0.08  & & & & & & \\
\hline
hCI1              & 50  & +1.07 & 1.16 & 1.39 & 0.89  & hCI1+EN2              & 50  & +0.65  & 0.65 & 0.84 & 0.53 \\
hCI1.5            & 50  & +1.95 & 1.95 & 2.04 & 0.59  & hCI1.5+EN2            & 50  & +0.62  & 0.63 & 0.76 & 0.44 \\
hCI2              & 16  & +2.99 & 3.53 & 3.61 & 0.76  & & & & & & \\
hCI2.5            & 16  & +1.95 & 1.95 & 2.06 & 0.66  & & & & & & \\
hCI3              & 16  & +0.19 & 0.19 & 0.21 & 0.08  & & & & & & \\
hCI3.5            & 16  & +0.11 & 0.11 & 0.13 & 0.07  & & & & & & \\
\hline
$\Delta$CSF       & 50  & -0.71 & 0.77 & 0.91 & 0.58  & & & & & & \\
$\Delta$CISD      & 50  & -0.12 & 0.17 & 0.22 & 0.18  & $\Delta$CISD+EN2      & 50  & +0.02  & 0.06 & 0.09 & 0.09 \\
$\Delta$CISDT     & 16  & -0.20 & 0.20 & 0.22 & 0.11  & & & & & & \\
$\Delta$CISDTQ    & 16  & -0.02 & 0.02 & 0.02 & 0.02  & & & & & & \\
\hline
$\Delta$hCI1      & 50  & -1.40 & 1.40 & 1.55 & 0.67  & $\Delta$hCI1+EN2      & 50  & +0.12  & 0.24 & 0.30 & 0.27 \\
$\Delta$hCI1.5    & 50  & -2.80 & 2.80 & 3.03 & 1.15  & $\Delta$hCI1.5+EN2    & 50  & -0.01  & 0.13 & 0.19 & 0.19 \\
$\Delta$hCI2      & 50  & -0.18 & 0.20 & 0.25 & 0.16  & $\Delta$hCI2+EN2      & 50  & +0.01  & 0.07 & 0.10 & 0.10 \\
$\Delta$hCI2.5    & 16  & -0.27 & 0.27 & 0.30 & 0.13  & & & & & & \\
$\Delta$hCI3      & 16  & -0.22 & 0.22 & 0.24 & 0.10  & & & & & & \\
$\Delta$hCI3.5    & 16  & -0.08 & 0.08 & 0.09 & 0.05  & & & & & & \\
\hline
sCI2/sCI2         & 50  & +1.35 & 1.35 & 1.51 & 0.68  & sCI2/sCI2+EN2         & 50  & +0.60  & 0.60 & 0.79 & 0.51 \\
sCI2/sCI0         & 50  & -0.34 & 0.55 & 0.74 & 0.66  & sCI2/sCI0+EN2         & 50  & +0.51  & 0.52 & 0.69 & 0.47 \\
$\Delta$sCI2/sCI2 & 50  & +0.66 & 0.78 & 0.89 & 0.60  & $\Delta$sCI2/sCI2+EN2 & 50  & +0.21  & 0.26 & 0.35 & 0.28 \\
$\Delta$sCI2/sCI0 & 50  & -1.04 & 1.04 & 1.21 & 0.62  & $\Delta$sCI2/sCI0+EN2 & 50  & +0.12  & 0.23 & 0.27 & 0.24 \\
\end{tabular}
\end{ruledtabular}
\end{table*}

As discussed above, the errors on excitation energies obtained with the hCI models first increase as one augments the hierarchy parameter $h$, reaching a maximum at hCI2, and then decrease towards higher orders.
This behavior parallels what is well established for the excitation energies computed with eCI, where CISD is far worse than CIS, which in turn is inferior to CISDT, whereas CISDTQ is considerably more accurate.
This can be understood based on the role of the excited determinants for ground and excited states.
The excited determinants of the low-order hCI models (hCI1 and hCI1.5) already account for some correlation for the ground state,
but represent mostly orbital relaxation of the excited state, which thus remains less correlated.
This favored description of the ground state explains the overestimated excitation energies at these orders.
The effect becomes exaggerated at the hCI2 (and CISD) level since it captures most of the ground state correlation through the unpaired double excitations,
which in turn just start to describe correlation for the excited state.
Higher-order models, starting at hCI3 and CISDT, are needed to recover a large fraction of the excited-state correlation, and at this point, the errors on the excitation energies decline progressively.
Because the excited determinants are accessed from the ground-state determinant and due to ground-state HF orbitals, the description of the ground state is always favored.
This explains why hCI systematically overestimates the excitation energies, just as eCI does.

Employing state-specific orbitals would be expected to suppress the bias toward the ground state and lead to improved results, as observed, for instance, when going from CISD to $\Delta$CISD \cite{Kossoski_2023}.
However, for the low orders of state-specific hCI ($\Delta$hCI1 and $\Delta$hCI1.5), we actually found larger errors than with the ground-state-based approach, this time by underestimating the excitation energies.
Besides the set of orbitals, the classes of determinants included at each order play an equally important role and explains our observation.
Still considering the excitations of the closed-shell systems,
different classes of determinants are accessed from the ground-state reference (a closed-shell determinant) and from the excited-state reference (a single open-shell CSF),
for a given hierarchy parameter $h$.
The case of $\Delta$hCI1 can be understood by comparing the hCI1 determinants for these two references, shown in Fig.~\ref{fig:determinants}.
There is far more diversity in the classes of determinants employed for the excited-state calculation than for the ground-state one, a consequence of the different references.
Because of that, low orders of hCI may be expected to capture a larger fraction of correlation for the excited state than for ground state.
This is further supported by the more accurate results obtained for excitations of open-shell systems (discussed in detail later), where the same type of reference is employed.
Although hCI is constructed to account for all classes of determinants whose number share the same computational scaling,
at the lower orders, this procedure does not lead to a balanced description of correlation, at least for excitations of closed-shell systems and the present choice of minimal references.
An important point to realize from the present results is that a state-specific approach is not necessarily more accurate than a ground-state-based approach.
Both the orbitals and the classes of determinants of a given CI model control the fine balance of ground- and excited-state correlation effects.

At the $\Delta$hCI2 level, the unbalance associated with the state-specific classes of determinants is reduced to a great extent.
At this level, the state-specific advantage clearly manifests, with $\Delta$hCI2 having a MAE of \SI{0.20}{\eV}, substantially smaller than the MAE of \SI{3.53}{\eV} obtained from hCI2.
This comparison between ground-state-based and state-specific approaches is analogous to our previous finding on the eCI models of the same order, CISD, and $\Delta$CISD. \cite{Kossoski_2023}
Even though hCI2 is somewhat more accurate than CISD (both have large errors), their state-specific versions share a comparable performance.
Actually, $\Delta$hCI2 is slightly less accurate than $\Delta$CISD, with MAEs of \SI{0.20}{\eV} and \SI{0.17}{\eV}.
Moreover, $\Delta$hCI2 has a more negative MSE than $\Delta$CISD (\SI{-0.18}{\eV} against \SI{-0.12}{\eV}).
Albeit small, these differences are statistically significant, and $\Delta$CISD would still be preferable to $\Delta$hCI2 by some margin.
The slightly worse performance of $\Delta$hCI2 probably stems from a residual unbalance associated with the state-specific classes of determinants.
$\Delta$hCI2 is more accurate for singlets than for triplets (MAEs of \SI{0.16}{\eV} and \SI{0.25}{\eV}),
whereas Rydberg states are better described than valence states (MAEs of \SI{0.14}{\eV} and \SI{0.24}{\eV}).
The same trends are found for the low-order models ($\Delta$hCI1 and $\Delta$hCI1.5), and also for $\Delta$CISD, which slightly outperforms $\Delta$hCI2 for each type of excitation.
The specific MAEs can be found in the {\SupInf}.

Moving to $\Delta$hCI2.5 increases the errors (MAE of \SI{0.27}{\eV}) when compared to $\Delta$hCI2 (MAE of \SI{0.20}{\eV}).
This could be due to another set of unbalanced state-specific determinants that enters at this stage, likewise to what was observed between $\Delta$hCI1 and $\Delta$hCI1.5.
Despite the differing number of states considered, the observed variation in the accuracy of $\Delta$hCI2 and $\Delta$hCI2.5 is statistically significant,
which is confirmed by calculating the statistical errors for the same subset of excited states.
Even though $\Delta$hCI2.5 is much more accurate than hCI2.5, $\Delta$hCI2 or $\Delta$CISD remain cheaper and more accurate.

The situation improves at $\Delta$hCI3 (MAE of \SI{0.22}{\eV}), though it still remains less accurate than $\Delta$hCI2,
only surpassed at the $\Delta$hCI3.5 level (MAE of \SI{0.08}{\eV}).
There is no gain in going from hCI3 (MAE of \SI{0.19}{\eV}) to $\Delta$hCI3 and from hCI3.5 (MAE of \SI{0.11}{\eV}) to $\Delta$hCI3.5,
likewise to what had been found for CISDT and $\Delta$CISDT. \cite{Kossoski_2023}
We notice, however, that the state-specific route presents negative MSEs, in contrast to the positive values obtained with the ground-state-based route, for both hCI and eCI.
Furthermore, eCI and hCI present comparable performances at this order, with an arguable preference for the former,
as also discussed above for $\Delta$CISD and $\Delta$hCI2.

We now shift to the discussion of excitations for open-shell systems.
The statistical errors are shown in Table \ref{tab:stat2}.
The key difference on the calculations for open-shell excitations is that the same type of reference was employed for ground- and excited-state calculations, namely,
a single open-shell determinant, shown in the center panel of Fig.~\ref{fig:hCI}.
In contrast, the excited states from the closed-shell systems relied on different classes of reference determinants.
This accounts for the more accurate results observed for open-shell excitations and explains most of the differences with respect to excitations from closed-shell systems, as discussed in detail in the following.

\begin{table*}[ht!]
\caption{Mean Signed Error (MSE), Mean Absolute Error (MAE), Root-Mean Square Error (RMSE), and Standard Deviation of the Errors (SDE) in Units of eV, with Respect to Reference Theoretical Values, for the Set of
Singly-Excited States from Open-Shell Doublets Listed in the {\SupInf}.
}
\label{tab:stat2}
\begin{ruledtabular}
\begin{tabular}{ldddddlddddd}
method            & \mc{1}{c}{count} & \mc{1}{c}{MSE} & \mc{1}{c}{MAE} & \mc{1}{c}{RMSE} & \mc{1}{c}{SDE} & method & \mc{1}{c}{count} & \mc{1}{c}{MSE} & \mc{1}{c}{MAE} & \mc{1}{c}{RMSE} & \mc{1}{c}{SDE} \\
\hline
CIS               & 19 & +0.38 & 0.41 & 0.63 & 0.50  & & & & & & \\
CISD              &  6 & +2.97 & 2.97 & 3.19 & 1.17  & & & & & & \\
CISDT             &  6 & +0.06 & 0.07 & 0.09 & 0.06  & & & & & & \\
CISDTQ            &  6 & +0.08 & 0.08 & 0.11 & 0.08  & & & & & & \\
\hline
hCI1              & 19 & +0.65 & 0.73 & 0.91 & 0.64  & hCI1+EN2           & 19 & +0.27 & 0.28 & 0.36 & 0.24 \\
hCI1.5            & 19 & +1.07 & 1.07 & 1.26 & 0.68  & hCI1.5+EN2         & 19 & +0.22 & 0.25 & 0.33 & 0.24 \\
hCI2              &  6 & +1.32 & 1.32 & 1.60 & 0.90  & & & & & & \\
hCI2.5            &  6 & +0.03 & 0.05 & 0.06 & 0.05  & & & & & & \\
hCI3              &  6 & +0.00 & 0.04 & 0.05 & 0.05  & & & & & & \\
hCI3.5            &  6 & +0.01 & 0.03 & 0.05 & 0.04  & & & & & & \\
\hline
$\Delta$CSF       & 19 & -0.04 & 0.43 & 0.59 & 0.59  & & & & & & \\
$\Delta$CISD      & 19 & +0.00 & 0.12 & 0.21 & 0.21  & $\Delta$CISD+EN2   & 19 & -0.02 & 0.04 & 0.06 & 0.06 \\
$\Delta$CISDT     &  6 & -0.07 & 0.07 & 0.10 & 0.08  & & & & & & \\
$\Delta$CISDTQ    &  6 & -0.02 & 0.02 & 0.02 & 0.02  & & & & & & \\
\hline
$\Delta$hCI1      & 19 & -0.21 & 0.35 & 0.49 & 0.44  & $\Delta$hCI1+EN2   & 19 & +0.02 & 0.17 & 0.29 & 0.29 \\
$\Delta$hCI1.5    & 19 & -0.14 & 0.45 & 0.70 & 0.69  & $\Delta$hCI1.5+EN2 & 19 & +0.00 & 0.10 & 0.15 & 0.15 \\
$\Delta$hCI2      & 19 & -0.02 & 0.10 & 0.16 & 0.16  & $\Delta$hCI2+EN2   & 19 & -0.00 & 0.05 & 0.08 & 0.08 \\
$\Delta$hCI2.5    &  6 & -0.09 & 0.09 & 0.14 & 0.10  & & & & & & \\
$\Delta$hCI3      &  6 & -0.06 & 0.06 & 0.08 & 0.05  & & & & & & \\
$\Delta$hCI3.5    &  6 & -0.02 & 0.02 & 0.03 & 0.02  & & & & & & \\
\hline
sCI1              & 19 & +0.20 & 0.35 & 0.57 & 0.53  & sCI1+EN2           & 19 & +0.23 & 0.25 & 0.32 & 0.21 \\
$\Delta$sCI1      & 19 & -0.12 & 0.28 & 0.39 & 0.37  & $\Delta$sCI1+EN2   & 19 & +0.06 & 0.12 & 0.20 & 0.19 \\
\end{tabular}
\end{ruledtabular}
\end{table*}

When using ground-state orbitals, we found the performance initially degrades and later improves as the order increases, for both eCI and hCI.
This is similar to what was observed for the closed-shell excitations
and can be traced back to a biased description of the ground state related to the choice of ground-state orbitals.
An important difference, however, is that the maximum error, also at hCI2, is significantly smaller (MAE of \SI{1.32}{\eV}) for open-shells
than for closed-shells (MAE of \SI{3.53}{\eV}).
This clearly reflects the choice of reference, as both ground and excited states of the open shells considered here are qualitatively described with the same type of determinant.
For the same reason, the MSEs are systematically smaller for the open-shell excitations (though still positive because of the ground-state orbitals).

Once state-specific orbitals are employed, there is no remaining bias toward the ground state.
In stark contrast to the closed-shell case, the lower orders of hCI, $\Delta$hCI1 (MAE of \SI{0.35}{\eV}) and $\Delta$hCI1.5 (MAE of \SI{0.45}{\eV}),
become more accurate than their ground-state-based counterparts, hCI1 (MAE of \SI{0.73}{\eV}) and hCI1.5 (MAE of \SI{1.07}{\eV}).
Once again, this is thanks to the same type of reference in ground- and excited-state calculations.
The case of $\Delta$hCI1 for open-shell excitations is depicted in Fig.~\ref{fig:determinants}, in comparison to the more unbalanced case of excitations from closed-shells.
We notice, however, an apparent residual bias from the somewhat negative MSEs (\SI{-0.21}{\eV} with $\Delta$hCI1 and \SI{-0.14}{\eV} with $\Delta$hCI1.5),
and that $\Delta$hCI1.5 is slightly less accurate than $\Delta$hCI1.

This residual bias virtually disappears at the next order, $\Delta$hCI2 (MSE of \SI{-0.02}{\eV}),
whereas for the closed-shell systems even $\Delta$hCI3.5 significantly underestimated (MSE of \SI{-0.08}{\eV}) the excitation energies.
The same comparison between open- and closed-shells holds for eCI models.
For the open-shell transitions, $\Delta$hCI2 is slightly more accurate (MAE of \SI{0.10}{\eV} and RMSE of \SI{0.16}{\eV}) than $\Delta$CISD (MAE of \SI{0.12}{\eV} and RMSE of \SI{0.21}{\eV}).
In further contrast to the closed-shell case,
$\Delta$hCI2.5 is as accurate as $\Delta$hCI2, producing a MAE of \SI{0.09}{\eV}.
(Accounting only for the 6 states considered for $\Delta$hCI2.5, $\Delta$hCI2 would also have a MAE of \SI{0.09}{\eV}.)
Likewise, higher-order hCI models become progressively more accurate.
Even though the statistics become more limited,
they suggest both a small advantage of the hCI models against eCI
and that the ground-state routes perform slightly better than the state-specific ones.

For the state-specific models, we have further assessed the impact of not imposing spin eigenstates (the results and statistics can be found in the {\SupInf}).
This amounts to not including the appropriate spin-flip determinants lying above the hierarchy or excitation degree of choice.
The effect is minimal for the excitation energies of both closed- and open-shell systems, in the $\Delta$hCI1.5, $\Delta$hCI2, and $\Delta$CISD models.
The mean difference on the individual excitation energies lies between \SI{0.01}{\eV} and \SI{0.02}{\eV} for the latter two and one order of magnitude less for $\Delta$hCI1.5.
Similarly, small effects are seen in the global statistics.
In turn, $\Delta$hCI1 is more affected, which would be expected given it is a low order model.
Average individual excitations of closed- and open-shell systems vary by \SI{0.26}{\eV} and \SI{0.06}{\eV}, respectively, always in the sense of improving the energies in the former case,
though not enough to cause a considerable reduction of the large absolute errors.
Overall, not constraining the CI solutions to have well-defined spin brings practical complications (as more roots have to be calculated)
and does not improve the computed excitation energies obtained with the more competitive $\Delta$CISD and $\Delta$hCI2 models.
The sCI models, discussed in the next subsection, are naturally spin eigenstates, and there is no need to enforce that.

\subsection{sCI for excited states}
\label{sec:res_D}

We refer back to Table \ref{tab:stat1} to discuss the performance of sCI for excited states of closed-shell systems.
The first model, sCI2/sCI2, systematically overestimates the excitation energies, producing a MAE of \SI{1.35}{\eV}.
The same two factors discussed above for hCI explain such large errors.
First, the ground-state description is favored due to the use of ground-state HF orbitals.
Indeed, with state-specific orbitals, the $\Delta$sCI2/sCI2 model reduces the MAE to \SI{0.78}{\eV}, though still overestimating the excitation energies.
Second, there is an unbalanced description of the correlation for ground and excited states, associated with the classes of determinants.
While sCI2/sCI2 accounts for an additional $s=2$ sector for the ground state (which can be qualitatively described in the $s=0$ sector),
no additional pairs of electrons are allowed to become unpaired in the excited state calculation, even though the state is qualitatively described by a determinant that is contained in the $s=2$ sector.
In other words, unpaired excitations are allowed to correlate the ground state, but not the excited state, given their respective closed-shell and open-shell characters,
thus creating a bias towards the former.

A possible solution to this unbalance is to restrict the determinants to a maximum seniority number of $s=0$ for the ground state and $s=2$ for the excited state calculation.
This is precisely the sCI2/sCI0 model, which delivers a MAE of \SI{0.55}{\eV}, compared to \SI{1.35}{\eV} of sCI2/sCI2.
However, when going to state-specific orbitals, $\Delta$sCI2/sCI0 systematically undershoots the excitation energies and provides a MAE of \SI{1.04}{\eV}.
Clearly, the seniority-two sector captures more correlation for the excited states than the seniority-zero does for the ground state.
Ultimately, we did not find a combination of sCI models and orbitals that produced reasonable excitation energies for closed-shell systems.

The situation for open-shell systems is quite different, in close analogy to the previous discussion regarding hCI.
As shown in Table \ref{tab:stat2}, sCI1 provides considerably more accurate excitation energies for open-shell than for closed-shell systems.
The reason should not be surprising at this point.
Both ground and excited states of open-shells can be qualitatively described by the same type of reference (a single $s=1$ open-shell determinant).
Moreover, the state-specific approach is superior.
sCI1 presents a MAE of \SI{0.35}{\eV} and overestimates the excitation energies (MSE of \SI{+0.20}{\eV}), in view of the bias introduced by the ground-state orbitals.
With state-specific orbitals, $\Delta$sCI1 reduces the MAE to \SI{0.28}{\eV} and the MSE to \SI{-0.12}{\eV}, which is smaller in absolute value than found for sCI1.

Even though $\Delta$sCI1 is less accurate and has less favorable computational scaling than CI models like $\Delta$CISD and $\Delta$hCI2,
its decent errors are encouraging for another reason:
the development of polynomial scaling CC methods based on the concept of seniority.
For closed-shell systems, DOCI (here referred to as sCI0) energies can be very well reproduced with state-specific pair coupled-cluster doubles (pCCD), a method that has mean-field cost,
for both ground \cite{Bytautas_2011,Allen_1962,Smith_1965,Veillard_1967} and excited \cite{Kossoski_2021,Marie_2021} states.
Likewise, a formulation of pCCD to open-shell systems might share an analogous connection to low-order sCI models like sCI1.
If that is the case, then a state-specific approach of such pCCD formulation adapted to open-shells may approach the accuracy of $\Delta$sCI1 (MAE of \SI{0.28}{\eV}) at a mean-field cost.
This method could then provide an improved starting point to recover the remaining weak correlation than if starting from the (also mean-field though less accurate) $\Delta$CSF model (MAE of \SI{0.43}{\eV}).
To the best of our knowledge, there has been a single proposed extension of pCCD to open-shell systems, \cite{Boguslawski_2021}
based on the ionization-potential equation-of-motion CC (EOM-CC) formalism. \cite{Stanton_1994,Muneaki_2006,Bomble_2005}

As pointed out before, hCI and sCI models are not invariant under rotations within the subspaces of occupied and virtual orbitals.
For the state-specific calculations, such rotations for the excited-state orbitals represent additional degrees of freedom to those of the ground-state orbitals. \cite{Kossoski_2022}
As discussed in the end of Sec.~\ref{sec:res_A}, these degrees of freedom could be exploited,
by localizing occupied and virtual subspaces or by variationally optimizing all orbitals at a correlated level, for instance.
We did not pursue these ideas here, and their impact on the excitation energies computed with hCI and sCI remains an open question.

\subsection{hCI and sCI plus EN2 correction for excited states}
\label{sec:res_E}

The renormalized EN2 correction significantly reduces the errors across most CI models, for both closed- and open-shell systems.
The statistical measures for the CI models corrected with the EN2 correction can be seen in
Table~\ref{tab:stat1} (closed-shell systems) and Table~\ref{tab:stat2} (open-shell systems),
along with the results for the bare (unperturbed) CI models.
We have not considered the EN2 correction for the higher-order models.

For a given scaling (see Table~\ref{tab:scaling}), whether the bare CI model or the lower-order CI model plus EN2 correction produces the smaller errors depends on the truncation order,
and also on the type of excitations (from closed- or open-shells).
For instance, $\Delta$hCI2 and $\Delta$hCI1+EN2 have the same $\order*{N^6}$ computational cost
but the former has smaller MAEs, for closed-shells (\SI{0.20}{\eV} compared to \SI{0.24}{\eV}), and more significantly for open-shells (\SI{0.10}{\eV} compared to \SI{0.17}{\eV}).
In turn, $\Delta$hCI1.5+EN2 is significantly more accurate than $\Delta$hCI2.5 for closed-shells,
with MAEs of \SI{0.13}{\eV} and \SI{0.27}{\eV}, whereas for open-shells both models produce a MAE of \SI{0.10}{\eV}.
The same trend is found at the next order.
$\Delta$hCI2+EN2 is much more accurate than $\Delta$hCI3 for closed-shells, with MAEs of \SI{0.06}{\eV} and \SI{0.22}{\eV},
whereas for the open-shell transitions, they deliver comparable MAEs of \SIrange{0.05}{0.06}{\eV}.
We recall that the calculation of the EN2 correction has a small associated prefactor, which makes the above $\Delta$hCI+EN2 models attractive alternatives to their higher-order unperturbed $\Delta$hCI counterparts.

Among the different CI models, it is worth highlighting the huge improvement observed from $\Delta$hCI1.5 to $\Delta$hCI1.5+EN2.
For the closed-shells, the MAE drops from \SI{2.80}{\eV} to \SI{0.13}{\eV}, whereas a reduction from \SI{0.45}{\eV} to \SI{0.10}{\eV} is seen for the open-shells.
In both cases, the MSE is virtually zero.
In addition, the individual errors for singlets, triplets, Rydberg, and valence states are comparable, lying in the range \SIrange{0.11}{0.15}{\eV}.

We found a similar impact of the EN2 correction for $\Delta$CISD and $\Delta$hCI2.
As discussed above, $\Delta$CISD is somewhat more accurate than $\Delta$hCI2 for the closed-shells (MAEs of \SI{0.17}{\eV} and \SI{0.20}{\eV}),
whereas the EN2 correction decreases their respective MAE to \SI{0.06}{\eV} and \SI{0.07}{\eV}.
In turn, while $\Delta$CISD is slightly less accurate than $\Delta$hCI2 for the open-shells (MAEs of \SI{0.12}{\eV} and \SI{0.10}{\eV}),
the EN2 corrected models show a MAE of \SI{0.04}{\eV} and \SI{0.05}{\eV}, respectively.
Furthermore, the accuracy of $\Delta$hCI2+EN2 is comparable for singlets and triplets (MAEs of \SI{0.07}{\eV} and \SI{0.06}{\eV}),
and superior for Rydberg states when compared to valence states (MAE of \SI{0.04}{\eV} and \SI{0.08}{\eV}).
$\Delta$CISD+EN2 displays the same trends, slightly outperforming $\Delta$hCI2+EN2 for each of the four types of transitions mentioned above, just as discussed above for the unperturbed case.

In another contrast to eCI and hCI routes, the EN2 correction has a more limited effect on the sCI route.
There is little to no advantage for the models based on ground-state HF orbitals.
The outcome is more favorable with state-specific orbitals,
where, for the closed-shells, $\Delta$sCI2/sCI2+EN2 and $\Delta$sCI2/sCI0+EN2 show comparable MAEs (\SI{0.26}{\eV} and \SI{0.23}{\eV}, respectively).
For the open-shells, the MAE decreases from \SI{0.28}{\eV} ($\Delta$sCI1) to \SI{0.12}{\eV} ($\Delta$sCI1+EN2),
the latter error being comparable to that of the much less expensive $\Delta$hCI1.5+EN2 model (\SI{0.10}{\eV}).
Despite the improvement, the EN2 correction is not enough to render the sCI models competitive.

\section{Conclusion}
\label{sec:conclusion}

Here, we have generalized hCI \cite{Kossoski_2022} for an arbitrary reference determinant, thus extending its applicability to radical species and state-specific excited-state calculations.

By surveying the dissociation of four radicals,
we found that the hCI route outperforms or matches eCI, for both weakly correlated (equilibrium properties) and strongly correlated (dissociation) regimes.
These and previous \cite{Kossoski_2022} findings demonstrate the ability of hCI models to recover weak and strong correlations for both open- and closed-shell systems.
Meanwhile, sCI leads to far less accurate results in comparison to hCI or eCI, for a given computational cost.
For closed- and open-shell systems, the EN2 perturbative correction
substantially accelerates and stabilizes the convergence of hCI and eCI (while keeping the advantage of the former),
though it ameliorates the sCI route to a lesser extent.
The standard EN2 correction typically produced more accurate results than its renormalized form, except perhaps for describing single-bond breaking.
Overall, lower-order models (like hCI1+EN2) were found to be fairly accurate given their low computational cost.
At a given CI level, the perturbative correction is significantly more effective in recovering the missing correlation energy than variationally optimizing the orbitals. \cite{Kossoski_2022}
In the future, it may be worth combining orbital optimization at a lower level of hCI \cite{Kossoski_2022} with the correction provided by perturbation theory.

We further gauged the performance of hCI to describe excited states of closed- and open-shell systems, based on either HF ground-state orbitals or state-specific orbitals.
For excitations of closed-shell systems, $\Delta$hCI2 and $\Delta$hCI3 are comparable to their excitation-based counterparts, $\Delta$CISD and $\Delta$CISDT,
whereas $\Delta$hCI1, $\Delta$hCI1.5, and $\Delta$hCI2.5 are inaccurate bearing in mind their computational cost.
The poor performance at lower orders is ascribed to the different minimal references employed,
a single seniority-zero determinant for the ground state and a single seniority-two CSF for the excited state,
which introduces a strong bias on the classes of determinants accessed in the hCI calculations.
$\Delta$hCI performs significantly better for the doublet transitions when compared to the results for closed shells.
In this case, the reference of both ground and excited states comprises the same type of single seniority-one determinant.
The advantage of using state-specific orbitals over ground-state ones depends on the choice of reference and the order in which the CI is truncated.
When similar references are adopted (as for the doublet transitions), such advantage is already evident at $\Delta$hCI1.
In contrast, for unlike references (as for the closed-shell excitations), the state-specific approach only becomes advantageous at somewhat higher orders ($\Delta$hCI2).
The present findings highlight the challenge in describing, on an equal footing, states with qualitatively different characters, in particular ground and singly-excited states of closed-shell systems.
In this regard, rotations within occupied and virtual orbitals represent yet another factor worth exploring in the future developments of hCI models.
hCI also produced reasonable energies for the automerization barrier of cyclobutadiene.
It remains to be seen how these models perform for other strongly correlated systems, larger molecules, and charge transfer excited states.

The accuracy of the CI models is significantly enhanced thanks to the EN2 correction,
the case of $\Delta$hCI1.5 for closed-shells being the most striking, with a MAE plummeting from \SI{2.80}{\eV} to \SI{0.13}{\eV}.
Low-orders of state-specific CI supplemented with a perturbative correction can thus provide a cost-effective option for the computation of excitation energies.

We put forward the interesting perspective of developing hierarchy-based CC (hCC) methods (hCC1, hCC2, \ldots),
along with derived EOM-CC formulations to target excited states (EOM-hCC1, EOM-hCC2, \ldots).
EOM-CC with single and double excitations (EOM-CCSD) \cite{Rowe_1968,Monkhorst_1977,Koch_1990,Stanton_1993}
is very accurate for describing singly-excited states, \cite{Loos_2018,Loos_2020,Loos_2020a} albeit it slightly overestimates their excitation energies.
In contrast, CISD provides too large excitation energies, \cite{Koch_1990,Kossoski_2023}
even though both CISD and EOM-CCSD methods span the same excited determinants and rely on HF ground-state orbitals.
Here, we found that hCI2 is more accurate than CISD, despite their large absolute errors. In particular, the overestimated energies are less exaggerated in the former.
We ponder whether this improvement would be transferable to EOM-hCC2.
Along the same line, it would be interesting to develop and gauge the performance of cheaper methods like EOM-hCC1.
Importantly, it is not obvious what the computational scaling of such methods would be.
Alternatively, one can envision state-specific hCC methods to target excited states.

Finally, we employed different sCI models to compute excitation energies of closed- and open-shell systems.
Despite the four different models employed for the closed-shell excitations, the results are overall disappointing.
For the open-shell transitions, the outcome is more encouraging.
Despite the improvement brought about by the EN2 correction, the sCI models remain less accurate and more expensive than the eCI or hCI options.
The relevance of these sCI models lies in their possible connection with CC methods.
While this is already established for closed-shell systems (pCCD and DOCI deliver very similar energies), \cite{Bytautas_2011,Allen_1962,Smith_1965,Veillard_1967,Kossoski_2021,Marie_2021}
here we raised the question of whether there exists a polynomial scaling extension of pCCD to open-shell systems that matches sCI1 in terms of computed energies.
In this sense, the present results obtained with $\Delta$sCI1 for open-shell excitations are appealing and encourage the development of a generalized pCCD method.

\begin{acknowledgements}
This work was performed using HPC resources from CALMIP (Toulouse) under allocation 2023-18005.
This project has received funding from the European Research Council (ERC) under the European Union's Horizon 2020 research and innovation programme (Grant agreement No.~863481).
\end{acknowledgements}

\section*{Supporting information available}
\label{sec:SI}

Definition of the statistical measures;
equilibrium geometries of ethylene and vinyl; details about the fitting of potential energy curves;
number of determinants of each CI model;
potential energy curves for \ce{OH}, \ce{CN}, vinyl, and \ce{H7},
computed according to FCI and the various hCI, eCI, and sCI models considered here;
and convergence of the NPE, distance errors, and equilibrium properties of
open-shell (\ce{OH}, \ce{CN}, vinyl, and \ce{H7})
and closed-shell (\ce{HF}, \ce{F2}, ethylene, \ce{N2}, \ce{H4}, and \ce{H8}) systems,
as functions of the number of determinants, according to the various CI models considered here while
corrected with the standard and the renormalized EN2 perturbative energy.
The data associated with the PECs and derived properties can be found at \url{https://github.com/kossoski/open_shell_hCI}.

For the full set of 50 excited states for closed-shell molecules and 19 excited states for open-shell radicals,
total energies and excitation energies obtained with
the various hCI, eCI, and sCI models considered here, in both ground-state-based and state-specific approaches;
number of determinants in the reference; saddle point order associated with the $\Delta$CSF solutions;
reference excitation energies and corresponding methods;
and complementary statistical metrics.
For a subset of 27 excited states for closed-shell molecules and 15 excited states for radicals,
additional total energies and excitation energies obtained with low-order state-specific hCI and eCI models without enforcing spin pure states.
For a subset of 16 excited states for closed-shell molecules and 6 excited states for radicals,
additional total energies and excitation energies obtained with higher-order hCI and eCI models.


\appendix

\section{Number of determinants}
\label{app:appendix}

\begin{figure}[h!]
\includegraphics[width=1.0\linewidth]{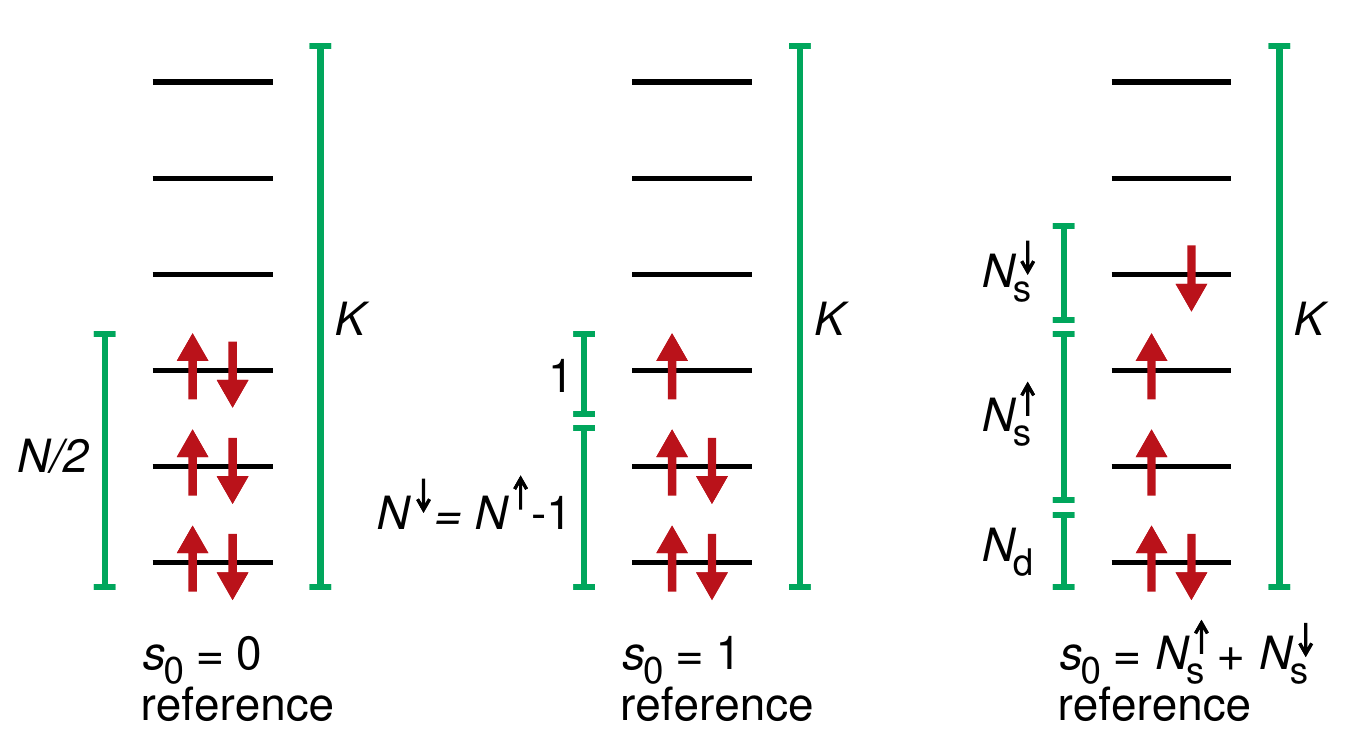}
\caption{
Closed-shell determinant ($s_0=0$) with $N$ electrons (left),
open-shell determinant ($s_0=1$) with $\Na$ spin-up electrons and $\Nb=\Na-1$ spin-down electrons (center),
and general determinant ($s_0=\Nas+\Nbs$), with $\Nd$ doubly-occupied orbitals, $\Nas$ singly-occupied spin-up electrons, and $\Nbs$ singly-occupied spin-down electrons (right).
In the three cases, there are $K$ spatial orbitals.
The number of determinants generated from these three reference determinants, from left to right, which have excitation degree $e$ and seniority $s$
are given in Eqs.~\eqref{eq:det_1}, \eqref{eq:det_2}, and \eqref{eq:det_3}.
}
\label{fig:appendix}
\end{figure}

What is the number of determinants in a given hCI model, defined by the hierarchy $h$ [see Eq.~\eqref{eq:h}] and the reference determinant?
Here, we address this question by first working out the simpler case of a closed-shell reference.
Then, we move to a slightly more complicated case of an open-shell reference with a single unpaired electron.
Finally, we deduce the general case.
The reference determinant for each case is shown in Fig.~\ref{fig:appendix}.

The gist of the deduction is as follows.
We start from the known number of determinants based only on the excitation degree $e$.
Then, we systematically decompose each term into specific contributions,
based on whether the excitation increases or reduces the seniority.
For the decomposition, we make use of Vandermonde's identity:
\begin{equation}
	\binom{n}{k} = \sum_{j}^{k} \binom{m}{j} \binom{n-m}{k-j}.
\end{equation}
Next, an incremental seniority can be assigned to each type of contribution.
Finally, the final answer can be obtained by summing only the contributions with the desired seniority $s$.

We start with the simpler case of a closed-shell determinant with $N$ electrons and $K$ spatial orbitals (left panel of Fig.~\ref{fig:appendix}).
From this reference determinant, the number of excited determinants generated by exciting $e$ electrons, is given by \cite{SzaboBook}
\begin{equation}
	\sum_{p=0}^{e} \binom{N/2}{p} \binom{K-N/2}{p} \binom{N/2}{e-p} \binom{K-N/2}{e-p},
\label{eq:e_closed}
\end{equation}
where the sum expresses the different combinations for exciting spin-up or spin-down electrons.
The first binomial term accounts for the number of possibilities for exciting $p$ spin-up electrons from the $N/2$ orbitals,
the second term represents the number of ways of placing these $p$ electrons into $K-N/2$ orbitals,
and similarly for the latter two terms and the $e-p$ remaining spin-down electrons.

To account for the seniority of the excited determinants, one must disentangle the excitations based on how they change the seniority number,
while keeping track of the previously excited electrons.
Starting from the closed-shell determinant, $p$ spin-up electrons are excited, which increases the seniority by $2p$
(a factor of $p$ due to the unpaired spin-up electrons just excited,
and another factor of $p$ due to the unpaired spin-down electrons left behind).
The first two binomial terms in Eq.~\eqref{eq:e_closed} are left untouched since they are always accompanied by the same change in seniority.
Next, the spin-down electrons can be excited from the same orbitals from which the spin-up electrons were excited (decreasing the seniority by one)
or instead from an orbital that remained doubly occupied (increasing the seniority by one).
The two possibilities are expressed by decomposing the third binomial term in Eq.~\eqref{eq:e_closed} as a sum over the two corresponding binomials, such as
\begin{equation}
	\binom{N/2}{e-p} = \sum_{q=0}^{e-p} \binom{p}{q} \binom{N/2-p}{e-p-q}.
\end{equation}
The first binomial counts the number of possibilities for exciting $q$ spin-down electrons from one of $p$ orbitals for which a spin-up electron was already excited.
By removing the unpaired spin-down electron left behind, the seniority thus decreases by a factor of $q$.
The second term accounts for the complementary excitations, where the $(e-p-q)$ spin-down electrons are chosen out of the $N/2-p$ orbitals that
remained doubly-occupied after exciting the $p$ spin-up electrons.
Therefore, this term increases the seniority by $(e-p-q)$.
We proceed similarly for the fourth binomial term of Eq.~\eqref{eq:e_closed}, by decomposing it into the sum of two other binomials,
where $r$ spin-down electron pairs with the previously spin-up electrons (thus reducing the seniority) and $(e-p-r)$ do not (which increases the seniority).
By collecting the seniority changes from each term: $(2p)$ from the unmodified binomials, $(-q)$ and $(e-p-q)$ from the third binomial, $(-r)$ and $(e-p-r)$ from the fourth one,
the seniority of the excited determinant is given by $s = 2 (e-q-r)$.
Finally, by combining the seniority specific binomials and imposing the desired seniority via a Kronecker delta,
the number of determinants with a given excitation degree $e$ and seniority $s$ produced from a closed-shell reference ($s_0=0$) is given by
\begin{multline}
	\sum_{p=0}^{e} \binom{N/2}{p} \binom{K-N/2}{p}
	\sum_{q=0}^{e-p} \binom{p}{q} \binom{N/2-p}{e-p-q}
	\\ \times
	\sum_{r=0}^{e-p} \binom{p}{r} \binom{K-N/2-p}{e-p-r}
 	\delta_{s,2(e-q-r)}.
\label{eq:det_1}
\end{multline}
To obtain the final number of determinants for a given hierarchy $h$, we simply sum over the allowed combinations of $e$ and $s$ according to Eq.~\eqref{eq:h}.

Moving to the case where the number of spin-up electrons ($\Na$) and spin-down electrons ($\Nb$) differ by one ($\Na=\Nb+1$), which is illustrated in the center panel of Fig.~\ref{fig:appendix},
the number of excited determinants generated exclusively by exciting $e$ electrons,
is analogous to Eq.~\eqref{eq:e_closed}, being given by
\begin{equation}
\sum_{p=0}^{e}
\binom{\Na}{p} \binom{K-\Na}{p} \binom{\Nb}{e-p} \binom{K-\Nb}{e-p}.
\label{eq:e_open}
\end{equation}

The deduction proceeds similarly to the closed-shell case.
The main differences are the following.
The first binomial term of Eq.~\eqref{eq:e_open} is decomposed into the sum of $t$ spin-up electrons excited from the singly-occupied orbital and $(p-t)$ that are excited from the $\Nb$ doubly-occupied ones.
The second binomial is left untouched, as before.
The third one is also decomposed as explained for the closed-shell case, with the difference that the $q$ spin-down electrons are excited from $(p-t)$ orbitals (instead of $p$)
from which spin-up electrons were excited.
Similarly, in the fourth binomial of Eq.~\eqref{eq:e_open}, $r$ spin-down electrons are placed into one of the $(1+p-t)$ orbitals (rather than $p$ in the closed-shell case) which contain an unpaired spin-up electron.
Collecting the individual contributions and setting the targeted seniority,
the number of determinants generated from a $s_0=1$ reference determinant, with excitation degree $e$ and seniority $s$ is given by
\begin{multline}
	\sum_{p=0}^{e}
	\sum_{t=0}^{p} \binom{1}{t} \binom{\Nb}{p-t}
	\binom{K-\Na}{p}
	\sum_{q=0}^{e-p} \binom{p-t}{q} \binom{\Nb-p+t}{e-p-q}
	\\ \times
	\sum_{r=0}^{e-p} \binom{1+p-t}{r} \binom{K-\Nb-1-p+t}{e-p-r}
 	\delta_{s,1+2(e-q-r-t)}.
\label{eq:det_2}
\end{multline}

In the general case, an arbitrary reference determinant is defined by three parameters, the number of doubly-occupied orbitals ($\Nd$),
the number of singly-occupied spin-up electrons ($\Nas = \Na - \Nd$),
and the number of singly-occupied spin-down electrons ($\Nbs = \Nb - \Nd$).
(Alternatively, one could employ the number of electrons $N$, $\Nd$, and the spin quantum number $S_z = (\Na-\Nb)/2$.)
For $K$ spatial orbitals, the number of virtual orbitals is $\Nv = K - \Nd - \Nas - \Nbs$.

The deduction is somewhat more involved but follows along the same lines as the two previously discussed cases.
The starting point is also given by Eq.~\eqref{eq:e_open}, with each binomial being written as a sum of two or more binomials that distinguish the change in seniority.
Here, we just state how each term is decomposed, where the details can be checked by inspection of the final result presented below [see Eq.~\eqref{eq:det_3}].
The first binomial of Eq.~\eqref{eq:e_open} is decomposed into the contributions of spin-up electrons excited from the $\Nd$ doubly- and from the $\Nas$ singly-occupied orbitals.
The second binomial is decomposed into the spin-up electrons being excited to the $\Nv$ virtual orbitals and to the $\Nbs$ singly-occupied orbitals.
The third one undergoes two decompositions and is thus expressed as a double sum over three binomials.
The spin-down electrons are distinguished between the $\Nbs$ singly-occupied orbitals, a subset of $\Nd$ doubly-occupied orbitals for which spin-up electrons were excited,
and the complementary subset for which no spin-up electrons were excited.
Finally, the fourth binomial of Eq.~\eqref{eq:e_open} counts the number of possibilities to place the excited spin-down electrons.
It is first decomposed into two terms, based on whether they are promoted to the $\Nv$ virtual orbitals or to the $\Nas$ singly-occupied orbitals.
Each term is further decomposed into two terms, accounting for the subset of $\Nv$ orbitals which now contain spin-up electrons,
and the subset of $\Nas$ orbitals from which spin-up electrons were removed,
thus leading to three sums over four binomials.
The final result for the number of determinants generated from an arbitrary reference determinant, constrained to have excitation degree $e$ and seniority $s$ is given by
\begin{widetext}
\begin{multline}
	\sum_{p=0}^{e}
	\sum_{t=0}^{p} \binom{\Nas}{t} \binom{\Nd}{p-t}
	\sum_{w=0}^{p} \binom{\Nv}{w} \binom{\Nbs}{p-w}
	\sum_{u=0}^{e-p} \binom{\Nbs}{u}
	\sum_{v=0}^{e-p-u} \binom{p-t}{v} \binom{\Nd-p+t}{e-p-u-v}
	\sum_{r=0}^{e-p} \sum_{m=0}^{r} \binom{w}{m} \binom{\Nv-w}{r-m}
	\\ \times
	\sum_{n=0}^{e-p-r} \binom{t}{n} \binom{\Nas-t}{e-p-r-n}
	\delta_{s,\Nas+\Nbs+2(e+w-p-t-u-v-m-n)}.
\label{eq:det_3}
\end{multline}
\end{widetext}

\bibliography{manuscript}

\begin{thebibliography}{108}%
\makeatletter
\providecommand \@ifxundefined [1]{%
 \@ifx{#1\undefined}
}%
\providecommand \@ifnum [1]{%
 \ifnum #1\expandafter \@firstoftwo
 \else \expandafter \@secondoftwo
 \fi
}%
\providecommand \@ifx [1]{%
 \ifx #1\expandafter \@firstoftwo
 \else \expandafter \@secondoftwo
 \fi
}%
\providecommand \natexlab [1]{#1}%
\providecommand \enquote  [1]{``#1''}%
\providecommand \bibnamefont  [1]{#1}%
\providecommand \bibfnamefont [1]{#1}%
\providecommand \citenamefont [1]{#1}%
\providecommand \href@noop [0]{\@secondoftwo}%
\providecommand \href [0]{\begingroup \@sanitize@url \@href}%
\providecommand \@href[1]{\@@startlink{#1}\@@href}%
\providecommand \@@href[1]{\endgroup#1\@@endlink}%
\providecommand \@sanitize@url [0]{\catcode `\\12\catcode `\$12\catcode
  `\&12\catcode `\#12\catcode `\^12\catcode `\_12\catcode `\%12\relax}%
\providecommand \@@startlink[1]{}%
\providecommand \@@endlink[0]{}%
\providecommand \url  [0]{\begingroup\@sanitize@url \@url }%
\providecommand \@url [1]{\endgroup\@href {#1}{\urlprefix }}%
\providecommand \urlprefix  [0]{URL }%
\providecommand \Eprint [0]{\href }%
\providecommand \doibase [0]{http://dx.doi.org/}%
\providecommand \selectlanguage [0]{\@gobble}%
\providecommand \bibinfo  [0]{\@secondoftwo}%
\providecommand \bibfield  [0]{\@secondoftwo}%
\providecommand \translation [1]{[#1]}%
\providecommand \BibitemOpen [0]{}%
\providecommand \bibitemStop [0]{}%
\providecommand \bibitemNoStop [0]{.\EOS\space}%
\providecommand \EOS [0]{\spacefactor3000\relax}%
\providecommand \BibitemShut  [1]{\csname bibitem#1\endcsname}%
\let\auto@bib@innerbib\@empty
\bibitem [{\citenamefont {Szabo}\ and\ \citenamefont
  {Ostlund}(1989)}]{SzaboBook}%
  \BibitemOpen
  \bibfield  {author} {\bibinfo {author} {\bibfnamefont {A.}~\bibnamefont
  {Szabo}}\ and\ \bibinfo {author} {\bibfnamefont {N.~S.}\ \bibnamefont
  {Ostlund}},\ }\href@noop {} {\emph {\bibinfo {title} {Modern quantum
  chemistry}}}\ (\bibinfo  {publisher} {McGraw-Hill},\ \bibinfo {address} {New
  York},\ \bibinfo {year} {1989})\BibitemShut {NoStop}%
\bibitem [{\citenamefont {Helgaker}, \citenamefont {J{\o}rgensen},\ and\
  \citenamefont {Olsen}(2013)}]{Helgakerbook}%
  \BibitemOpen
  \bibfield  {author} {\bibinfo {author} {\bibfnamefont {T.}~\bibnamefont
  {Helgaker}}, \bibinfo {author} {\bibfnamefont {P.}~\bibnamefont
  {J{\o}rgensen}}, \ and\ \bibinfo {author} {\bibfnamefont {J.}~\bibnamefont
  {Olsen}},\ }\href@noop {} {\emph {\bibinfo {title} {Molecular
  Electronic-Structure Theory}}}\ (\bibinfo  {publisher} {John Wiley \& Sons,
  Inc.},\ \bibinfo {year} {2013})\BibitemShut {NoStop}%
\bibitem [{\citenamefont {Bytautas}\ \emph {et~al.}(2011)\citenamefont
  {Bytautas}, \citenamefont {Henderson}, \citenamefont {{Jim{\'e}nez-Hoyos}},
  \citenamefont {Ellis},\ and\ \citenamefont {Scuseria}}]{Bytautas_2011}%
  \BibitemOpen
  \bibfield  {author} {\bibinfo {author} {\bibfnamefont {L.}~\bibnamefont
  {Bytautas}}, \bibinfo {author} {\bibfnamefont {T.~M.}\ \bibnamefont
  {Henderson}}, \bibinfo {author} {\bibfnamefont {C.~A.}\ \bibnamefont
  {{Jim{\'e}nez-Hoyos}}}, \bibinfo {author} {\bibfnamefont {J.~K.}\
  \bibnamefont {Ellis}}, \ and\ \bibinfo {author} {\bibfnamefont {G.~E.}\
  \bibnamefont {Scuseria}},\ }\href {\doibase 10.1063/1.3613706} {\bibfield
  {journal} {\bibinfo  {journal} {J. Chem. Phys.}\ }\textbf {\bibinfo {volume}
  {135}},\ \bibinfo {pages} {044119} (\bibinfo {year} {2011})}\BibitemShut
  {NoStop}%
\bibitem [{\citenamefont {Allen}\ and\ \citenamefont
  {Shull}(1962)}]{Allen_1962}%
  \BibitemOpen
  \bibfield  {author} {\bibinfo {author} {\bibfnamefont {T.~L.}\ \bibnamefont
  {Allen}}\ and\ \bibinfo {author} {\bibfnamefont {H.}~\bibnamefont {Shull}},\
  }\href {\doibase 10.1021/j100818a001} {\bibfield  {journal} {\bibinfo
  {journal} {J. Phys. Chem.}\ }\textbf {\bibinfo {volume} {66}},\ \bibinfo
  {pages} {2281} (\bibinfo {year} {1962})}\BibitemShut {NoStop}%
\bibitem [{\citenamefont {Smith}\ and\ \citenamefont
  {Fogel}(1965)}]{Smith_1965}%
  \BibitemOpen
  \bibfield  {author} {\bibinfo {author} {\bibfnamefont {D.~W.}\ \bibnamefont
  {Smith}}\ and\ \bibinfo {author} {\bibfnamefont {S.~J.}\ \bibnamefont
  {Fogel}},\ }\href {\doibase 10.1063/1.1701519} {\bibfield  {journal}
  {\bibinfo  {journal} {J. Chem. Phys.}\ }\textbf {\bibinfo {volume} {43}},\
  \bibinfo {pages} {S91} (\bibinfo {year} {1965})}\BibitemShut {NoStop}%
\bibitem [{\citenamefont {Veillard}\ and\ \citenamefont
  {Clementi}(1967)}]{Veillard_1967}%
  \BibitemOpen
  \bibfield  {author} {\bibinfo {author} {\bibfnamefont {A.}~\bibnamefont
  {Veillard}}\ and\ \bibinfo {author} {\bibfnamefont {E.}~\bibnamefont
  {Clementi}},\ }\href {\doibase 10.1007/BF01151915} {\bibfield  {journal}
  {\bibinfo  {journal} {Theoret. Chim. Acta}\ }\textbf {\bibinfo {volume}
  {7}},\ \bibinfo {pages} {133} (\bibinfo {year} {1967})}\BibitemShut {NoStop}%
\bibitem [{\citenamefont {Bytautas}, \citenamefont {Scuseria},\ and\
  \citenamefont {Ruedenberg}(2015)}]{Bytautas_2015}%
  \BibitemOpen
  \bibfield  {author} {\bibinfo {author} {\bibfnamefont {L.}~\bibnamefont
  {Bytautas}}, \bibinfo {author} {\bibfnamefont {G.~E.}\ \bibnamefont
  {Scuseria}}, \ and\ \bibinfo {author} {\bibfnamefont {K.}~\bibnamefont
  {Ruedenberg}},\ }\href {\doibase 10.1063/1.4929904} {\bibfield  {journal}
  {\bibinfo  {journal} {J. Chem. Phys.}\ }\textbf {\bibinfo {volume} {143}},\
  \bibinfo {pages} {094105} (\bibinfo {year} {2015})}\BibitemShut {NoStop}%
\bibitem [{\citenamefont {Alcoba}\ \emph
  {et~al.}(2014{\natexlab{a}})\citenamefont {Alcoba}, \citenamefont {Torre},
  \citenamefont {Lain}, \citenamefont {O{\~{n}}a}, \citenamefont {Capuzzi},
  \citenamefont {{Van Raemdonck}}, \citenamefont {Bultinck},\ and\
  \citenamefont {{Van Neck}}}]{Alcoba_2014}%
  \BibitemOpen
  \bibfield  {author} {\bibinfo {author} {\bibfnamefont {D.~R.}\ \bibnamefont
  {Alcoba}}, \bibinfo {author} {\bibfnamefont {A.}~\bibnamefont {Torre}},
  \bibinfo {author} {\bibfnamefont {L.}~\bibnamefont {Lain}}, \bibinfo {author}
  {\bibfnamefont {O.~B.}\ \bibnamefont {O{\~{n}}a}}, \bibinfo {author}
  {\bibfnamefont {P.}~\bibnamefont {Capuzzi}}, \bibinfo {author} {\bibfnamefont
  {M.}~\bibnamefont {{Van Raemdonck}}}, \bibinfo {author} {\bibfnamefont
  {P.}~\bibnamefont {Bultinck}}, \ and\ \bibinfo {author} {\bibfnamefont
  {D.}~\bibnamefont {{Van Neck}}},\ }\href {\doibase 10.1063/1.4904755}
  {\bibfield  {journal} {\bibinfo  {journal} {J. Chem. Phys.}\ }\textbf
  {\bibinfo {volume} {141}},\ \bibinfo {pages} {244118} (\bibinfo {year}
  {2014}{\natexlab{a}})}\BibitemShut {NoStop}%
\bibitem [{\citenamefont {Alcoba}\ \emph
  {et~al.}(2014{\natexlab{b}})\citenamefont {Alcoba}, \citenamefont {Torre},
  \citenamefont {Lain}, \citenamefont {Massaccesi},\ and\ \citenamefont
  {O{\~{n}}a}}]{Alcoba_2014b}%
  \BibitemOpen
  \bibfield  {author} {\bibinfo {author} {\bibfnamefont {D.~R.}\ \bibnamefont
  {Alcoba}}, \bibinfo {author} {\bibfnamefont {A.}~\bibnamefont {Torre}},
  \bibinfo {author} {\bibfnamefont {L.}~\bibnamefont {Lain}}, \bibinfo {author}
  {\bibfnamefont {G.~E.}\ \bibnamefont {Massaccesi}}, \ and\ \bibinfo {author}
  {\bibfnamefont {O.~B.}\ \bibnamefont {O{\~{n}}a}},\ }\href {\doibase
  10.1063/1.4882881} {\bibfield  {journal} {\bibinfo  {journal} {J. Chem.
  Phys.}\ }\textbf {\bibinfo {volume} {140}},\ \bibinfo {pages} {234103}
  (\bibinfo {year} {2014}{\natexlab{b}})}\BibitemShut {NoStop}%
\bibitem [{\citenamefont {Kossoski}, \citenamefont {Damour},\ and\
  \citenamefont {Loos}(2022)}]{Kossoski_2022}%
  \BibitemOpen
  \bibfield  {author} {\bibinfo {author} {\bibfnamefont {F.}~\bibnamefont
  {Kossoski}}, \bibinfo {author} {\bibfnamefont {Y.}~\bibnamefont {Damour}}, \
  and\ \bibinfo {author} {\bibfnamefont {P.-F.}\ \bibnamefont {Loos}},\ }\href
  {\doibase 10.1021/acs.jpclett.2c00730} {\bibfield  {journal} {\bibinfo
  {journal} {J. Phys. Chem. Lett.}\ }\textbf {\bibinfo {volume} {13}},\
  \bibinfo {pages} {4342} (\bibinfo {year} {2022})}\BibitemShut {NoStop}%
\bibitem [{\citenamefont {Van~Raemdonck}\ \emph {et~al.}(2015)\citenamefont
  {Van~Raemdonck}, \citenamefont {Alcoba}, \citenamefont {Poelmans},
  \citenamefont {De~Baerdemacker}, \citenamefont {Torre}, \citenamefont {Lain},
  \citenamefont {Massaccesi}, \citenamefont {Van~Neck},\ and\ \citenamefont
  {Bultinck}}]{Raemdonck_2015}%
  \BibitemOpen
  \bibfield  {author} {\bibinfo {author} {\bibfnamefont {M.}~\bibnamefont
  {Van~Raemdonck}}, \bibinfo {author} {\bibfnamefont {D.~R.}\ \bibnamefont
  {Alcoba}}, \bibinfo {author} {\bibfnamefont {W.}~\bibnamefont {Poelmans}},
  \bibinfo {author} {\bibfnamefont {S.}~\bibnamefont {De~Baerdemacker}},
  \bibinfo {author} {\bibfnamefont {A.}~\bibnamefont {Torre}}, \bibinfo
  {author} {\bibfnamefont {L.}~\bibnamefont {Lain}}, \bibinfo {author}
  {\bibfnamefont {G.~E.}\ \bibnamefont {Massaccesi}}, \bibinfo {author}
  {\bibfnamefont {D.}~\bibnamefont {Van~Neck}}, \ and\ \bibinfo {author}
  {\bibfnamefont {P.}~\bibnamefont {Bultinck}},\ }\href {\doibase
  10.1063/1.4930260} {\bibfield  {journal} {\bibinfo  {journal} {J. Chem.
  Phys.}\ }\textbf {\bibinfo {volume} {143}},\ \bibinfo {pages} {104106}
  (\bibinfo {year} {2015})}\BibitemShut {NoStop}%
\bibitem [{\citenamefont {Alcoba}\ \emph {et~al.}(2018)\citenamefont {Alcoba},
  \citenamefont {Torre}, \citenamefont {Lain}, \citenamefont {O{\~n}a},
  \citenamefont {Massaccesi},\ and\ \citenamefont {Capuzzi}}]{Alcoba_2018}%
  \BibitemOpen
  \bibfield  {author} {\bibinfo {author} {\bibfnamefont {D.~R.}\ \bibnamefont
  {Alcoba}}, \bibinfo {author} {\bibfnamefont {A.}~\bibnamefont {Torre}},
  \bibinfo {author} {\bibfnamefont {L.}~\bibnamefont {Lain}}, \bibinfo {author}
  {\bibfnamefont {O.~B.}\ \bibnamefont {O{\~n}a}}, \bibinfo {author}
  {\bibfnamefont {G.~E.}\ \bibnamefont {Massaccesi}}, \ and\ \bibinfo {author}
  {\bibfnamefont {P.}~\bibnamefont {Capuzzi}},\ }in\ \href {\doibase
  https://doi.org/10.1016/bs.aiq.2017.05.003} {\emph {\bibinfo {booktitle}
  {Novel Electronic Structure Theory: General Innovations and Strongly
  Correlated Systems}}},\ \bibinfo {series} {Advances in Quantum Chemistry},
  Vol.~\bibinfo {volume} {76},\ \bibinfo {editor} {edited by\ \bibinfo {editor}
  {\bibfnamefont {P.~E.}\ \bibnamefont {Hoggan}}}\ (\bibinfo  {publisher}
  {Academic Press},\ \bibinfo {year} {2018})\ pp.\ \bibinfo {pages}
  {315--332}\BibitemShut {NoStop}%
\bibitem [{\citenamefont {Hurley}, \citenamefont {Lennard-Jones},\ and\
  \citenamefont {Pople}(1953)}]{Hurley_1953}%
  \BibitemOpen
  \bibfield  {author} {\bibinfo {author} {\bibfnamefont {A.~C.}\ \bibnamefont
  {Hurley}}, \bibinfo {author} {\bibfnamefont {J.~E.}\ \bibnamefont
  {Lennard-Jones}}, \ and\ \bibinfo {author} {\bibfnamefont {J.~A.}\
  \bibnamefont {Pople}},\ }\href {\doibase 10.1098/rspa.1953.0198} {\bibfield
  {journal} {\bibinfo  {journal} {Proc. R. Soc. A}\ }\textbf {\bibinfo {volume}
  {220}},\ \bibinfo {pages} {446} (\bibinfo {year} {1953})}\BibitemShut
  {NoStop}%
\bibitem [{\citenamefont {Cullen}(1996)}]{Cullen_1996}%
  \BibitemOpen
  \bibfield  {author} {\bibinfo {author} {\bibfnamefont {J.}~\bibnamefont
  {Cullen}},\ }\href {\doibase https://doi.org/10.1016/0301-0104(95)00321-5}
  {\bibfield  {journal} {\bibinfo  {journal} {Chem. Phys.}\ }\textbf {\bibinfo
  {volume} {202}},\ \bibinfo {pages} {217} (\bibinfo {year}
  {1996})}\BibitemShut {NoStop}%
\bibitem [{\citenamefont {{Van Voorhis}}\ and\ \citenamefont
  {Head-Gordon}(2000)}]{VanVoorhis_2000}%
  \BibitemOpen
  \bibfield  {author} {\bibinfo {author} {\bibfnamefont {T.}~\bibnamefont {{Van
  Voorhis}}}\ and\ \bibinfo {author} {\bibfnamefont {M.}~\bibnamefont
  {Head-Gordon}},\ }\href {\doibase
  https://doi.org/10.1016/S0009-2614(99)01413-X} {\bibfield  {journal}
  {\bibinfo  {journal} {Chem. Phys. Lett.}\ }\textbf {\bibinfo {volume}
  {317}},\ \bibinfo {pages} {575} (\bibinfo {year} {2000})}\BibitemShut
  {NoStop}%
\bibitem [{\citenamefont {Parkhill}, \citenamefont {Lawler},\ and\
  \citenamefont {Head-Gordon}(2009)}]{Parkhill_2009}%
  \BibitemOpen
  \bibfield  {author} {\bibinfo {author} {\bibfnamefont {J.~A.}\ \bibnamefont
  {Parkhill}}, \bibinfo {author} {\bibfnamefont {K.}~\bibnamefont {Lawler}}, \
  and\ \bibinfo {author} {\bibfnamefont {M.}~\bibnamefont {Head-Gordon}},\
  }\href {\doibase 10.1063/1.3086027} {\bibfield  {journal} {\bibinfo
  {journal} {J. Chem. Phys.}\ }\textbf {\bibinfo {volume} {130}},\ \bibinfo
  {pages} {084101} (\bibinfo {year} {2009})}\BibitemShut {NoStop}%
\bibitem [{\citenamefont {Parkhill}\ and\ \citenamefont
  {Head-Gordon}(2010)}]{Parkhill_2010}%
  \BibitemOpen
  \bibfield  {author} {\bibinfo {author} {\bibfnamefont {J.~A.}\ \bibnamefont
  {Parkhill}}\ and\ \bibinfo {author} {\bibfnamefont {M.}~\bibnamefont
  {Head-Gordon}},\ }\href {\doibase 10.1063/1.3456001} {\bibfield  {journal}
  {\bibinfo  {journal} {J. Chem. Phys.}\ }\textbf {\bibinfo {volume} {133}},\
  \bibinfo {pages} {024103} (\bibinfo {year} {2010})}\BibitemShut {NoStop}%
\bibitem [{\citenamefont {Lehtola}, \citenamefont {Parkhill},\ and\
  \citenamefont {Head-Gordon}(2016)}]{Lehtola_2016}%
  \BibitemOpen
  \bibfield  {author} {\bibinfo {author} {\bibfnamefont {S.}~\bibnamefont
  {Lehtola}}, \bibinfo {author} {\bibfnamefont {J.}~\bibnamefont {Parkhill}}, \
  and\ \bibinfo {author} {\bibfnamefont {M.}~\bibnamefont {Head-Gordon}},\
  }\href {\doibase 10.1063/1.4964317} {\bibfield  {journal} {\bibinfo
  {journal} {J. Chem. Phys.}\ }\textbf {\bibinfo {volume} {145}},\ \bibinfo
  {pages} {134110} (\bibinfo {year} {2016})}\BibitemShut {NoStop}%
\bibitem [{\citenamefont {Lehtola}, \citenamefont {Parkhill},\ and\
  \citenamefont {Head-Gordon}(2018)}]{Lehtola_2018}%
  \BibitemOpen
  \bibfield  {author} {\bibinfo {author} {\bibfnamefont {S.}~\bibnamefont
  {Lehtola}}, \bibinfo {author} {\bibfnamefont {J.}~\bibnamefont {Parkhill}}, \
  and\ \bibinfo {author} {\bibfnamefont {M.}~\bibnamefont {Head-Gordon}},\
  }\href {\doibase 10.1080/00268976.2017.1342009} {\bibfield  {journal}
  {\bibinfo  {journal} {Mol. Phys.}\ }\textbf {\bibinfo {volume} {116}},\
  \bibinfo {pages} {547} (\bibinfo {year} {2018})}\BibitemShut {NoStop}%
\bibitem [{\citenamefont {Ziegler}, \citenamefont {Rauk},\ and\ \citenamefont
  {Baerends}(1977)}]{Ziegler_1977}%
  \BibitemOpen
  \bibfield  {author} {\bibinfo {author} {\bibfnamefont {T.}~\bibnamefont
  {Ziegler}}, \bibinfo {author} {\bibfnamefont {A.}~\bibnamefont {Rauk}}, \
  and\ \bibinfo {author} {\bibfnamefont {E.~J.}\ \bibnamefont {Baerends}},\
  }\href {\doibase 10.1007/BF00551551} {\bibfield  {journal} {\bibinfo
  {journal} {Theor. Chim. Acta}\ }\textbf {\bibinfo {volume} {43}},\ \bibinfo
  {pages} {261} (\bibinfo {year} {1977})}\BibitemShut {NoStop}%
\bibitem [{\citenamefont {Burton}\ and\ \citenamefont
  {Wales}(2021)}]{Burton_2021}%
  \BibitemOpen
  \bibfield  {author} {\bibinfo {author} {\bibfnamefont {H.~G.~A.}\
  \bibnamefont {Burton}}\ and\ \bibinfo {author} {\bibfnamefont {D.~J.}\
  \bibnamefont {Wales}},\ }\href {\doibase 10.1021/acs.jctc.0c00772} {\bibfield
   {journal} {\bibinfo  {journal} {J. Chem. Theory Comput.}\ }\textbf {\bibinfo
  {volume} {17}},\ \bibinfo {pages} {151} (\bibinfo {year} {2021})}\BibitemShut
  {NoStop}%
\bibitem [{\citenamefont {Shea}\ and\ \citenamefont
  {Neuscamman}(2018)}]{Shea_2018}%
  \BibitemOpen
  \bibfield  {author} {\bibinfo {author} {\bibfnamefont {J.~A.~R.}\
  \bibnamefont {Shea}}\ and\ \bibinfo {author} {\bibfnamefont {E.}~\bibnamefont
  {Neuscamman}},\ }\href {\doibase 10.1063/1.5045056} {\bibfield  {journal}
  {\bibinfo  {journal} {J. Chem. Phys.}\ }\textbf {\bibinfo {volume} {149}},\
  \bibinfo {pages} {081101} (\bibinfo {year} {2018})}\BibitemShut {NoStop}%
\bibitem [{\citenamefont {Tran}, \citenamefont {Shea},\ and\ \citenamefont
  {Neuscamman}(2019)}]{Tran_2019}%
  \BibitemOpen
  \bibfield  {author} {\bibinfo {author} {\bibfnamefont {L.~N.}\ \bibnamefont
  {Tran}}, \bibinfo {author} {\bibfnamefont {J.~A.~R.}\ \bibnamefont {Shea}}, \
  and\ \bibinfo {author} {\bibfnamefont {E.}~\bibnamefont {Neuscamman}},\
  }\href {\doibase 10.1021/acs.jctc.9b00351} {\bibfield  {journal} {\bibinfo
  {journal} {J. Chem. Theory Comput.}\ }\textbf {\bibinfo {volume} {15}},\
  \bibinfo {pages} {4790} (\bibinfo {year} {2019})}\BibitemShut {NoStop}%
\bibitem [{\citenamefont {Tran}\ and\ \citenamefont
  {Neuscamman}(2020)}]{Tran_2020}%
  \BibitemOpen
  \bibfield  {author} {\bibinfo {author} {\bibfnamefont {L.~N.}\ \bibnamefont
  {Tran}}\ and\ \bibinfo {author} {\bibfnamefont {E.}~\bibnamefont
  {Neuscamman}},\ }\href {\doibase 10.1021/acs.jpca.0c07593} {\bibfield
  {journal} {\bibinfo  {journal} {J. Phys. Chem. A}\ }\textbf {\bibinfo
  {volume} {124}},\ \bibinfo {pages} {8273} (\bibinfo {year}
  {2020})}\BibitemShut {NoStop}%
\bibitem [{\citenamefont {Hardikar}\ and\ \citenamefont
  {Neuscamman}(2020)}]{Hardikar_2020}%
  \BibitemOpen
  \bibfield  {author} {\bibinfo {author} {\bibfnamefont {T.~S.}\ \bibnamefont
  {Hardikar}}\ and\ \bibinfo {author} {\bibfnamefont {E.}~\bibnamefont
  {Neuscamman}},\ }\href {\doibase 10.1063/5.0019557} {\bibfield  {journal}
  {\bibinfo  {journal} {J. Chem. Phys.}\ }\textbf {\bibinfo {volume} {153}},\
  \bibinfo {pages} {164108} (\bibinfo {year} {2020})}\BibitemShut {NoStop}%
\bibitem [{\citenamefont {Shea}, \citenamefont {Gwin},\ and\ \citenamefont
  {Neuscamman}(2020)}]{Shea_2020}%
  \BibitemOpen
  \bibfield  {author} {\bibinfo {author} {\bibfnamefont {J.~A.~R.}\
  \bibnamefont {Shea}}, \bibinfo {author} {\bibfnamefont {E.}~\bibnamefont
  {Gwin}}, \ and\ \bibinfo {author} {\bibfnamefont {E.}~\bibnamefont
  {Neuscamman}},\ }\href {\doibase 10.1021/acs.jctc.9b01105} {\bibfield
  {journal} {\bibinfo  {journal} {J. Chem. Theory Comput.}\ }\textbf {\bibinfo
  {volume} {16}},\ \bibinfo {pages} {1526} (\bibinfo {year}
  {2020})}\BibitemShut {NoStop}%
\bibitem [{\citenamefont {Burton}(2022)}]{Burton_2022}%
  \BibitemOpen
  \bibfield  {author} {\bibinfo {author} {\bibfnamefont {H.~G.}\ \bibnamefont
  {Burton}},\ }\href {\doibase 10.1021/acs.jctc.1c01089} {\bibfield  {journal}
  {\bibinfo  {journal} {J. Chem. Theory Comput.}\ }\textbf {\bibinfo {volume}
  {18}},\ \bibinfo {pages} {1512} (\bibinfo {year} {2022})}\BibitemShut
  {NoStop}%
\bibitem [{\citenamefont {Hanscam}\ and\ \citenamefont
  {Neuscamman}(2022)}]{Hanscam_2022}%
  \BibitemOpen
  \bibfield  {author} {\bibinfo {author} {\bibfnamefont {R.}~\bibnamefont
  {Hanscam}}\ and\ \bibinfo {author} {\bibfnamefont {E.}~\bibnamefont
  {Neuscamman}},\ }\href {\doibase 10.1021/acs.jctc.2c00639} {\bibfield
  {journal} {\bibinfo  {journal} {J. Chem. Theory Comput.}\ }\textbf {\bibinfo
  {volume} {18}},\ \bibinfo {pages} {6608} (\bibinfo {year}
  {2022})}\BibitemShut {NoStop}%
\bibitem [{\citenamefont {Kossoski}\ and\ \citenamefont
  {Loos}(2023)}]{Kossoski_2023}%
  \BibitemOpen
  \bibfield  {author} {\bibinfo {author} {\bibfnamefont {F.}~\bibnamefont
  {Kossoski}}\ and\ \bibinfo {author} {\bibfnamefont {P.-F.}\ \bibnamefont
  {Loos}},\ }\href {\doibase 10.1021/acs.jctc.3c00057} {\bibfield  {journal}
  {\bibinfo  {journal} {J. Chem. Theory Comput.}\ }\textbf {\bibinfo {volume}
  {19}},\ \bibinfo {pages} {2258} (\bibinfo {year} {2023})}\BibitemShut
  {NoStop}%
\bibitem [{\citenamefont {Marie}\ and\ \citenamefont
  {Burton}(2023)}]{Marie_2023}%
  \BibitemOpen
  \bibfield  {author} {\bibinfo {author} {\bibfnamefont {A.}~\bibnamefont
  {Marie}}\ and\ \bibinfo {author} {\bibfnamefont {H.~G.~A.}\ \bibnamefont
  {Burton}},\ }\href {\doibase 10.1021/acs.jpca.3c00603} {\bibfield  {journal}
  {\bibinfo  {journal} {J. Phys. Chem. A}\ }\textbf {\bibinfo {volume} {127}},\
  \bibinfo {pages} {4538} (\bibinfo {year} {2023})}\BibitemShut {NoStop}%
\bibitem [{\citenamefont {Tran}\ and\ \citenamefont
  {Neuscamman}(2023)}]{Tran_2023}%
  \BibitemOpen
  \bibfield  {author} {\bibinfo {author} {\bibfnamefont {L.~N.}\ \bibnamefont
  {Tran}}\ and\ \bibinfo {author} {\bibfnamefont {E.}~\bibnamefont
  {Neuscamman}},\ }\href {\doibase 10.1021/acs.jpclett.3c01308} {\bibfield
  {journal} {\bibinfo  {journal} {J. Phys. Chem. Lett.}\ }\textbf {\bibinfo
  {volume} {14}},\ \bibinfo {pages} {7454} (\bibinfo {year}
  {2023})}\BibitemShut {NoStop}%
\bibitem [{\citenamefont {Filatov}\ and\ \citenamefont
  {Shaik}(1999)}]{Filatov_1999}%
  \BibitemOpen
  \bibfield  {author} {\bibinfo {author} {\bibfnamefont {M.}~\bibnamefont
  {Filatov}}\ and\ \bibinfo {author} {\bibfnamefont {S.}~\bibnamefont
  {Shaik}},\ }\href {\doibase https://doi.org/10.1016/S0009-2614(99)00336-X}
  {\bibfield  {journal} {\bibinfo  {journal} {Chem. Phys. Lett.}\ }\textbf
  {\bibinfo {volume} {304}},\ \bibinfo {pages} {429} (\bibinfo {year}
  {1999})}\BibitemShut {NoStop}%
\bibitem [{\citenamefont {Kowalczyk}, \citenamefont {Yost},\ and\ \citenamefont
  {Voorhis}(2011)}]{Kowalczyk_2011}%
  \BibitemOpen
  \bibfield  {author} {\bibinfo {author} {\bibfnamefont {T.}~\bibnamefont
  {Kowalczyk}}, \bibinfo {author} {\bibfnamefont {S.~R.}\ \bibnamefont {Yost}},
  \ and\ \bibinfo {author} {\bibfnamefont {T.~V.}\ \bibnamefont {Voorhis}},\
  }\href {\doibase 10.1063/1.3530801} {\bibfield  {journal} {\bibinfo
  {journal} {J. Chem. Phys.}\ }\textbf {\bibinfo {volume} {134}},\ \bibinfo
  {pages} {054128} (\bibinfo {year} {2011})}\BibitemShut {NoStop}%
\bibitem [{\citenamefont {Kowalczyk}\ \emph {et~al.}(2013)\citenamefont
  {Kowalczyk}, \citenamefont {Tsuchimochi}, \citenamefont {Chen}, \citenamefont
  {Top},\ and\ \citenamefont {Van~Voorhis}}]{Kowalczyk_2013}%
  \BibitemOpen
  \bibfield  {author} {\bibinfo {author} {\bibfnamefont {T.}~\bibnamefont
  {Kowalczyk}}, \bibinfo {author} {\bibfnamefont {T.}~\bibnamefont
  {Tsuchimochi}}, \bibinfo {author} {\bibfnamefont {P.-T.}\ \bibnamefont
  {Chen}}, \bibinfo {author} {\bibfnamefont {L.}~\bibnamefont {Top}}, \ and\
  \bibinfo {author} {\bibfnamefont {T.}~\bibnamefont {Van~Voorhis}},\ }\href
  {\doibase 10.1063/1.4801790} {\bibfield  {journal} {\bibinfo  {journal} {J.
  Chem. Phys.}\ }\textbf {\bibinfo {volume} {138}},\ \bibinfo {pages} {164101}
  (\bibinfo {year} {2013})}\BibitemShut {NoStop}%
\bibitem [{\citenamefont {Gilbert}, \citenamefont {Besley},\ and\ \citenamefont
  {Gill}(2008)}]{Gilbert_2008}%
  \BibitemOpen
  \bibfield  {author} {\bibinfo {author} {\bibfnamefont {A.~T.}\ \bibnamefont
  {Gilbert}}, \bibinfo {author} {\bibfnamefont {N.~A.}\ \bibnamefont {Besley}},
  \ and\ \bibinfo {author} {\bibfnamefont {P.~M.}\ \bibnamefont {Gill}},\
  }\href {\doibase 10.1021/jp801738f} {\bibfield  {journal} {\bibinfo
  {journal} {J. Phys. Chem. A}\ }\textbf {\bibinfo {volume} {112}},\ \bibinfo
  {pages} {13164} (\bibinfo {year} {2008})}\BibitemShut {NoStop}%
\bibitem [{\citenamefont {Barca}, \citenamefont {Gilbert},\ and\ \citenamefont
  {Gill}(2018)}]{Barca_2018}%
  \BibitemOpen
  \bibfield  {author} {\bibinfo {author} {\bibfnamefont {G.~M.}\ \bibnamefont
  {Barca}}, \bibinfo {author} {\bibfnamefont {A.~T.}\ \bibnamefont {Gilbert}},
  \ and\ \bibinfo {author} {\bibfnamefont {P.~M.}\ \bibnamefont {Gill}},\
  }\href {\doibase 10.1021/acs.jctc.7b00994} {\bibfield  {journal} {\bibinfo
  {journal} {J. Chem. Theory Comput.}\ }\textbf {\bibinfo {volume} {14}},\
  \bibinfo {pages} {1501} (\bibinfo {year} {2018})}\BibitemShut {NoStop}%
\bibitem [{\citenamefont {Hait}\ and\ \citenamefont
  {Head-Gordon}(2020)}]{Hait_2020}%
  \BibitemOpen
  \bibfield  {author} {\bibinfo {author} {\bibfnamefont {D.}~\bibnamefont
  {Hait}}\ and\ \bibinfo {author} {\bibfnamefont {M.}~\bibnamefont
  {Head-Gordon}},\ }\href {\doibase 10.1021/acs.jctc.9b01127} {\bibfield
  {journal} {\bibinfo  {journal} {J. Chem. Theory Comput.}\ }\textbf {\bibinfo
  {volume} {16}},\ \bibinfo {pages} {1699} (\bibinfo {year}
  {2020})}\BibitemShut {NoStop}%
\bibitem [{\citenamefont {Hait}\ and\ \citenamefont
  {Head-Gordon}(2021)}]{Hait_2021}%
  \BibitemOpen
  \bibfield  {author} {\bibinfo {author} {\bibfnamefont {D.}~\bibnamefont
  {Hait}}\ and\ \bibinfo {author} {\bibfnamefont {M.}~\bibnamefont
  {Head-Gordon}},\ }\href {\doibase 10.1021/acs.jpclett.1c00744} {\bibfield
  {journal} {\bibinfo  {journal} {J. Phys. Chem. Lett.}\ }\textbf {\bibinfo
  {volume} {12}},\ \bibinfo {pages} {4517} (\bibinfo {year}
  {2021})}\BibitemShut {NoStop}%
\bibitem [{\citenamefont {Zhao}\ and\ \citenamefont
  {Neuscamman}(2020{\natexlab{a}})}]{Zhao_2020}%
  \BibitemOpen
  \bibfield  {author} {\bibinfo {author} {\bibfnamefont {L.}~\bibnamefont
  {Zhao}}\ and\ \bibinfo {author} {\bibfnamefont {E.}~\bibnamefont
  {Neuscamman}},\ }\href {\doibase 10.1021/acs.jctc.9b00530} {\bibfield
  {journal} {\bibinfo  {journal} {J. Chem. Theory Comput.}\ }\textbf {\bibinfo
  {volume} {16}},\ \bibinfo {pages} {164} (\bibinfo {year}
  {2020}{\natexlab{a}})}\BibitemShut {NoStop}%
\bibitem [{\citenamefont {Levi}, \citenamefont {Ivanov},\ and\ \citenamefont
  {J{\'{o}}nsson}(2020)}]{Levi_2020}%
  \BibitemOpen
  \bibfield  {author} {\bibinfo {author} {\bibfnamefont {G.}~\bibnamefont
  {Levi}}, \bibinfo {author} {\bibfnamefont {A.~V.}\ \bibnamefont {Ivanov}}, \
  and\ \bibinfo {author} {\bibfnamefont {H.}~\bibnamefont {J{\'{o}}nsson}},\
  }\href {\doibase 10.1021/acs.jctc.0c00597} {\bibfield  {journal} {\bibinfo
  {journal} {J. Chem. Theory Comput.}\ }\textbf {\bibinfo {volume} {16}},\
  \bibinfo {pages} {6968} (\bibinfo {year} {2020})}\BibitemShut {NoStop}%
\bibitem [{\citenamefont {Carter-Fenk}\ and\ \citenamefont
  {Herbert}(2020)}]{Carter-Fenk_2020}%
  \BibitemOpen
  \bibfield  {author} {\bibinfo {author} {\bibfnamefont {K.}~\bibnamefont
  {Carter-Fenk}}\ and\ \bibinfo {author} {\bibfnamefont {J.~M.}\ \bibnamefont
  {Herbert}},\ }\href {\doibase 10.1021/acs.jctc.0c00502} {\bibfield  {journal}
  {\bibinfo  {journal} {J. Chem. Theory Comput.}\ }\textbf {\bibinfo {volume}
  {16}},\ \bibinfo {pages} {5067} (\bibinfo {year} {2020})}\BibitemShut
  {NoStop}%
\bibitem [{\citenamefont {Toffoli}\ \emph {et~al.}(2022)\citenamefont
  {Toffoli}, \citenamefont {Quarin}, \citenamefont {Fronzoni},\ and\
  \citenamefont {Stener}}]{Toffoli_2022}%
  \BibitemOpen
  \bibfield  {author} {\bibinfo {author} {\bibfnamefont {D.}~\bibnamefont
  {Toffoli}}, \bibinfo {author} {\bibfnamefont {M.}~\bibnamefont {Quarin}},
  \bibinfo {author} {\bibfnamefont {G.}~\bibnamefont {Fronzoni}}, \ and\
  \bibinfo {author} {\bibfnamefont {M.}~\bibnamefont {Stener}},\ }\href
  {\doibase 10.1021/acs.jpca.2c04473} {\bibfield  {journal} {\bibinfo
  {journal} {J. Phys. Chem. A}\ }\textbf {\bibinfo {volume} {126}},\ \bibinfo
  {pages} {7137} (\bibinfo {year} {2022})}\BibitemShut {NoStop}%
\bibitem [{\citenamefont {Schmerwitz}\ \emph {et~al.}(2022)\citenamefont
  {Schmerwitz}, \citenamefont {Ivanov}, \citenamefont {Jonsson}, \citenamefont
  {Jonsson},\ and\ \citenamefont {Levi}}]{Schmerwitz_2022}%
  \BibitemOpen
  \bibfield  {author} {\bibinfo {author} {\bibfnamefont {Y.~L.~A.}\
  \bibnamefont {Schmerwitz}}, \bibinfo {author} {\bibfnamefont {A.~V.}\
  \bibnamefont {Ivanov}}, \bibinfo {author} {\bibfnamefont {E.~O.}\
  \bibnamefont {Jonsson}}, \bibinfo {author} {\bibfnamefont {H.}~\bibnamefont
  {Jonsson}}, \ and\ \bibinfo {author} {\bibfnamefont {G.}~\bibnamefont
  {Levi}},\ }\href {\doibase 10.1021/acs.jpclett.2c00741} {\bibfield  {journal}
  {\bibinfo  {journal} {J. Phys. Chem. Lett.}\ }\textbf {\bibinfo {volume}
  {13}},\ \bibinfo {pages} {3990} (\bibinfo {year} {2022})}\BibitemShut
  {NoStop}%
\bibitem [{\citenamefont {Schmerwitz}, \citenamefont {Levi},\ and\
  \citenamefont {J{\'o}nsson}(2023)}]{Schmerwitz_2023}%
  \BibitemOpen
  \bibfield  {author} {\bibinfo {author} {\bibfnamefont {Y.~L.~A.}\
  \bibnamefont {Schmerwitz}}, \bibinfo {author} {\bibfnamefont
  {G.}~\bibnamefont {Levi}}, \ and\ \bibinfo {author} {\bibfnamefont
  {H.}~\bibnamefont {J{\'o}nsson}},\ }\href {\doibase 10.1021/acs.jctc.3c00178}
  {\bibfield  {journal} {\bibinfo  {journal} {J. Chem. Theory Comput.}\
  }\textbf {\bibinfo {volume} {19}},\ \bibinfo {pages} {3634} (\bibinfo {year}
  {2023})}\BibitemShut {NoStop}%
\bibitem [{\citenamefont {Clune}, \citenamefont {Shea},\ and\ \citenamefont
  {Neuscamman}(2020)}]{Clune_2020}%
  \BibitemOpen
  \bibfield  {author} {\bibinfo {author} {\bibfnamefont {R.}~\bibnamefont
  {Clune}}, \bibinfo {author} {\bibfnamefont {J.~A.~R.}\ \bibnamefont {Shea}},
  \ and\ \bibinfo {author} {\bibfnamefont {E.}~\bibnamefont {Neuscamman}},\
  }\href {\doibase 10.1021/acs.jctc.0c00308} {\bibfield  {journal} {\bibinfo
  {journal} {J. Chem. Theory Comput.}\ }\textbf {\bibinfo {volume} {16}},\
  \bibinfo {pages} {6132} (\bibinfo {year} {2020})}\BibitemShut {NoStop}%
\bibitem [{\citenamefont {Zhao}\ and\ \citenamefont
  {Neuscamman}(2020{\natexlab{b}})}]{Zhao_2020b}%
  \BibitemOpen
  \bibfield  {author} {\bibinfo {author} {\bibfnamefont {L.}~\bibnamefont
  {Zhao}}\ and\ \bibinfo {author} {\bibfnamefont {E.}~\bibnamefont
  {Neuscamman}},\ }\href {\doibase 10.1063/5.0003438} {\bibfield  {journal}
  {\bibinfo  {journal} {J. Chem. Phys.}\ }\textbf {\bibinfo {volume} {152}},\
  \bibinfo {pages} {204112} (\bibinfo {year} {2020}{\natexlab{b}})}\BibitemShut
  {NoStop}%
\bibitem [{\citenamefont {Clune}\ \emph {et~al.}(2023)\citenamefont {Clune},
  \citenamefont {Shea}, \citenamefont {Hardikar}, \citenamefont {Tuckman},\
  and\ \citenamefont {Neuscamman}}]{Clune_2023}%
  \BibitemOpen
  \bibfield  {author} {\bibinfo {author} {\bibfnamefont {R.}~\bibnamefont
  {Clune}}, \bibinfo {author} {\bibfnamefont {J.~A.~R.}\ \bibnamefont {Shea}},
  \bibinfo {author} {\bibfnamefont {T.~S.}\ \bibnamefont {Hardikar}}, \bibinfo
  {author} {\bibfnamefont {H.}~\bibnamefont {Tuckman}}, \ and\ \bibinfo
  {author} {\bibfnamefont {E.}~\bibnamefont {Neuscamman}},\ }\href {\doibase
  10.1063/5.0146975} {\bibfield  {journal} {\bibinfo  {journal} {J. Chem.
  Phys.}\ }\textbf {\bibinfo {volume} {158}},\ \bibinfo {pages} {224113}
  (\bibinfo {year} {2023})}\BibitemShut {NoStop}%
\bibitem [{\citenamefont {Scemama}\ \emph
  {et~al.}(2018{\natexlab{a}})\citenamefont {Scemama}, \citenamefont
  {Garniron}, \citenamefont {Caffarel},\ and\ \citenamefont
  {Loos}}]{Scemama_2018a}%
  \BibitemOpen
  \bibfield  {author} {\bibinfo {author} {\bibfnamefont {A.}~\bibnamefont
  {Scemama}}, \bibinfo {author} {\bibfnamefont {Y.}~\bibnamefont {Garniron}},
  \bibinfo {author} {\bibfnamefont {M.}~\bibnamefont {Caffarel}}, \ and\
  \bibinfo {author} {\bibfnamefont {P.~F.}\ \bibnamefont {Loos}},\ }\href
  {\doibase 10.1021/acs.jctc.7b01250} {\bibfield  {journal} {\bibinfo
  {journal} {J. Chem. Theory Comput.}\ }\textbf {\bibinfo {volume} {14}},\
  \bibinfo {pages} {1395} (\bibinfo {year} {2018}{\natexlab{a}})}\BibitemShut
  {NoStop}%
\bibitem [{\citenamefont {Scemama}\ \emph
  {et~al.}(2018{\natexlab{b}})\citenamefont {Scemama}, \citenamefont {Benali},
  \citenamefont {Jacquemin}, \citenamefont {Caffarel},\ and\ \citenamefont
  {Loos}}]{Scemama_2018b}%
  \BibitemOpen
  \bibfield  {author} {\bibinfo {author} {\bibfnamefont {A.}~\bibnamefont
  {Scemama}}, \bibinfo {author} {\bibfnamefont {A.}~\bibnamefont {Benali}},
  \bibinfo {author} {\bibfnamefont {D.}~\bibnamefont {Jacquemin}}, \bibinfo
  {author} {\bibfnamefont {M.}~\bibnamefont {Caffarel}}, \ and\ \bibinfo
  {author} {\bibfnamefont {P.~F.}\ \bibnamefont {Loos}},\ }\href {\doibase
  10.1063/1.5041327} {\bibfield  {journal} {\bibinfo  {journal} {J. Chem.
  Phys.}\ }\textbf {\bibinfo {volume} {149}},\ \bibinfo {pages} {034108}
  (\bibinfo {year} {2018}{\natexlab{b}})}\BibitemShut {NoStop}%
\bibitem [{\citenamefont {Dash}\ \emph {et~al.}(2018)\citenamefont {Dash},
  \citenamefont {Moroni}, \citenamefont {Scemama},\ and\ \citenamefont
  {Filippi}}]{Dash_2018}%
  \BibitemOpen
  \bibfield  {author} {\bibinfo {author} {\bibfnamefont {M.}~\bibnamefont
  {Dash}}, \bibinfo {author} {\bibfnamefont {S.}~\bibnamefont {Moroni}},
  \bibinfo {author} {\bibfnamefont {A.}~\bibnamefont {Scemama}}, \ and\
  \bibinfo {author} {\bibfnamefont {C.}~\bibnamefont {Filippi}},\ }\href
  {\doibase 10.1021/acs.jctc.8b00393} {\bibfield  {journal} {\bibinfo
  {journal} {J. Chem. Theory Comput.}\ }\textbf {\bibinfo {volume} {14}},\
  \bibinfo {pages} {4176} (\bibinfo {year} {2018})}\BibitemShut {NoStop}%
\bibitem [{\citenamefont {Dash}\ \emph {et~al.}(2019)\citenamefont {Dash},
  \citenamefont {Feldt}, \citenamefont {Moroni}, \citenamefont {Scemama},\ and\
  \citenamefont {Filippi}}]{Dash_2019}%
  \BibitemOpen
  \bibfield  {author} {\bibinfo {author} {\bibfnamefont {M.}~\bibnamefont
  {Dash}}, \bibinfo {author} {\bibfnamefont {J.}~\bibnamefont {Feldt}},
  \bibinfo {author} {\bibfnamefont {S.}~\bibnamefont {Moroni}}, \bibinfo
  {author} {\bibfnamefont {A.}~\bibnamefont {Scemama}}, \ and\ \bibinfo
  {author} {\bibfnamefont {C.}~\bibnamefont {Filippi}},\ }\href {\doibase
  10.1021/acs.jctc.9b00476} {\bibfield  {journal} {\bibinfo  {journal} {J.
  Chem. Theory Comput.}\ }\textbf {\bibinfo {volume} {15}},\ \bibinfo {pages}
  {4896} (\bibinfo {year} {2019})}\BibitemShut {NoStop}%
\bibitem [{\citenamefont {Dash}\ \emph {et~al.}(2021)\citenamefont {Dash},
  \citenamefont {Moroni}, \citenamefont {Filippi},\ and\ \citenamefont
  {Scemama}}]{Dash_2021}%
  \BibitemOpen
  \bibfield  {author} {\bibinfo {author} {\bibfnamefont {M.}~\bibnamefont
  {Dash}}, \bibinfo {author} {\bibfnamefont {S.}~\bibnamefont {Moroni}},
  \bibinfo {author} {\bibfnamefont {C.}~\bibnamefont {Filippi}}, \ and\
  \bibinfo {author} {\bibfnamefont {A.}~\bibnamefont {Scemama}},\ }\href
  {\doibase 10.1021/acs.jctc.1c00212} {\bibfield  {journal} {\bibinfo
  {journal} {J. Chem. Theory Comput.}\ }\textbf {\bibinfo {volume} {17}},\
  \bibinfo {pages} {3426} (\bibinfo {year} {2021})}\BibitemShut {NoStop}%
\bibitem [{\citenamefont {Cuzzocrea}\ \emph {et~al.}(2022)\citenamefont
  {Cuzzocrea}, \citenamefont {Moroni}, \citenamefont {Scemama},\ and\
  \citenamefont {Filippi}}]{Cuzzocrea_2022}%
  \BibitemOpen
  \bibfield  {author} {\bibinfo {author} {\bibfnamefont {A.}~\bibnamefont
  {Cuzzocrea}}, \bibinfo {author} {\bibfnamefont {S.}~\bibnamefont {Moroni}},
  \bibinfo {author} {\bibfnamefont {A.}~\bibnamefont {Scemama}}, \ and\
  \bibinfo {author} {\bibfnamefont {C.}~\bibnamefont {Filippi}},\ }\href
  {\doibase 10.1021/acs.jctc.1c01162} {\bibfield  {journal} {\bibinfo
  {journal} {J. Chem. Theory Comput.}\ }\textbf {\bibinfo {volume} {18}},\
  \bibinfo {pages} {1089} (\bibinfo {year} {2022})}\BibitemShut {NoStop}%
\bibitem [{\citenamefont {Shepard}\ \emph {et~al.}(2022)\citenamefont
  {Shepard}, \citenamefont {Panad{\'e}s-Barrueta}, \citenamefont {Moroni},
  \citenamefont {Scemama},\ and\ \citenamefont {Filippi}}]{Shepard_2022}%
  \BibitemOpen
  \bibfield  {author} {\bibinfo {author} {\bibfnamefont {S.}~\bibnamefont
  {Shepard}}, \bibinfo {author} {\bibfnamefont {R.~L.}\ \bibnamefont
  {Panad{\'e}s-Barrueta}}, \bibinfo {author} {\bibfnamefont {S.}~\bibnamefont
  {Moroni}}, \bibinfo {author} {\bibfnamefont {A.}~\bibnamefont {Scemama}}, \
  and\ \bibinfo {author} {\bibfnamefont {C.}~\bibnamefont {Filippi}},\ }\href
  {\doibase 10.1021/acs.jctc.2c00769} {\bibfield  {journal} {\bibinfo
  {journal} {J. Chem. Theory Comput.}\ }\textbf {\bibinfo {volume} {18}},\
  \bibinfo {pages} {6722} (\bibinfo {year} {2022})}\BibitemShut {NoStop}%
\bibitem [{\citenamefont {Otis}, \citenamefont {Craig},\ and\ \citenamefont
  {Neuscamman}(2020)}]{Otis_2020}%
  \BibitemOpen
  \bibfield  {author} {\bibinfo {author} {\bibfnamefont {L.}~\bibnamefont
  {Otis}}, \bibinfo {author} {\bibfnamefont {I.~M.}\ \bibnamefont {Craig}}, \
  and\ \bibinfo {author} {\bibfnamefont {E.}~\bibnamefont {Neuscamman}},\
  }\href {\doibase 10.1063/5.0024572} {\bibfield  {journal} {\bibinfo
  {journal} {J. Chem. Phys.}\ }\textbf {\bibinfo {volume} {153}},\ \bibinfo
  {pages} {234105} (\bibinfo {year} {2020})}\BibitemShut {NoStop}%
\bibitem [{\citenamefont {Otis}\ and\ \citenamefont
  {Neuscamman}(2023)}]{Otis_2023}%
  \BibitemOpen
  \bibfield  {author} {\bibinfo {author} {\bibfnamefont {L.}~\bibnamefont
  {Otis}}\ and\ \bibinfo {author} {\bibfnamefont {E.}~\bibnamefont
  {Neuscamman}},\ }\href {\doibase 10.1002/wcms.1659} {\bibfield  {journal}
  {\bibinfo  {journal} {WIREs Comput. Mol. Sci.}\ } (\bibinfo {year} {2023}),\
  10.1002/wcms.1659}\BibitemShut {NoStop}%
\bibitem [{\citenamefont {Piecuch}\ and\ \citenamefont
  {Kowalski}(2000)}]{Piecuch_2000}%
  \BibitemOpen
  \bibfield  {author} {\bibinfo {author} {\bibfnamefont {P.}~\bibnamefont
  {Piecuch}}\ and\ \bibinfo {author} {\bibfnamefont {K.}~\bibnamefont
  {Kowalski}},\ }in\ \href {\doibase 10.1142/9789812792501_0001} {\emph
  {\bibinfo {booktitle} {Computational {{Chemistry}}: {{Reviews}} of {{Current
  Trends}}}}},\ \bibinfo {series} {Computational {{Chemistry}}: {{Reviews}} of
  {{Current Trends}}}, Vol.\ \bibinfo {volume} {Volume 5}\ (\bibinfo
  {publisher} {{World Scientific}},\ \bibinfo {year} {2000})\ pp.\ \bibinfo
  {pages} {1--104}\BibitemShut {NoStop}%
\bibitem [{\citenamefont {Mayhall}\ and\ \citenamefont
  {Raghavachari}(2010)}]{Mayhall_2010}%
  \BibitemOpen
  \bibfield  {author} {\bibinfo {author} {\bibfnamefont {N.~J.}\ \bibnamefont
  {Mayhall}}\ and\ \bibinfo {author} {\bibfnamefont {K.}~\bibnamefont
  {Raghavachari}},\ }\href {\doibase 10.1021/ct100321k} {\bibfield  {journal}
  {\bibinfo  {journal} {J. Chem. Theory Comput.}\ }\textbf {\bibinfo {volume}
  {6}},\ \bibinfo {pages} {2714} (\bibinfo {year} {2010})}\BibitemShut
  {NoStop}%
\bibitem [{\citenamefont {Lee}, \citenamefont {Small},\ and\ \citenamefont
  {{Head-Gordon}}(2019)}]{Lee_2019}%
  \BibitemOpen
  \bibfield  {author} {\bibinfo {author} {\bibfnamefont {J.}~\bibnamefont
  {Lee}}, \bibinfo {author} {\bibfnamefont {D.~W.}\ \bibnamefont {Small}}, \
  and\ \bibinfo {author} {\bibfnamefont {M.}~\bibnamefont {{Head-Gordon}}},\
  }\href {\doibase 10.1063/1.5128795} {\bibfield  {journal} {\bibinfo
  {journal} {J. Chem. Phys.}\ }\textbf {\bibinfo {volume} {151}},\ \bibinfo
  {pages} {214103} (\bibinfo {year} {2019})}\BibitemShut {NoStop}%
\bibitem [{\citenamefont {Kossoski}\ \emph {et~al.}(2021)\citenamefont
  {Kossoski}, \citenamefont {Marie}, \citenamefont {Scemama}, \citenamefont
  {Caffarel},\ and\ \citenamefont {Loos}}]{Kossoski_2021}%
  \BibitemOpen
  \bibfield  {author} {\bibinfo {author} {\bibfnamefont {F.}~\bibnamefont
  {Kossoski}}, \bibinfo {author} {\bibfnamefont {A.}~\bibnamefont {Marie}},
  \bibinfo {author} {\bibfnamefont {A.}~\bibnamefont {Scemama}}, \bibinfo
  {author} {\bibfnamefont {M.}~\bibnamefont {Caffarel}}, \ and\ \bibinfo
  {author} {\bibfnamefont {P.-F.}\ \bibnamefont {Loos}},\ }\href {\doibase
  10.1021/acs.jctc.1c00348} {\bibfield  {journal} {\bibinfo  {journal} {J.
  Chem. Theory Comput.}\ }\textbf {\bibinfo {volume} {17}},\ \bibinfo {pages}
  {4756} (\bibinfo {year} {2021})}\BibitemShut {NoStop}%
\bibitem [{\citenamefont {Marie}, \citenamefont {Kossoski},\ and\ \citenamefont
  {Loos}(2021)}]{Marie_2021}%
  \BibitemOpen
  \bibfield  {author} {\bibinfo {author} {\bibfnamefont {A.}~\bibnamefont
  {Marie}}, \bibinfo {author} {\bibfnamefont {F.}~\bibnamefont {Kossoski}}, \
  and\ \bibinfo {author} {\bibfnamefont {P.-F.}\ \bibnamefont {Loos}},\ }\href
  {\doibase 10.1063/5.0060698} {\bibfield  {journal} {\bibinfo  {journal} {J.
  Chem. Phys.}\ }\textbf {\bibinfo {volume} {155}},\ \bibinfo {pages} {104105}
  (\bibinfo {year} {2021})}\BibitemShut {NoStop}%
\bibitem [{\citenamefont {Rishi}\ \emph {et~al.}(2023)\citenamefont {Rishi},
  \citenamefont {Ravi}, \citenamefont {Perera},\ and\ \citenamefont
  {Bartlett}}]{Rishi_2023}%
  \BibitemOpen
  \bibfield  {author} {\bibinfo {author} {\bibfnamefont {V.}~\bibnamefont
  {Rishi}}, \bibinfo {author} {\bibfnamefont {M.}~\bibnamefont {Ravi}},
  \bibinfo {author} {\bibfnamefont {A.}~\bibnamefont {Perera}}, \ and\ \bibinfo
  {author} {\bibfnamefont {R.~J.}\ \bibnamefont {Bartlett}},\ }\href {\doibase
  10.1021/acs.jpca.2c07697} {\bibfield  {journal} {\bibinfo  {journal} {J.
  Phys. Chem. A}\ }\textbf {\bibinfo {volume} {127}},\ \bibinfo {pages} {828}
  (\bibinfo {year} {2023})}\BibitemShut {NoStop}%
\bibitem [{\citenamefont {Tuckman}\ and\ \citenamefont
  {Neuscamman}(2023)}]{Tuckman_2023}%
  \BibitemOpen
  \bibfield  {author} {\bibinfo {author} {\bibfnamefont {H.}~\bibnamefont
  {Tuckman}}\ and\ \bibinfo {author} {\bibfnamefont {E.}~\bibnamefont
  {Neuscamman}},\ }\href@noop {} {\enquote {\bibinfo {title} {An
  excited-state-specific projected coupled-cluster theory},}\ } (\bibinfo
  {year} {2023}),\ \Eprint {http://arxiv.org/abs/2302.06731} {arXiv:2302.06731
  [physics.chem-ph]} \BibitemShut {NoStop}%
\bibitem [{\citenamefont {Garniron}\ \emph {et~al.}(2019)\citenamefont
  {Garniron}, \citenamefont {Gasperich}, \citenamefont {Applencourt},
  \citenamefont {Benali}, \citenamefont {Fert{\'e}}, \citenamefont {Paquier},
  \citenamefont {Pradines}, \citenamefont {Assaraf}, \citenamefont {Reinhardt},
  \citenamefont {Toulouse}, \citenamefont {Barbaresco}, \citenamefont {Renon},
  \citenamefont {David}, \citenamefont {Malrieu}, \citenamefont {V{\'e}ril},
  \citenamefont {Caffarel}, \citenamefont {Loos}, \citenamefont {Giner},\ and\
  \citenamefont {Scemama}}]{Garniron_2019}%
  \BibitemOpen
  \bibfield  {author} {\bibinfo {author} {\bibfnamefont {Y.}~\bibnamefont
  {Garniron}}, \bibinfo {author} {\bibfnamefont {K.}~\bibnamefont {Gasperich}},
  \bibinfo {author} {\bibfnamefont {T.}~\bibnamefont {Applencourt}}, \bibinfo
  {author} {\bibfnamefont {A.}~\bibnamefont {Benali}}, \bibinfo {author}
  {\bibfnamefont {A.}~\bibnamefont {Fert{\'e}}}, \bibinfo {author}
  {\bibfnamefont {J.}~\bibnamefont {Paquier}}, \bibinfo {author} {\bibfnamefont
  {B.}~\bibnamefont {Pradines}}, \bibinfo {author} {\bibfnamefont
  {R.}~\bibnamefont {Assaraf}}, \bibinfo {author} {\bibfnamefont
  {P.}~\bibnamefont {Reinhardt}}, \bibinfo {author} {\bibfnamefont
  {J.}~\bibnamefont {Toulouse}}, \bibinfo {author} {\bibfnamefont
  {P.}~\bibnamefont {Barbaresco}}, \bibinfo {author} {\bibfnamefont
  {N.}~\bibnamefont {Renon}}, \bibinfo {author} {\bibfnamefont
  {G.}~\bibnamefont {David}}, \bibinfo {author} {\bibfnamefont {J.~P.}\
  \bibnamefont {Malrieu}}, \bibinfo {author} {\bibfnamefont {M.}~\bibnamefont
  {V{\'e}ril}}, \bibinfo {author} {\bibfnamefont {M.}~\bibnamefont {Caffarel}},
  \bibinfo {author} {\bibfnamefont {P.~F.}\ \bibnamefont {Loos}}, \bibinfo
  {author} {\bibfnamefont {E.}~\bibnamefont {Giner}}, \ and\ \bibinfo {author}
  {\bibfnamefont {A.}~\bibnamefont {Scemama}},\ }\href {\doibase
  10.1021/acs.jctc.9b00176} {\bibfield  {journal} {\bibinfo  {journal} {J.
  Chem. Theory Comput.}\ }\textbf {\bibinfo {volume} {15}},\ \bibinfo {pages}
  {3591} (\bibinfo {year} {2019})}\BibitemShut {NoStop}%
\bibitem [{\citenamefont {Limacher}\ \emph {et~al.}(2013)\citenamefont
  {Limacher}, \citenamefont {Ayers}, \citenamefont {Johnson}, \citenamefont
  {De~Baerdemacker}, \citenamefont {Van~Neck},\ and\ \citenamefont
  {Bultinck}}]{Limacher_2013}%
  \BibitemOpen
  \bibfield  {author} {\bibinfo {author} {\bibfnamefont {P.~A.}\ \bibnamefont
  {Limacher}}, \bibinfo {author} {\bibfnamefont {P.~W.}\ \bibnamefont {Ayers}},
  \bibinfo {author} {\bibfnamefont {P.~A.}\ \bibnamefont {Johnson}}, \bibinfo
  {author} {\bibfnamefont {S.}~\bibnamefont {De~Baerdemacker}}, \bibinfo
  {author} {\bibfnamefont {D.}~\bibnamefont {Van~Neck}}, \ and\ \bibinfo
  {author} {\bibfnamefont {P.}~\bibnamefont {Bultinck}},\ }\href {\doibase
  10.1021/ct300902c} {\bibfield  {journal} {\bibinfo  {journal} {J. Chem.
  Theory Comput.}\ }\textbf {\bibinfo {volume} {9}},\ \bibinfo {pages} {1394}
  (\bibinfo {year} {2013})}\BibitemShut {NoStop}%
\bibitem [{\citenamefont {Limacher}\ \emph {et~al.}(2014)\citenamefont
  {Limacher}, \citenamefont {Kim}, \citenamefont {Ayers}, \citenamefont
  {Johnson}, \citenamefont {Baerdemacker}, \citenamefont {Neck},\ and\
  \citenamefont {Bultinck}}]{Limacher_2014}%
  \BibitemOpen
  \bibfield  {author} {\bibinfo {author} {\bibfnamefont {P.~A.}\ \bibnamefont
  {Limacher}}, \bibinfo {author} {\bibfnamefont {T.~D.}\ \bibnamefont {Kim}},
  \bibinfo {author} {\bibfnamefont {P.~W.}\ \bibnamefont {Ayers}}, \bibinfo
  {author} {\bibfnamefont {P.~A.}\ \bibnamefont {Johnson}}, \bibinfo {author}
  {\bibfnamefont {S.~D.}\ \bibnamefont {Baerdemacker}}, \bibinfo {author}
  {\bibfnamefont {D.~V.}\ \bibnamefont {Neck}}, \ and\ \bibinfo {author}
  {\bibfnamefont {P.}~\bibnamefont {Bultinck}},\ }\href {\doibase
  10.1080/00268976.2013.874600} {\bibfield  {journal} {\bibinfo  {journal}
  {Mol. Phys.}\ }\textbf {\bibinfo {volume} {112}},\ \bibinfo {pages} {853}
  (\bibinfo {year} {2014})}\BibitemShut {NoStop}%
\bibitem [{\citenamefont {Tecmer}\ \emph {et~al.}(2014)\citenamefont {Tecmer},
  \citenamefont {Boguslawski}, \citenamefont {Johnson}, \citenamefont
  {Limacher}, \citenamefont {Chan}, \citenamefont {Verstraelen},\ and\
  \citenamefont {Ayers}}]{Tecmer_2014}%
  \BibitemOpen
  \bibfield  {author} {\bibinfo {author} {\bibfnamefont {P.}~\bibnamefont
  {Tecmer}}, \bibinfo {author} {\bibfnamefont {K.}~\bibnamefont {Boguslawski}},
  \bibinfo {author} {\bibfnamefont {P.~A.}\ \bibnamefont {Johnson}}, \bibinfo
  {author} {\bibfnamefont {P.~A.}\ \bibnamefont {Limacher}}, \bibinfo {author}
  {\bibfnamefont {M.}~\bibnamefont {Chan}}, \bibinfo {author} {\bibfnamefont
  {T.}~\bibnamefont {Verstraelen}}, \ and\ \bibinfo {author} {\bibfnamefont
  {P.~W.}\ \bibnamefont {Ayers}},\ }\href {\doibase 10.1021/jp502127v}
  {\bibfield  {journal} {\bibinfo  {journal} {J. Phys. Chem. A}\ }\textbf
  {\bibinfo {volume} {118}},\ \bibinfo {pages} {9058} (\bibinfo {year}
  {2014})}\BibitemShut {NoStop}%
\bibitem [{\citenamefont {Boguslawski}\ \emph
  {et~al.}(2014{\natexlab{a}})\citenamefont {Boguslawski}, \citenamefont
  {Tecmer}, \citenamefont {Ayers}, \citenamefont {Bultinck}, \citenamefont
  {De~Baerdemacker},\ and\ \citenamefont {Van~Neck}}]{Boguslawski_2014a}%
  \BibitemOpen
  \bibfield  {author} {\bibinfo {author} {\bibfnamefont {K.}~\bibnamefont
  {Boguslawski}}, \bibinfo {author} {\bibfnamefont {P.}~\bibnamefont {Tecmer}},
  \bibinfo {author} {\bibfnamefont {P.~W.}\ \bibnamefont {Ayers}}, \bibinfo
  {author} {\bibfnamefont {P.}~\bibnamefont {Bultinck}}, \bibinfo {author}
  {\bibfnamefont {S.}~\bibnamefont {De~Baerdemacker}}, \ and\ \bibinfo {author}
  {\bibfnamefont {D.}~\bibnamefont {Van~Neck}},\ }\href {\doibase
  10.1103/PhysRevB.89.201106} {\bibfield  {journal} {\bibinfo  {journal} {Phys.
  Rev. B}\ }\textbf {\bibinfo {volume} {89}},\ \bibinfo {pages} {201106}
  (\bibinfo {year} {2014}{\natexlab{a}})}\BibitemShut {NoStop}%
\bibitem [{\citenamefont {Boguslawski}\ and\ \citenamefont
  {Ayers}(2015)}]{Boguslawski_2015}%
  \BibitemOpen
  \bibfield  {author} {\bibinfo {author} {\bibfnamefont {K.}~\bibnamefont
  {Boguslawski}}\ and\ \bibinfo {author} {\bibfnamefont {P.~W.}\ \bibnamefont
  {Ayers}},\ }\href {\doibase 10.1021/acs.jctc.5b00776} {\bibfield  {journal}
  {\bibinfo  {journal} {J. Chem. Theory Comput.}\ }\textbf {\bibinfo {volume}
  {11}},\ \bibinfo {pages} {5252} (\bibinfo {year} {2015})}\BibitemShut
  {NoStop}%
\bibitem [{\citenamefont {Boguslawski}\ \emph
  {et~al.}(2014{\natexlab{b}})\citenamefont {Boguslawski}, \citenamefont
  {Tecmer}, \citenamefont {Bultinck}, \citenamefont {De~Baerdemacker},
  \citenamefont {Van~Neck},\ and\ \citenamefont {Ayers}}]{Boguslawski_2014b}%
  \BibitemOpen
  \bibfield  {author} {\bibinfo {author} {\bibfnamefont {K.}~\bibnamefont
  {Boguslawski}}, \bibinfo {author} {\bibfnamefont {P.}~\bibnamefont {Tecmer}},
  \bibinfo {author} {\bibfnamefont {P.}~\bibnamefont {Bultinck}}, \bibinfo
  {author} {\bibfnamefont {S.}~\bibnamefont {De~Baerdemacker}}, \bibinfo
  {author} {\bibfnamefont {D.}~\bibnamefont {Van~Neck}}, \ and\ \bibinfo
  {author} {\bibfnamefont {P.~W.}\ \bibnamefont {Ayers}},\ }\href {\doibase
  10.1021/ct500759q} {\bibfield  {journal} {\bibinfo  {journal} {J. Chem.
  Theory Comput.}\ }\textbf {\bibinfo {volume} {10}},\ \bibinfo {pages} {4873}
  (\bibinfo {year} {2014}{\natexlab{b}})}\BibitemShut {NoStop}%
\bibitem [{\citenamefont {Boguslawski}\ \emph
  {et~al.}(2014{\natexlab{c}})\citenamefont {Boguslawski}, \citenamefont
  {Tecmer}, \citenamefont {Limacher}, \citenamefont {Johnson}, \citenamefont
  {Ayers}, \citenamefont {Bultinck}, \citenamefont {De~Baerdemacker},\ and\
  \citenamefont {Van~Neck}}]{Boguslawski_2014c}%
  \BibitemOpen
  \bibfield  {author} {\bibinfo {author} {\bibfnamefont {K.}~\bibnamefont
  {Boguslawski}}, \bibinfo {author} {\bibfnamefont {P.}~\bibnamefont {Tecmer}},
  \bibinfo {author} {\bibfnamefont {P.~A.}\ \bibnamefont {Limacher}}, \bibinfo
  {author} {\bibfnamefont {P.~A.}\ \bibnamefont {Johnson}}, \bibinfo {author}
  {\bibfnamefont {P.~W.}\ \bibnamefont {Ayers}}, \bibinfo {author}
  {\bibfnamefont {P.}~\bibnamefont {Bultinck}}, \bibinfo {author}
  {\bibfnamefont {S.}~\bibnamefont {De~Baerdemacker}}, \ and\ \bibinfo {author}
  {\bibfnamefont {D.}~\bibnamefont {Van~Neck}},\ }\href {\doibase
  10.1063/1.4880820} {\bibfield  {journal} {\bibinfo  {journal} {J. Chem.
  Phys.}\ }\textbf {\bibinfo {volume} {140}},\ \bibinfo {pages} {214114}
  (\bibinfo {year} {2014}{\natexlab{c}})}\BibitemShut {NoStop}%
\bibitem [{\citenamefont {Johnson}\ \emph {et~al.}(2020)\citenamefont
  {Johnson}, \citenamefont {Fecteau}, \citenamefont {Berthiaume}, \citenamefont
  {Cloutier}, \citenamefont {Carrier}, \citenamefont {Gratton}, \citenamefont
  {Bultinck}, \citenamefont {De~Baerdemacker}, \citenamefont {Van~Neck},
  \citenamefont {Limacher},\ and\ \citenamefont {Ayers}}]{Johnson_2020}%
  \BibitemOpen
  \bibfield  {author} {\bibinfo {author} {\bibfnamefont {P.~A.}\ \bibnamefont
  {Johnson}}, \bibinfo {author} {\bibfnamefont {C.-{\'E}.}\ \bibnamefont
  {Fecteau}}, \bibinfo {author} {\bibfnamefont {F.}~\bibnamefont {Berthiaume}},
  \bibinfo {author} {\bibfnamefont {S.}~\bibnamefont {Cloutier}}, \bibinfo
  {author} {\bibfnamefont {L.}~\bibnamefont {Carrier}}, \bibinfo {author}
  {\bibfnamefont {M.}~\bibnamefont {Gratton}}, \bibinfo {author} {\bibfnamefont
  {P.}~\bibnamefont {Bultinck}}, \bibinfo {author} {\bibfnamefont
  {S.}~\bibnamefont {De~Baerdemacker}}, \bibinfo {author} {\bibfnamefont
  {D.}~\bibnamefont {Van~Neck}}, \bibinfo {author} {\bibfnamefont
  {P.}~\bibnamefont {Limacher}}, \ and\ \bibinfo {author} {\bibfnamefont
  {P.~W.}\ \bibnamefont {Ayers}},\ }\href {\doibase 10.1063/5.0022189}
  {\bibfield  {journal} {\bibinfo  {journal} {J. Chem. Phys.}\ }\textbf
  {\bibinfo {volume} {153}},\ \bibinfo {pages} {104110} (\bibinfo {year}
  {2020})}\BibitemShut {NoStop}%
\bibitem [{\citenamefont {Henderson}\ \emph {et~al.}(2014)\citenamefont
  {Henderson}, \citenamefont {Bulik}, \citenamefont {Stein},\ and\
  \citenamefont {Scuseria}}]{Henderson_2014}%
  \BibitemOpen
  \bibfield  {author} {\bibinfo {author} {\bibfnamefont {T.~M.}\ \bibnamefont
  {Henderson}}, \bibinfo {author} {\bibfnamefont {I.~W.}\ \bibnamefont
  {Bulik}}, \bibinfo {author} {\bibfnamefont {T.}~\bibnamefont {Stein}}, \ and\
  \bibinfo {author} {\bibfnamefont {G.~E.}\ \bibnamefont {Scuseria}},\ }\href
  {\doibase 10.1063/1.4904384} {\bibfield  {journal} {\bibinfo  {journal} {J.
  Chem. Phys.}\ }\textbf {\bibinfo {volume} {141}},\ \bibinfo {pages} {244104}
  (\bibinfo {year} {2014})}\BibitemShut {NoStop}%
\bibitem [{\citenamefont {Stein}, \citenamefont {Henderson},\ and\
  \citenamefont {Scuseria}(2014)}]{Stein_2014}%
  \BibitemOpen
  \bibfield  {author} {\bibinfo {author} {\bibfnamefont {T.}~\bibnamefont
  {Stein}}, \bibinfo {author} {\bibfnamefont {T.~M.}\ \bibnamefont
  {Henderson}}, \ and\ \bibinfo {author} {\bibfnamefont {G.~E.}\ \bibnamefont
  {Scuseria}},\ }\href {\doibase 10.1063/1.4880819} {\bibfield  {journal}
  {\bibinfo  {journal} {J. Chem. Phys.}\ }\textbf {\bibinfo {volume} {140}},\
  \bibinfo {pages} {214113} (\bibinfo {year} {2014})}\BibitemShut {NoStop}%
\bibitem [{\citenamefont {Henderson}, \citenamefont {Bulik},\ and\
  \citenamefont {Scuseria}(2015)}]{Henderson_2015}%
  \BibitemOpen
  \bibfield  {author} {\bibinfo {author} {\bibfnamefont {T.~M.}\ \bibnamefont
  {Henderson}}, \bibinfo {author} {\bibfnamefont {I.~W.}\ \bibnamefont
  {Bulik}}, \ and\ \bibinfo {author} {\bibfnamefont {G.~E.}\ \bibnamefont
  {Scuseria}},\ }\href {\doibase 10.1063/1.4921986} {\bibfield  {journal}
  {\bibinfo  {journal} {J. Chem. Phys.}\ }\textbf {\bibinfo {volume} {142}},\
  \bibinfo {pages} {214116} (\bibinfo {year} {2015})}\BibitemShut {NoStop}%
\bibitem [{\citenamefont {Chen}, \citenamefont {Zhou},\ and\ \citenamefont
  {Wu}(2015)}]{Chen_2015}%
  \BibitemOpen
  \bibfield  {author} {\bibinfo {author} {\bibfnamefont {Z.}~\bibnamefont
  {Chen}}, \bibinfo {author} {\bibfnamefont {C.}~\bibnamefont {Zhou}}, \ and\
  \bibinfo {author} {\bibfnamefont {W.}~\bibnamefont {Wu}},\ }\href {\doibase
  10.1021/acs.jctc.5b00416} {\bibfield  {journal} {\bibinfo  {journal} {J.
  Chem. Theory Comput.}\ }\textbf {\bibinfo {volume} {11}},\ \bibinfo {pages}
  {4102} (\bibinfo {year} {2015})}\BibitemShut {NoStop}%
\bibitem [{\citenamefont {Bytautas}\ and\ \citenamefont
  {Dukelsky}(2018)}]{Bytautas_2018}%
  \BibitemOpen
  \bibfield  {author} {\bibinfo {author} {\bibfnamefont {L.}~\bibnamefont
  {Bytautas}}\ and\ \bibinfo {author} {\bibfnamefont {J.}~\bibnamefont
  {Dukelsky}},\ }\href {\doibase https://doi.org/10.1016/j.comptc.2018.08.011}
  {\bibfield  {journal} {\bibinfo  {journal} {Comput. Theor. Chem.}\ }\textbf
  {\bibinfo {volume} {1141}},\ \bibinfo {pages} {74} (\bibinfo {year}
  {2018})}\BibitemShut {NoStop}%
\bibitem [{\citenamefont {Boguslawski}(2021)}]{Boguslawski_2021}%
  \BibitemOpen
  \bibfield  {author} {\bibinfo {author} {\bibfnamefont {K.}~\bibnamefont
  {Boguslawski}},\ }\href {\doibase 10.1039/D1CC04539C} {\bibfield  {journal}
  {\bibinfo  {journal} {Chem. Commun.}\ }\textbf {\bibinfo {volume} {57}},\
  \bibinfo {pages} {12277} (\bibinfo {year} {2021})}\BibitemShut {NoStop}%
\bibitem [{\citenamefont {Tecmer}\ and\ \citenamefont
  {Boguslawski}(2022)}]{Tecmer_2022}%
  \BibitemOpen
  \bibfield  {author} {\bibinfo {author} {\bibfnamefont {P.}~\bibnamefont
  {Tecmer}}\ and\ \bibinfo {author} {\bibfnamefont {K.}~\bibnamefont
  {Boguslawski}},\ }\href {\doibase 10.1039/D2CP02528K} {\bibfield  {journal}
  {\bibinfo  {journal} {Phys. Chem. Chem. Phys.}\ }\textbf {\bibinfo {volume}
  {24}},\ \bibinfo {pages} {23026} (\bibinfo {year} {2022})}\BibitemShut
  {NoStop}%
\bibitem [{\citenamefont {Mamache}, \citenamefont {Ga\l{}y\'nska},\ and\
  \citenamefont {Boguslawski}(2023)}]{Mamache_2023}%
  \BibitemOpen
  \bibfield  {author} {\bibinfo {author} {\bibfnamefont {S.}~\bibnamefont
  {Mamache}}, \bibinfo {author} {\bibfnamefont {M.}~\bibnamefont
  {Ga\l{}y\'nska}}, \ and\ \bibinfo {author} {\bibfnamefont {K.}~\bibnamefont
  {Boguslawski}},\ }\href {\doibase 10.1039/D3CP01963B} {\bibfield  {journal}
  {\bibinfo  {journal} {Phys. Chem. Chem. Phys.}\ }\textbf {\bibinfo {volume}
  {25}},\ \bibinfo {pages} {18023} (\bibinfo {year} {2023})}\BibitemShut
  {NoStop}%
\bibitem [{\citenamefont {Fecteau}\ \emph {et~al.}(2022)\citenamefont
  {Fecteau}, \citenamefont {Cloutier}, \citenamefont {Moisset}, \citenamefont
  {Boulay}, \citenamefont {Bultinck}, \citenamefont {Faribault},\ and\
  \citenamefont {Johnson}}]{Fecteau_2022}%
  \BibitemOpen
  \bibfield  {author} {\bibinfo {author} {\bibfnamefont {C.-{\'{E}}.}\
  \bibnamefont {Fecteau}}, \bibinfo {author} {\bibfnamefont {S.}~\bibnamefont
  {Cloutier}}, \bibinfo {author} {\bibfnamefont {J.-D.}\ \bibnamefont
  {Moisset}}, \bibinfo {author} {\bibfnamefont {J.}~\bibnamefont {Boulay}},
  \bibinfo {author} {\bibfnamefont {P.}~\bibnamefont {Bultinck}}, \bibinfo
  {author} {\bibfnamefont {A.}~\bibnamefont {Faribault}}, \ and\ \bibinfo
  {author} {\bibfnamefont {P.~A.}\ \bibnamefont {Johnson}},\ }\href {\doibase
  10.1063/5.0091338} {\bibfield  {journal} {\bibinfo  {journal} {J. Chem.
  Phys.}\ }\textbf {\bibinfo {volume} {156}},\ \bibinfo {pages} {194103}
  (\bibinfo {year} {2022})}\BibitemShut {NoStop}%
\bibitem [{\citenamefont {Boguslawski}(2016)}]{Boguslawski_2016b}%
  \BibitemOpen
  \bibfield  {author} {\bibinfo {author} {\bibfnamefont {K.}~\bibnamefont
  {Boguslawski}},\ }\href {\doibase 10.1063/1.4972053} {\bibfield  {journal}
  {\bibinfo  {journal} {J. Chem. Phys.}\ }\textbf {\bibinfo {volume} {145}},\
  \bibinfo {pages} {234105} (\bibinfo {year} {2016})}\BibitemShut {NoStop}%
\bibitem [{\citenamefont {Boguslawski}(2017)}]{Boguslawski_2016c}%
  \BibitemOpen
  \bibfield  {author} {\bibinfo {author} {\bibfnamefont {K.}~\bibnamefont
  {Boguslawski}},\ }\href {\doibase 10.1063/1.5006124} {\bibfield  {journal}
  {\bibinfo  {journal} {J. Chem. Phys.}\ }\textbf {\bibinfo {volume} {147}},\
  \bibinfo {pages} {139901} (\bibinfo {year} {2017})}\BibitemShut {NoStop}%
\bibitem [{\citenamefont {Boguslawski}(2019)}]{Boguslawski_2019}%
  \BibitemOpen
  \bibfield  {author} {\bibinfo {author} {\bibfnamefont {K.}~\bibnamefont
  {Boguslawski}},\ }\href {\doibase 10.1021/acs.jctc.8b01053} {\bibfield
  {journal} {\bibinfo  {journal} {J. Chem. Theory Comput.}\ }\textbf {\bibinfo
  {volume} {15}},\ \bibinfo {pages} {18} (\bibinfo {year} {2019})}\BibitemShut
  {NoStop}%
\bibitem [{\citenamefont {Nowak}, \citenamefont {Tecmer},\ and\ \citenamefont
  {Boguslawski}(2019)}]{Nowak_2019}%
  \BibitemOpen
  \bibfield  {author} {\bibinfo {author} {\bibfnamefont {A.}~\bibnamefont
  {Nowak}}, \bibinfo {author} {\bibfnamefont {P.}~\bibnamefont {Tecmer}}, \
  and\ \bibinfo {author} {\bibfnamefont {K.}~\bibnamefont {Boguslawski}},\
  }\href {\doibase 10.1039/C9CP03678D} {\bibfield  {journal} {\bibinfo
  {journal} {Phys. Chem. Chem. Phys.}\ }\textbf {\bibinfo {volume} {21}},\
  \bibinfo {pages} {19039} (\bibinfo {year} {2019})}\BibitemShut {NoStop}%
\bibitem [{\citenamefont {Nowak}\ and\ \citenamefont
  {Boguslawski}(2023)}]{Nowak_2023}%
  \BibitemOpen
  \bibfield  {author} {\bibinfo {author} {\bibfnamefont {A.}~\bibnamefont
  {Nowak}}\ and\ \bibinfo {author} {\bibfnamefont {K.}~\bibnamefont
  {Boguslawski}},\ }\href {\doibase 10.1039/D2CP05171K} {\bibfield  {journal}
  {\bibinfo  {journal} {Phys. Chem. Chem. Phys.}\ }\textbf {\bibinfo {volume}
  {25}},\ \bibinfo {pages} {7289} (\bibinfo {year} {2023})}\BibitemShut
  {NoStop}%
\bibitem [{\citenamefont {Krylov}(2006)}]{Krylov_2006}%
  \BibitemOpen
  \bibfield  {author} {\bibinfo {author} {\bibfnamefont {A.~I.}\ \bibnamefont
  {Krylov}},\ }\href {\doibase 10.1021/ar0402006} {\bibfield  {journal}
  {\bibinfo  {journal} {Acc. Chem. Res.}\ }\textbf {\bibinfo {volume} {39}},\
  \bibinfo {pages} {83} (\bibinfo {year} {2006})}\BibitemShut {NoStop}%
\bibitem [{\citenamefont {Horbatenko}\ \emph {et~al.}(2021)\citenamefont
  {Horbatenko}, \citenamefont {Sadiq}, \citenamefont {Lee}, \citenamefont
  {Filatov},\ and\ \citenamefont {Choi}}]{Horbatenko_2021}%
  \BibitemOpen
  \bibfield  {author} {\bibinfo {author} {\bibfnamefont {Y.}~\bibnamefont
  {Horbatenko}}, \bibinfo {author} {\bibfnamefont {S.}~\bibnamefont {Sadiq}},
  \bibinfo {author} {\bibfnamefont {S.}~\bibnamefont {Lee}}, \bibinfo {author}
  {\bibfnamefont {M.}~\bibnamefont {Filatov}}, \ and\ \bibinfo {author}
  {\bibfnamefont {C.~H.}\ \bibnamefont {Choi}},\ }\href {\doibase
  10.1021/acs.jctc.0c01074} {\bibfield  {journal} {\bibinfo  {journal} {J.
  Chem. Theory Comput.}\ }\textbf {\bibinfo {volume} {17}},\ \bibinfo {pages}
  {848} (\bibinfo {year} {2021})}\BibitemShut {NoStop}%
\bibitem [{\citenamefont {Monino}\ \emph {et~al.}(2022)\citenamefont {Monino},
  \citenamefont {Boggio-Pasqua}, \citenamefont {Scemama}, \citenamefont
  {Jacquemin},\ and\ \citenamefont {Loos}}]{Monino_2022}%
  \BibitemOpen
  \bibfield  {author} {\bibinfo {author} {\bibfnamefont {E.}~\bibnamefont
  {Monino}}, \bibinfo {author} {\bibfnamefont {M.}~\bibnamefont
  {Boggio-Pasqua}}, \bibinfo {author} {\bibfnamefont {A.}~\bibnamefont
  {Scemama}}, \bibinfo {author} {\bibfnamefont {D.}~\bibnamefont {Jacquemin}},
  \ and\ \bibinfo {author} {\bibfnamefont {P.-F.}\ \bibnamefont {Loos}},\
  }\href {\doibase 10.1021/acs.jpca.2c02480} {\bibfield  {journal} {\bibinfo
  {journal} {J. Phys. Chem. A}\ }\textbf {\bibinfo {volume} {126}},\ \bibinfo
  {pages} {4664} (\bibinfo {year} {2022})}\BibitemShut {NoStop}%
\bibitem [{\citenamefont {Chilkuri}\ \emph {et~al.}(2021)\citenamefont
  {Chilkuri}, \citenamefont {Applencourt}, \citenamefont {Gasperich},
  \citenamefont {Loos},\ and\ \citenamefont {Scemama}}]{Chilkuri_2021}%
  \BibitemOpen
  \bibfield  {author} {\bibinfo {author} {\bibfnamefont {V.~G.}\ \bibnamefont
  {Chilkuri}}, \bibinfo {author} {\bibfnamefont {T.}~\bibnamefont
  {Applencourt}}, \bibinfo {author} {\bibfnamefont {K.}~\bibnamefont
  {Gasperich}}, \bibinfo {author} {\bibfnamefont {P.-F.}\ \bibnamefont {Loos}},
  \ and\ \bibinfo {author} {\bibfnamefont {A.}~\bibnamefont {Scemama}},\ }\href
  {\doibase 10.1016/bs.aiq.2021.04.001} {\bibfield  {journal} {\bibinfo
  {journal} {Adv. Quantum Chem.}\ }\textbf {\bibinfo {volume} {83}},\ \bibinfo
  {pages} {65} (\bibinfo {year} {2021})}\BibitemShut {NoStop}%
\bibitem [{\citenamefont {Maurice}\ and\ \citenamefont
  {Head-Gordon}(1996)}]{Maurice_1996}%
  \BibitemOpen
  \bibfield  {author} {\bibinfo {author} {\bibfnamefont {D.}~\bibnamefont
  {Maurice}}\ and\ \bibinfo {author} {\bibfnamefont {M.}~\bibnamefont
  {Head-Gordon}},\ }\href {\doibase 10.1021/jp952754j} {\bibfield  {journal}
  {\bibinfo  {journal} {J. Phys. Chem.}\ }\textbf {\bibinfo {volume} {100}},\
  \bibinfo {pages} {6131} (\bibinfo {year} {1996})}\BibitemShut {NoStop}%
\bibitem [{\citenamefont {Huron}, \citenamefont {Malrieu},\ and\ \citenamefont
  {Rancurel}(1973)}]{Huron_1973}%
  \BibitemOpen
  \bibfield  {author} {\bibinfo {author} {\bibfnamefont {B.}~\bibnamefont
  {Huron}}, \bibinfo {author} {\bibfnamefont {J.~P.}\ \bibnamefont {Malrieu}},
  \ and\ \bibinfo {author} {\bibfnamefont {P.}~\bibnamefont {Rancurel}},\
  }\href {\doibase 10.1063/1.1679199} {\bibfield  {journal} {\bibinfo
  {journal} {J. Chem. Phys.}\ }\textbf {\bibinfo {volume} {58}},\ \bibinfo
  {pages} {5745} (\bibinfo {year} {1973})}\BibitemShut {NoStop}%
\bibitem [{\citenamefont {Giner}, \citenamefont {Scemama},\ and\ \citenamefont
  {Caffarel}(2013)}]{Giner_2013}%
  \BibitemOpen
  \bibfield  {author} {\bibinfo {author} {\bibfnamefont {E.}~\bibnamefont
  {Giner}}, \bibinfo {author} {\bibfnamefont {A.}~\bibnamefont {Scemama}}, \
  and\ \bibinfo {author} {\bibfnamefont {M.}~\bibnamefont {Caffarel}},\ }\href
  {\doibase 10.1139/cjc-2013-0017} {\bibfield  {journal} {\bibinfo  {journal}
  {Can. J. Chem.}\ }\textbf {\bibinfo {volume} {91}},\ \bibinfo {pages} {879}
  (\bibinfo {year} {2013})}\BibitemShut {NoStop}%
\bibitem [{\citenamefont {Giner}, \citenamefont {Scemama},\ and\ \citenamefont
  {Caffarel}(2015)}]{Giner_2015}%
  \BibitemOpen
  \bibfield  {author} {\bibinfo {author} {\bibfnamefont {E.}~\bibnamefont
  {Giner}}, \bibinfo {author} {\bibfnamefont {A.}~\bibnamefont {Scemama}}, \
  and\ \bibinfo {author} {\bibfnamefont {M.}~\bibnamefont {Caffarel}},\ }\href
  {\doibase 10.1063/1.4905528} {\bibfield  {journal} {\bibinfo  {journal} {J.
  Chem. Phys.}\ }\textbf {\bibinfo {volume} {142}},\ \bibinfo {pages} {044115}
  (\bibinfo {year} {2015})}\BibitemShut {NoStop}%
\bibitem [{\citenamefont {Garniron}\ \emph {et~al.}(2018)\citenamefont
  {Garniron}, \citenamefont {Scemama}, \citenamefont {Giner}, \citenamefont
  {Caffarel},\ and\ \citenamefont {Loos}}]{Garniron_2018}%
  \BibitemOpen
  \bibfield  {author} {\bibinfo {author} {\bibfnamefont {Y.}~\bibnamefont
  {Garniron}}, \bibinfo {author} {\bibfnamefont {A.}~\bibnamefont {Scemama}},
  \bibinfo {author} {\bibfnamefont {E.}~\bibnamefont {Giner}}, \bibinfo
  {author} {\bibfnamefont {M.}~\bibnamefont {Caffarel}}, \ and\ \bibinfo
  {author} {\bibfnamefont {P.~F.}\ \bibnamefont {Loos}},\ }\href {\doibase
  10.1063/1.5044503} {\bibfield  {journal} {\bibinfo  {journal} {J. Chem.
  Phys.}\ }\textbf {\bibinfo {volume} {149}},\ \bibinfo {pages} {064103}
  (\bibinfo {year} {2018})}\BibitemShut {NoStop}%
\bibitem [{\citenamefont {Garniron}\ \emph {et~al.}(2017)\citenamefont
  {Garniron}, \citenamefont {Scemama}, \citenamefont {Loos},\ and\
  \citenamefont {Caffarel}}]{Garniron_2017}%
  \BibitemOpen
  \bibfield  {author} {\bibinfo {author} {\bibfnamefont {Y.}~\bibnamefont
  {Garniron}}, \bibinfo {author} {\bibfnamefont {A.}~\bibnamefont {Scemama}},
  \bibinfo {author} {\bibfnamefont {P.-F.}\ \bibnamefont {Loos}}, \ and\
  \bibinfo {author} {\bibfnamefont {M.}~\bibnamefont {Caffarel}},\ }\href
  {\doibase 10.1063/1.4992127} {\bibfield  {journal} {\bibinfo  {journal} {J.
  Chem. Phys.}\ }\textbf {\bibinfo {volume} {147}},\ \bibinfo {pages} {034101}
  (\bibinfo {year} {2017})}\BibitemShut {NoStop}%
\bibitem [{\citenamefont {Davidson}(1975)}]{Davidson_1975}%
  \BibitemOpen
  \bibfield  {author} {\bibinfo {author} {\bibfnamefont {E.~R.}\ \bibnamefont
  {Davidson}},\ }\href {\doibase 10.1016/0021-9991(75)90065-0} {\bibfield
  {journal} {\bibinfo  {journal} {J. Comput. Phys.}\ }\textbf {\bibinfo
  {volume} {17}},\ \bibinfo {pages} {87} (\bibinfo {year} {1975})}\BibitemShut
  {NoStop}%
\bibitem [{\citenamefont {Loos}\ \emph
  {et~al.}(2020{\natexlab{a}})\citenamefont {Loos}, \citenamefont {Scemama},
  \citenamefont {Boggio-Pasqua},\ and\ \citenamefont {Jacquemin}}]{Loos_2020}%
  \BibitemOpen
  \bibfield  {author} {\bibinfo {author} {\bibfnamefont {P.-F.}\ \bibnamefont
  {Loos}}, \bibinfo {author} {\bibfnamefont {A.}~\bibnamefont {Scemama}},
  \bibinfo {author} {\bibfnamefont {M.}~\bibnamefont {Boggio-Pasqua}}, \ and\
  \bibinfo {author} {\bibfnamefont {D.}~\bibnamefont {Jacquemin}},\ }\href
  {\doibase 10.1021/acs.jctc.0c00227} {\bibfield  {journal} {\bibinfo
  {journal} {J. Chem. Theory Comput.}\ }\textbf {\bibinfo {volume} {16}},\
  \bibinfo {pages} {3720} (\bibinfo {year} {2020}{\natexlab{a}})}\BibitemShut
  {NoStop}%
\bibitem [{\citenamefont {V{\'e}ril}\ \emph {et~al.}(2021)\citenamefont
  {V{\'e}ril}, \citenamefont {Scemama}, \citenamefont {Caffarel}, \citenamefont
  {Lipparini}, \citenamefont {Boggio-Pasqua}, \citenamefont {Jacquemin},\ and\
  \citenamefont {Loos}}]{Veril_2021}%
  \BibitemOpen
  \bibfield  {author} {\bibinfo {author} {\bibfnamefont {M.}~\bibnamefont
  {V{\'e}ril}}, \bibinfo {author} {\bibfnamefont {A.}~\bibnamefont {Scemama}},
  \bibinfo {author} {\bibfnamefont {M.}~\bibnamefont {Caffarel}}, \bibinfo
  {author} {\bibfnamefont {F.}~\bibnamefont {Lipparini}}, \bibinfo {author}
  {\bibfnamefont {M.}~\bibnamefont {Boggio-Pasqua}}, \bibinfo {author}
  {\bibfnamefont {D.}~\bibnamefont {Jacquemin}}, \ and\ \bibinfo {author}
  {\bibfnamefont {P.-F.}\ \bibnamefont {Loos}},\ }\href {\doibase
  10.1002/wcms.1517} {\bibfield  {journal} {\bibinfo  {journal} {WIREs Comput.
  Mol. Sci.}\ ,\ \bibinfo {pages} {e1517}} (\bibinfo {year}
  {2021})}\BibitemShut {NoStop}%
\bibitem [{\citenamefont {Stanton}\ and\ \citenamefont
  {Gauss}(1994)}]{Stanton_1994}%
  \BibitemOpen
  \bibfield  {author} {\bibinfo {author} {\bibfnamefont {J.~F.}\ \bibnamefont
  {Stanton}}\ and\ \bibinfo {author} {\bibfnamefont {J.}~\bibnamefont
  {Gauss}},\ }\href {\doibase 10.1063/1.468022} {\bibfield  {journal} {\bibinfo
   {journal} {J. Chem. Phys.}\ }\textbf {\bibinfo {volume} {101}},\ \bibinfo
  {pages} {8938} (\bibinfo {year} {1994})}\BibitemShut {NoStop}%
\bibitem [{\citenamefont {Kamiya}\ and\ \citenamefont
  {Hirata}(2006)}]{Muneaki_2006}%
  \BibitemOpen
  \bibfield  {author} {\bibinfo {author} {\bibfnamefont {M.}~\bibnamefont
  {Kamiya}}\ and\ \bibinfo {author} {\bibfnamefont {S.}~\bibnamefont
  {Hirata}},\ }\href {\doibase 10.1063/1.2244570} {\bibfield  {journal}
  {\bibinfo  {journal} {J. Chem. Phys.}\ }\textbf {\bibinfo {volume} {125}},\
  \bibinfo {pages} {074111} (\bibinfo {year} {2006})}\BibitemShut {NoStop}%
\bibitem [{\citenamefont {Bomble}\ \emph {et~al.}(2005)\citenamefont {Bomble},
  \citenamefont {Saeh}, \citenamefont {Stanton}, \citenamefont {Szalay},
  \citenamefont {K{\'a}llay},\ and\ \citenamefont {Gauss}}]{Bomble_2005}%
  \BibitemOpen
  \bibfield  {author} {\bibinfo {author} {\bibfnamefont {Y.~J.}\ \bibnamefont
  {Bomble}}, \bibinfo {author} {\bibfnamefont {J.~C.}\ \bibnamefont {Saeh}},
  \bibinfo {author} {\bibfnamefont {J.~F.}\ \bibnamefont {Stanton}}, \bibinfo
  {author} {\bibfnamefont {P.~G.}\ \bibnamefont {Szalay}}, \bibinfo {author}
  {\bibfnamefont {M.}~\bibnamefont {K{\'a}llay}}, \ and\ \bibinfo {author}
  {\bibfnamefont {J.}~\bibnamefont {Gauss}},\ }\href {\doibase
  10.1063/1.1884600} {\bibfield  {journal} {\bibinfo  {journal} {J. Chem.
  Phys.}\ }\textbf {\bibinfo {volume} {122}},\ \bibinfo {pages} {154107}
  (\bibinfo {year} {2005})}\BibitemShut {NoStop}%
\bibitem [{\citenamefont {Rowe}(1968)}]{Rowe_1968}%
  \BibitemOpen
  \bibfield  {author} {\bibinfo {author} {\bibfnamefont {D.~J.}\ \bibnamefont
  {Rowe}},\ }\href {\doibase 10.1103/RevModPhys.40.153} {\bibfield  {journal}
  {\bibinfo  {journal} {Rev. Mod. Phys.}\ }\textbf {\bibinfo {volume} {40}},\
  \bibinfo {pages} {153} (\bibinfo {year} {1968})}\BibitemShut {NoStop}%
\bibitem [{\citenamefont {Monkhorst}(1977)}]{Monkhorst_1977}%
  \BibitemOpen
  \bibfield  {author} {\bibinfo {author} {\bibfnamefont {H.~J.}\ \bibnamefont
  {Monkhorst}},\ }\href {\doibase https://doi.org/10.1002/qua.560120850}
  {\bibfield  {journal} {\bibinfo  {journal} {Int. J. Quantum Chem.}\ }\textbf
  {\bibinfo {volume} {12}},\ \bibinfo {pages} {421} (\bibinfo {year}
  {1977})}\BibitemShut {NoStop}%
\bibitem [{\citenamefont {Koch}\ \emph {et~al.}(1990)\citenamefont {Koch},
  \citenamefont {Jensen}, \citenamefont {Jorgensen},\ and\ \citenamefont
  {Helgaker}}]{Koch_1990}%
  \BibitemOpen
  \bibfield  {author} {\bibinfo {author} {\bibfnamefont {H.}~\bibnamefont
  {Koch}}, \bibinfo {author} {\bibfnamefont {H.~J.~A.}\ \bibnamefont {Jensen}},
  \bibinfo {author} {\bibfnamefont {P.}~\bibnamefont {Jorgensen}}, \ and\
  \bibinfo {author} {\bibfnamefont {T.}~\bibnamefont {Helgaker}},\ }\href
  {\doibase 10.1063/1.458815} {\bibfield  {journal} {\bibinfo  {journal} {J.
  Chem. Phys.}\ }\textbf {\bibinfo {volume} {93}},\ \bibinfo {pages} {3345}
  (\bibinfo {year} {1990})}\BibitemShut {NoStop}%
\bibitem [{\citenamefont {Stanton}\ and\ \citenamefont
  {Bartlett}(1993)}]{Stanton_1993}%
  \BibitemOpen
  \bibfield  {author} {\bibinfo {author} {\bibfnamefont {J.~F.}\ \bibnamefont
  {Stanton}}\ and\ \bibinfo {author} {\bibfnamefont {R.~J.}\ \bibnamefont
  {Bartlett}},\ }\href {\doibase 10.1063/1.464746} {\bibfield  {journal}
  {\bibinfo  {journal} {J. Chem. Phys.}\ }\textbf {\bibinfo {volume} {98}},\
  \bibinfo {pages} {7029} (\bibinfo {year} {1993})}\BibitemShut {NoStop}%
\bibitem [{\citenamefont {Loos}\ \emph {et~al.}(2018)\citenamefont {Loos},
  \citenamefont {Scemama}, \citenamefont {Blondel}, \citenamefont {Garniron},
  \citenamefont {Caffarel},\ and\ \citenamefont {Jacquemin}}]{Loos_2018}%
  \BibitemOpen
  \bibfield  {author} {\bibinfo {author} {\bibfnamefont {P.~F.}\ \bibnamefont
  {Loos}}, \bibinfo {author} {\bibfnamefont {A.}~\bibnamefont {Scemama}},
  \bibinfo {author} {\bibfnamefont {A.}~\bibnamefont {Blondel}}, \bibinfo
  {author} {\bibfnamefont {Y.}~\bibnamefont {Garniron}}, \bibinfo {author}
  {\bibfnamefont {M.}~\bibnamefont {Caffarel}}, \ and\ \bibinfo {author}
  {\bibfnamefont {D.}~\bibnamefont {Jacquemin}},\ }\href {\doibase
  10.1021/acs.jctc.8b00406} {\bibfield  {journal} {\bibinfo  {journal} {J.
  Chem. Theory Comput.}\ }\textbf {\bibinfo {volume} {14}},\ \bibinfo {pages}
  {4360} (\bibinfo {year} {2018})}\BibitemShut {NoStop}%
\bibitem [{\citenamefont {Loos}\ \emph
  {et~al.}(2020{\natexlab{b}})\citenamefont {Loos}, \citenamefont {Lipparini},
  \citenamefont {Boggio-Pasqua}, \citenamefont {Scemama},\ and\ \citenamefont
  {Jacquemin}}]{Loos_2020a}%
  \BibitemOpen
  \bibfield  {author} {\bibinfo {author} {\bibfnamefont {P.~F.}\ \bibnamefont
  {Loos}}, \bibinfo {author} {\bibfnamefont {F.}~\bibnamefont {Lipparini}},
  \bibinfo {author} {\bibfnamefont {M.}~\bibnamefont {Boggio-Pasqua}}, \bibinfo
  {author} {\bibfnamefont {A.}~\bibnamefont {Scemama}}, \ and\ \bibinfo
  {author} {\bibfnamefont {D.}~\bibnamefont {Jacquemin}},\ }\href {\doibase
  10.1021/acs.jctc.9b01216} {\bibfield  {journal} {\bibinfo  {journal} {J.
  Chem. Theory Comput.}\ }\textbf {\bibinfo {volume} {16}},\ \bibinfo {pages}
  {1711} (\bibinfo {year} {2020}{\natexlab{b}})}\BibitemShut {NoStop}%
\end{thebibliography}%

\end{document}